\documentclass[preprint,12pt,authoryear]{elsarticle}

\usepackage{amssymb}
\usepackage{amsmath}

\usepackage[utf8]{inputenc} 
\usepackage[T1]{fontenc}    
\usepackage{hyperref}       
\usepackage{url}            
\usepackage{booktabs}       
\usepackage{amsfonts}       
\usepackage{nicefrac}       
\usepackage{microtype}      
\usepackage{xcolor}         
\definecolor{darkgreen}{rgb}{0.0, 0.5, 0.0}
\definecolor{darkgray}{rgb}{0.35, 0.35, 0.35}  
\definecolor{verylightgray}{rgb}{0.925, 0.925, 0.925}  

\usepackage{amsmath,amssymb,amsfonts}
\usepackage{multirow}
\usepackage{multicol}
\usepackage{makecell}
\usepackage{pifont}
\usepackage{stfloats}
\usepackage{longtable}
\usepackage{pdflscape}
\usepackage{afterpage}
\usepackage{amsmath}
\usepackage{bm}
\usepackage{xurl}

\usepackage[most]{tcolorbox}

\usepackage{algorithm}
\usepackage{algpseudocode}

\usepackage{lipsum}
\makeatletter
\newenvironment{breakablealgorithm}
{
		\begin{center}
			\refstepcounter{algorithm}
			\hrule height.8pt depth0pt \kern2pt
			\renewcommand{\caption}[2][\relax]{
				{\raggedright\textbf{\fname@algorithm~\thealgorithm} ##2\par}%
				\ifx\relax##1\relax 
				\addcontentsline{loa}{algorithm}{\protect\numberline{\thealgorithm}##2}%
				\else 
				\addcontentsline{loa}{algorithm}{\protect\numberline{\thealgorithm}##1}%
				\fi
				\kern2pt\hrule\kern2pt
			}
		}{
		\kern2pt\hrule\relax
	\end{center}
}
\makeatother

\usepackage{graphicx}
\usepackage{textcomp}

\usepackage{subcaption}
\usepackage{caption}

\usepackage{arydshln}
\usepackage{array}
\newcolumntype{C}[1]{>{\centering\arraybackslash}p{#1}}

\usepackage[most]{tcolorbox}

\journal{Knowledge-Based Systems}

\begin{document}

\begin{frontmatter}

\title{Enhancing Large Language Model-Based Systems for End-to-End Circuit Analysis Problem Solving}

\author{Liangliang Chen\fnref{cofirst}}
\author{Weiyu Sun\fnref{cofirst}}
\author{Huiru Xie\fnref{cofirst}}
\author{Yongnuo Cai}
\author{Ying Zhang\corref{mycorrespondingauthor}}
\address{School of Electrical and Computer Engineering,
	Georgia Institute of Technology,\\ Atlanta, GA 30332 USA}
\fntext[cofirst]{These authors contributed equally to this work.}
\cortext[mycorrespondingauthor]{Corresponding author}
\ead{yzhang@gatech.edu}

\begin{abstract}
Large language models (LLMs) have demonstrated strong performance in data-rich domains such as programming, yet their reliability in engineering tasks remains limited. Circuit analysis, requiring multimodal understanding and precise mathematical reasoning, highlights these challenges. Although Gemini 2.5 Pro has improved capabilities in diagram interpretation and analog-circuit reasoning, it still struggles to consistently solve problems containing both textual descriptions and circuit diagrams. This paper presents an enhanced end-to-end circuit problem-solving framework using Gemini 2.5 Pro as the backbone model for scalable engineering-education applications. We systematically evaluate Gemini 2.5 Pro on undergraduate circuit-analysis problems and characterize two key failure modes: 1) circuit-recognition hallucinations, particularly incorrect source polarity detection, and 2) reasoning-process hallucinations, such as incorrect current direction assumptions. To address recognition errors, we integrate a fine-tuned YOLO detector and OpenCV-based processing to isolate voltage and current sources for polarity re-identification. To mitigate reasoning errors, we introduce an \texttt{ngspice}-driven verification loop that triggers iterative solution refinement with optional human-in-the-loop feedback. On 83 problems, the proposed pipeline achieves 97.59\% accuracy versus Gemini's 79.52\% baseline. On four hand-drawn diagram variations, accuracy improves from {\}60.61\%---71.21\% to 89.39\%---92.42\% with statistically significant gains ($p<0.005$)}. An additional evaluation on 43 problems from a different textbook improves accuracy from 58.14\% to 83.72\%, providing further evidence of cross-textbook generalizability. Error analysis indicates that circuit recognition remains the dominant source of remaining failures, particularly for different diagram representations. These results demonstrate the robustness, scalability, generalizability, and practical applicability of the proposed framework for engineering education and real-world circuit analysis.
\end{abstract}



\end{frontmatter}

\section{Introduction}

Large language models (LLMs), with their rapid recent development \citep{achiam2023gpt, liu2024deepseek, comanici2025gemini}, are revolutionizing many fields, including programming \citep{austin2021program, guo2024deepseek}, mathematics \citep{shao2024deepseekmath, ahn2024large}, robotics \citep{chen2024rlingua}, clinical medicine \citep{li2025reviewing}, and education \citep{milano2023large, kasneci2023chatgpt, kumar2025math}. In education, supported by multi-modal inputs and outputs---such as text and audio---and enriched with abundant prior knowledge, LLMs have the potential to provide personalized support to students \citep{wang2024large, sharma2025role}, thereby reducing teachers' workload \citep{sun2025data}.

Reliability is a critical concern when applying LLMs in educational contexts, as incorrect responses due to hallucinations may mislead students in their learning \citep{wang2024large, huang2025survey}. Consequently, LLMs have seen wider adoption in educational disciplines where they were trained on abundant data, such as programming \citep{raihan2025large}, mathematics \citep{kumar2025math}, language teaching \citep{caines2023application, gao2024exploring}, and K-12 subjects \citep{wu2024analyzing}, in which LLMs tend to perform well intrinsically. However, there remains a lack of research exploring the use of LLMs in broader engineering domains for college students---circuit analysis being one such example. A recent benchmarking study \citep{skelic2025circuit} reports that GPT-4o achieved only 48.04\% accuracy when evaluated on the final numerical answers across 510 undergraduate circuit problems, consisting of 5 problems each from 102 templates. Notably, this accuracy was measured using the pass@3/5 criterion, meaning that at least 3 out of 5 answers had to be correct for a given template to count as a pass. These results suggest that GPT-4o currently cannot solve undergraduate circuit analysis problems with sufficient reliability. 

Although GPT-4o cannot independently solve circuit analysis problems with high accuracy, it demonstrates more reliable performance in assessing students' submitted homework when provided with sufficient context, such as detailed problem descriptions and reference solutions \citep{chen2025benchmarking, chen2025enhancing}. Building on this observation, an AI-enabled tutor was developed and deployed in an undergraduate circuit analysis course to deliver personalized homework feedback and answer student queries \citep{chen2025wip}. However, the manual preparation of the required contextual data is labor-intensive and limits the scalability of this approach to other problems and subjects. This limitation underscores the need for a reliable automated problem solver, which is the focus of this study.

In engineering disciplines, leveraging LLMs for problem-solving poses significant challenges due to the high demands for multimodal understanding, precise reasoning, and domain-specific knowledge. For instance, solving a circuit analysis problem \citep{svoboda2013introduction} requires an LLM to interpret circuit diagrams, select appropriate analytical methods, and carry out accurate mathematical and circuit-based reasoning. A correct solution hinges on the successful completion of each of these steps. Recent circuit-specific benchmarking further illustrates these challenges. \cite{akbari2026circuitsense} introduced CircuitSense, a hierarchical benchmark that evaluates eight state-of-the-art MLLMs across circuit perception, analysis, and design tasks. Among the evaluated models, Gemini 2.5 Pro demonstrated the strongest overall performance, tying or leading the perception subtasks---including component detection, connection identification, and function classification---and outperforming the other models across all reported circuit-analysis subcategories. Nevertheless, its performance decreased substantially on synthetic problems requiring symbolic equation derivation, revealing a considerable gap between circuit-diagram perception and reliable mathematical reasoning. These findings motivate our selection of Gemini 2.5 Pro as a strong foundation model while simultaneously highlighting the need for additional mechanisms to improve the reliability of end-to-end circuit problem solving. In our preliminary evaluation on undergraduate textbook problems, Gemini 2.5 Pro achieves a zero-shot accuracy of 79.52\%, providing a sufficiently capable baseline for systematically identifying and addressing its remaining failure modes.

The evaluation results indicate that Gemini 2.5 Pro tends to exhibit hallucinations due to incorrect recognition of source polarities, misinterpretation of current directions and element connections, and errors in circuit analysis. These issues can be broadly categorized into circuit diagram recognition-related and circuit analysis-related errors. This separation between perception and analytical reasoning is consistent with the perception--reasoning gap observed by \cite{akbari2026circuitsense}. When a vanilla LLM struggles with solving certain tasks, common strategies to enhance the performance of LLM-based systems include providing more detailed instructions in the prompts, leveraging external knowledge, or using auxiliary tools. These approaches are known respectively as prompt engineering \citep{sahoo2024systematic, wei2022chain, son2025multi}, retrieval-augmented generation (RAG) \citep{lewis2020retrieval, gao2023retrieval, siddharth2024retrieval}, and function calling \citep{li2024large}. Motivated by these general ideas, various techniques can be employed to improve solution accuracy, such as step-by-step prompting \citep{wei2022chain}, external image recognition tools like YOLO \citep{redmon2016you} and OpenCV \citep{bradski2008learning}, and SPICE-based circuit simulation tools \citep{tuinenga1988spice, vogt2021ngspice}. However, it remains unclear how to integrate these methods and tools into a maximally automated framework for solving undergraduate circuit analysis problems. \textit{To the best of our knowledge, we are the first to develop an end-to-end circuit problem-solving framework that covers diverse undergraduate-level circuit topics, accepts both circuit diagrams and textual problem descriptions, automatically integrates perception correction with simulation-based verification, and produces natural-language solutions with intermediate reasoning steps.}

To resolve the identified gaps and practical needs, this paper presents an end-to-end circuit problem solver based on Gemini 2.5 Pro that reliably generates both the step-by-step problem-solving process and final answers. The main contributions of this work are summarized as follows:

\begin{itemize}
	\item[i)] We first systematically characterize the failure modes of Gemini 2.5 Pro in solving undergraduate circuit analysis problems with textbook-quality diagrams. Although recent benchmarking studies have demonstrated its strong overall circuit-understanding capabilities, our problem-level evaluation identifies specific failure modes that limit reliable end-to-end problem solving. Based on the evaluation results, we identify and classify the sources of error in Gemini's responses into two main categories: 1) hallucinations in element connections, where incorrect recognition of source polarities is a primary issue; and 2) hallucinations in circuit analysis, in which the incorrect use of current directions during the problem-solving process is a key source of error. To address these issues, we propose enhancement methods categorized into circuit recognition enhancement and problem-solving enhancement, based on the identified hallucination types.
	
	\item[ii)] To mitigate vision hallucinations related to source polarities, we fine-tune a YOLO model \citep{redmon2016you} to detect the bounding box positions of independent voltage and current sources, and leverage OpenCV \citep{bradski2008learning} to extract this information for dependent sources. Using the bounding box data, sources are cropped from the circuit diagram, and their polarities are re-identified by Gemini. Since the cropped images contain only the sources---excluding other parts of the circuit diagram---Gemini 2.5 Pro is able to reliably recognize their polarities. This locally recognized polarity information is then used to correct the source polarities in Gemini's original global recognition, effectively eliminating hallucinations in source polarity identification.
	
	\item[iii)] To address errors in the circuit analysis process, we employ the simulation tool \texttt{ngspice} \citep{vogt2021ngspice} to generate solution benchmarks by simulating the circuit and extracting the required variables. For each circuit problem, Gemini 2.5 Pro generates a corresponding \texttt{.cir} file, which serves as the input to \texttt{ngspice}. Since \texttt{ngspice} operates purely on circuit structure, its results are reliable provided the circuit is correctly recognized and Gemini accurately translates the textual description into a valid \texttt{.cir} file. The simulation results are compared automatically with Gemini's outputs. If they match, Gemini's solution is accepted. In the case of a mismatch, both Gemini and \texttt{ngspice} are rerun to correct potential generation errors. For problems with persistent mismatches, a human-in-the-loop approach is employed to verify the accuracy of both circuit recognition and Gemini’s problem-solving process. Experimental results show that, in the case of textbook-quality circuit diagrams, the proposed pipeline successfully solves 81 out of 83 problems (97.59\%) across a range of circuit analysis topics, representing a significant improvement over the original Gemini 2.5 Pro's 79.52\% success rate. Ablation analysis further demonstrates the complementary roles of the individual framework components, showing that localized vision preprocessing is important for preventing circuit-recognition errors from propagating into both LLM-based reasoning and \texttt{ngspice} simulation.\footnote{There are a total of 83 circuit problems considered in our problem set, of which 66 are suitable for \texttt{ngspice} simulation, as indicated in Table \ref{TableA1}. When describing the mismatches between Gemini- and \texttt{ngspice}-based solutions, we consider only the 66 problems suitable for both \texttt{ngspice} simulation and Gemini-based problem-solving.}
	
	\item[iv)] To demonstrate the generalizability and robustness of the proposed pipeline on problems with hand-drawn circuit diagrams, we constructed four variations of circuit diagrams for the 66 problems in our dataset that are suitable for both Gemini 2.5 Pro problem solving and \texttt{ngspice} verification. The test results indicate that, while the baseline Gemini 2.5 Pro with a simple problem-solving prompt achieves only 60.61\%---71.21\% accuracy on problems with hand-drawn circuit diagrams, the proposed multi-step framework significantly improves performance to 89.39\%---92.42\%, approaching that achieved with textbook-quality circuit diagrams. To further evaluate problem-level generalizability, we test the framework on an additional set of 43 circuit problems selected from a different textbook, where it improves problem-solving accuracy from 58.14\% with direct Gemini prompting to 83.72\%. Exact binomial tests further confirm that these performance gains are statistically significant across all six evaluation cases, with all $p$-values below 0.005 and five of the six below 0.001. A systematic error analysis further shows that circuit recognition remains the dominant source of the residual failures, particularly for problems with larger variation in circuit-diagram representation. The average time and LLM cost per problem further indicate that the proposed method is both time- and cost-efficient and thus more scalable than manual problem solving, highlighting its practical applicability in real-world circuit problem-solving tasks.
\end{itemize}

From another perspective, the rapid development of LLMs holds significant potential to transform the field of education. However, building a versatile AI-enabled educational tool requires access to a substantial amount of high-quality data, the creation of which is labor-intensive if generated or labeled manually. Unlike AI-driven tutoring systems in domains such as programming---where data formats are highly structured and large-scale, high-quality datasets are readily available---engineering disciplines face notable challenges due to the scarcity of such data. Developing a reliable tool for circuit analysis problem-solving can help bridge this gap by enabling the automated generation of high-quality circuit analysis solutions using LLMs in conjunction with external assistant tools. These generated solutions can support a range of educational applications, including question answering, problem recommendation, and evaluation of student-submitted work \citep{chen2025benchmarking, chen2025enhancing, chen2025wip, xie2026leveraging}. Moreover, the circuit analysis problem-solving framework presented in this paper serves as an early example of how LLMs can be effectively integrated into broader domains of engineering education. More broadly, it illustrates how existing external tools---widely developed across various engineering disciplines---can be leveraged to detect hallucinations in LLM outputs and enhance the reliability of LLM-based systems. Although Gemini 2.5 Pro is used as the backbone model in this study, the modular design of the proposed framework---combining specialized visual processing, structured circuit representation, physical simulation verification, and human-in-the-loop refinement---is fundamentally model-agnostic and can be extended to other multimodal LLMs as their circuit-recognition capabilities improve. Thus, this work extends the practical capabilities of LLMs through system-level augmentation and lays a promising foundation for their broader adoption in engineering education.

\section{Related Work}

This section reviews related work across several key areas: large language models in education, tutoring platforms for engineering education, LLM benchmarking in engineering domains, with particular emphasis on electrical engineering, tool-augmented LLMs, interactive critiquing and decision making, and multimodal diagram interpretation and modular collaboration.

\textbf{Large Language Models in Education:} Due to their rich encoded prior knowledge and natural language processing capabilities, LLMs have been widely discussed and employed as educational assistants \citep{kasneci2023chatgpt, wang2024large, jeon2023large, yan2024practical}. Applications of LLMs in education span various subjects and levels, from K-12 to higher education \citep{wu2024analyzing, yigci2025large}. For instance, \cite{dan2023educhat} developed an LLM-based chatbot system to support personalized education. The system offers functions such as question-answering and essay assessment, while also incorporating Socratic teaching methods and emotional support. \cite{song2024automated} leveraged open-source LLMs to achieve automated essay scoring and revision. Using a dataset of 600 human-scored essays, the authors found that open-source models with approximately 10B parameters can effectively perform essay scoring and revision when appropriately prompted. From the perspective of assisting educators, \cite{ravi2025co} explored how LLMs can support K-12 interdisciplinary teachers in implementing project-based learning by automating routine tasks and promoting personalized learning. At the college level, \cite{liffiton2023codehelp} designed an LLM-powered tool with built-in guardrails to provide scalable student support without directly revealing solutions. Deployment in a first-year undergraduate computer and data science course showed that the tool was well received by both students and instructors. \cite{gupta2025beyond} evaluated the capabilities of various LLMs---including GPT-3.5 Turbo, GPT-4, GPT-4o, OpenAI’s o1 mini, and o1 preview---in solving and tutoring college algebra problems, finding that GPT-4o achieved the highest accuracy in both final answers (97.3\%) and interactive prompting-based evaluations (93.3\%). However, it remains largely unexplored how LLMs can be leveraged to support college-level engineering courses---such as circuit analysis---while requiring \textit{minimal instructional material preparation}.

\textbf{Tutoring Platforms for Engineering Education:} Before the era of LLMs, numerous online tutoring platforms for engineering education had already been developed \citep{grey2020perceptions, blummer2009best}. However, due to the absence of intelligent natural language processing tools, these systems were largely rule-based, heavily dependent on pre-prepared data, and constrained by strict usage requirements. For example, a rule-based circuit tutoring system was developed \citep{skromme2015step, skromme2020interactive, skromme2024step}, covering a wide range of undergraduate circuit analysis topics. In this framework, users must type their solutions using a predefined template, limiting the naturalness of interaction. Moreover, problem creation does not support pixel-based diagram inputs; instead, users must construct circuit diagrams according to rigid formatting rules. The advent of LLMs has made natural language and image-based interactions possible, leading to more advanced and user-friendly tutoring systems in areas such as software engineering \citep{frankford2024ai}, chemical engineering \citep{caccavale2024fermentai}, and electrical engineering \citep{chen2025wip, knievel2025aitee,xie2026leveraging}. In electrical engineering education specifically, \cite{knievel2025aitee} designed a tutoring framework for circuit analysis in which circuit diagrams are first converted into netlists, then solved by an LLM and validated using PySpice. The authors also incorporated a RAG-based approach to strengthen the reliability of the LLM’s responses. However, their system is limited to resistor-based circuits containing only resistors and independent sources, representing a small subset of undergraduate circuit problems. \cite{chen2025wip} proposed an LLM-enabled smart tutor informed by their prior evaluations \citep{chen2025benchmarking} and enhancement methods \citep{chen2025enhancing}. While effective, this framework requires instructors to prepare extensive homework materials---including detailed reference solutions and final answers---to ensure reliable feedback and responses to open-ended questions. Such preparation is labor-intensive and imposes significant workload on instructors. Thus, in this paper, we develop an end-to-end circuit problem solver that accepts diagrams and problem statements as inputs and produces natural-language solutions with intermediate steps and detailed reasoning.

\textbf{LLM Benchmarking in Engineering Domains:} Unlike programming and mathematical reasoning, solving engineering problems often requires domain-specific knowledge, advanced reasoning capabilities, and diagram comprehension, making such tasks more challenging for LLMs \citep{zhou2025engibench}. Several studies have benchmarked LLMs across different engineering domains. For example, \cite{mudur2025feabench} introduced a benchmark to evaluate LLMs' abilities to simulate and solve problems in physics, mathematics, and engineering. They proposed a pipeline that integrates COMSOL Multiphysics\textsuperscript{\textregistered} via an LLM-enabled application programming interface. \cite{zhou2024elecbench} presented a benchmark specifically designed for assessing LLM performance in power-system operation tasks, structured around six primary metrics (i.e., factuality, logicality, stability, security, fairness, and expressiveness) and 24 secondary metrics. The authors constructed a mixed dataset of simulation-generated and real-world text data, applied it to multiple LLMs, and highlighted both the capabilities and limitations of these models in tasks such as grid dispatch, monitoring, and blackout recovery. \cite{Li_2025_CVPR} developed a comprehensive benchmark for evaluating LLMs and large multi-modal models across 10 electrical and electronics engineering domains, including electric circuits, analog circuits, control systems, etc. Their findings showed that LLMs struggle to solve problems in these areas. Notably, the evaluated models did not include more advanced models such as Gemini 2.5 Pro and GPT-5. \cite{skelic2025circuit} demonstrated that GPT-4o is unreliable in solving undergraduate-level circuit analysis problems, using a dataset derived and adapted from relevant MIT courses. More recently, \cite{akbari2026circuitsense} introduced CircuitSense, a hierarchical benchmark comprising more than 8,000 problems spanning circuit perception, analysis, and design across multiple abstraction levels, from resistor networks to system-level block diagrams. The benchmark explicitly evaluates visual-to-mathematical reasoning through tasks including component detection, connection identification, function classification, circuit analysis, and symbolic equation derivation. Among the eight evaluated MLLMs, Gemini 2.5 Pro achieved the strongest overall performance, exhibiting particularly strong circuit-perception capabilities and outperforming the other models across all reported analysis subcategories. However, the benchmark also revealed a substantial gap between visual perception and symbolic reasoning: while Gemini 2.5 Pro achieved strong performance on curated circuit problems, its overall accuracy decreased to 19.06\% on synthetic problems requiring explicit symbolic equation derivation. These findings provide circuit-specific evidence that even leading MLLMs remain insufficiently reliable for autonomous circuit reasoning and motivate the integration of external verification and error-correction mechanisms. Importantly, existing studies such as CircuitSense primarily focus on benchmarking and diagnosing MLLM circuit-understanding capabilities rather than developing an end-to-end educational problem-solving system that automatically detects and corrects perception and reasoning errors. Aside from \cite{knievel2025aitee}, which focused solely on resistor-based circuits, no other work has proposed LLM-enabled circuit solvers to the best of our knowledge. A robust problem-solving pipeline capable of handling circuit problems involving advanced topics and components---beyond just resistors and independent sources as in \cite{knievel2025aitee}---remains absent. 

\textbf{Tool-Augmented LLMs, Interactive Critiquing, and Decision Making:} While LLMs exhibit impressive zero-shot reasoning capabilities \citep{achiam2023gpt, liu2024deepseek, comanici2025gemini}, recent studies have shown that they are fundamentally limited when tasked with self-verification and complex, multi-step decision-making without external support \citep{gou2024critic, qin2024toolllm}. \cite{gou2024critic} introduced CRITIC, demonstrating that black-box LLMs struggle to evaluate their own correctness (i.e., ``they do not know what they know'') and that intrinsic self-correction often fails or even degrades performance unless grounded by external tool interactions (e.g., search engines, Python interpreters). Similarly, \cite{qin2024toolllm} developed ToolLLM, scaling LLM tool-use to over 16,000 real-world APIs and introducing depth-first search-based decision tree planning. Their work highlights that linear reasoning paths, such as standard CoT \citep{wei2022chain} and ReAct \citep{yao2023react}, suffer from error propagation, and that efficient tool-augmented problem solving requires structured exploration, execution feedback, and decision retraction. In scientific and engineering problem solving, where analytical derivations can be verified against rigorous mathematical and physical simulators, tool integration provides an essential grounding mechanism. Our work builds upon these insights by coupling Gemini 2.5 Pro with an external \texttt{ngspice} simulator in a closed-loop verification and retry architecture, turning simulation mismatches into actionable feedback for iterative solution refinement.

\textbf{Multimodal Diagram Interpretation and Modular Collaboration:} Interpreting domain-specific diagrams (e.g., geometric layouts, functional curves, and schematic diagrams) remains a major bottleneck for vision-language models (VLMs) \citep{zhang2024mathverse,zhou2024visual}. \cite{zhang2024mathverse} introduced MathVerse to evaluate whether multimodal LLMs truly comprehend visual diagrams, revealing that models often rely heavily on textual redundancies in the prompt rather than the input imagery. When forced to interpret diagrams with minimal textual cues, VLM accuracy drops substantially, with visual perception errors (e.g., misidentifying key visual components and structural properties) accounting for the vast majority of failures. To address cross-modal representation gaps in VLMs, \cite{zhou2024visual} proposed visual in-context learning, demonstrating that transforming complex visual representations into structured, intent-oriented linguistic summaries alleviates the cross-modal alignment bottleneck and significantly improves reasoning accuracy over raw image inputs. Furthermore, when dealing with complex, heterogeneous multimodal tasks, unified end-to-end models often lack the flexibility and granularity needed for domain-specific accuracy. Addressing this, \cite{zhou2025mam} proposed MAM, a modular multi-agent framework for multimodal medical diagnosis. MAM demonstrates that decomposing complex problem-solving workflows into role-specialized collaborative agents substantially outperforms monolithic multimodal LLMs. Inspired by these findings, our circuit problem-solving pipeline departs from a naive, single-pass VLM paradigm. We design a modular architecture that explicitly decouples local visual element extraction (using fine-tuned YOLO and OpenCV), structured circuit netlist description synthesis, SPICE-based physical simulation, analytical problem solving, and consensus-driven human-in-the-loop intervention.

\newpage
\section{Circuit Analysis Problem-Solving Framework Design}
\label{S3}
This section outlines the proposed pipeline for a reliable circuit analysis problem solver. Section \ref{S31} presents preliminary evaluation results of the original Gemini 2.5 Pro in solving circuit analysis problems. Section \ref{S32} details enhancements in circuit recognition using a fine-tuned YOLO model and OpenCV, along with a multi-step targeted prompting strategy for robust problem-solving. Section \ref{S33} describes the result verification process using \texttt{ngspice}. Finally, Section \ref{S34} explains the incorporation of human-in-the-loop interventions to correct errors in both circuit diagram recognition and problem-solving.

\subsection{Preliminaries --- Original Gemini 2.5 Pro's Results in Circuit Analysis Problem Solving}
\label{S31}
This section presents our preliminary evaluation of the original Gemini 2.5 Pro model on circuit analysis problem solving with textbook-quality diagrams, which motivates the development of our enhancement methods. The problem set is drawn from an undergraduate-level circuit analysis course in the School of Electrical and Computer Engineering at the Georgia Institute of Technology. It comprises 86 problems covering a range of topics, including electric variables and elements, operational amplifiers, energy storage elements, and frequency response. All problems are adapted from those in \cite{svoboda2013introduction}. For this study, we selected 83 of the 86 problems, excluding three related to Bode plots in the frequency response section. These were omitted because, unlike other problem types, Bode plot solutions are difficult to assess automatically using Gemini. Detailed information on the problem indices, the number of problems in each topic, and the minimal adaptations applied to the original problem statements is provided in \ref{SA}.

The preliminary test utilizes the Gemini API\footnote{See \url{https://ai.google.dev/gemini-api/docs}.}
to obtain solutions. For each problem, both the textual problem statement and the accompanying circuit diagram (if available) are directly provided to Gemini 2.5 Pro, which is prompted to solve the problem without any additional instructions. Out of 83 problems, Gemini 2.5 Pro correctly solved 66, resulting in an accuracy rate of 79.52\%. A solution is considered correct if the final answer generated by Gemini---whether in the form of numerical values, time-varying functions, or mathematical expressions---matches the reference final answer. Table \ref{table:1} lists the problems that Gemini 2.5 Pro answered incorrectly, categorized into four classes based on the reasons for incorrectness, as detailed below.

\begin{table}[!t]
	\centering
	\footnotesize
	\caption{Results of the original Gemini 2.5 Pro (\texttt{gemini-2.5-pro})}
	\label{table:1}
	\begin{tabular}{ccc}
		\hline
		Reason for Incorrectness & Ratio of Incorrect Problems & Incorrect Problems \\
		\hline
		\makecell{Incorrect recognitions\\ of source polarities} & 6/17 (35.29\%) & \makecell[l]{P 3.4-4, P 4.2-2, P 6.3-3,\\ P 6.4-1, \textit{P 6.4-6}, P 9.6-1} \\ \hline
		\makecell{Hallucinations of\\current directions} & 5/17 (29.41\%) & \makecell[l]{P 4.6-2, P 4.7-6, P 5.2-1, \\ P 6.3-2, P 10.8-1} \\ \hline
		\makecell{Hallucinations of\\element connections} & 2/17 (11.76\%) & \makecell[l]{P 5.2-3, P 6.4-5} \\ \hline
		\makecell{Hallucinations of\\circuit analyses} & 5/17 (29.41\%) & \makecell[l]{P 3.2-2, \textit{P 6.4-6}, P 8.3-4,\\ P 9.5-1, P 13.2-3} \\ 
		\hline\hline
		Overall Correct Ratio & \multicolumn{2}{c}{66/83 (79.52\%)} \\
		\hline
		\multicolumn{3}{l}{\parbox{13.15cm}{* The problem index in \textit{italic} font (i.e., \textit{P 6.4-6}) indicates that there are multiple reasons for the incorrectness.}}
	\end{tabular}
\end{table}

\begin{itemize}
	\item [--] \textit{Incorrect Recognition of Source Polarities:} Circuit sources have defined polarities, which are indicated in circuit diagrams. For instance, a voltage source uses “+” and “–” signs to denote its positive and negative terminals, respectively. Similarly, the arrow in a current source indicates the direction of current flow, thereby establishing the source's polarity. In our preliminary tests, we observed that Gemini 2.5 Pro may fail to correctly recognize source polarities, sometimes interpreting them in reverse when the circuit diagram is provided directly. Consequently, if the source polarities are misinterpreted, the problem is unlikely to be solved correctly.
	\item [--] \textit{Hallucinations of Current Directions:} In some circuit diagrams, one or more currents may be labeled, showing their names and directions. For example, certain circuit problems may involve mesh current analysis, and the associated diagrams would include labeled mesh currents. In such cases, Gemini 2.5 Pro sometimes hallucinates the current directions, preventing it from solving the problem correctly.
	\item [--] \textit{Hallucinations of Element Connections:} Although Gemini 2.5 Pro has relatively strong vision capabilities, it may fail to correctly recognize element connections when a circuit diagram contains many components or when the connection relationships are complex. Such misinterpretations often result in incorrect solutions.
	\item [--] \textit{Hallucinations in Circuit Analysis:} In addition to recognition-related hallucinations, Gemini 2.5 Pro may also exhibit hallucinations during the reasoning process when solving problems. These reasoning errors can likewise lead to incorrect solutions.
\end{itemize}

The proposed enhancement method for the LLM-based circuit solver is motivated by our observations of four types of hallucinations exhibited by Gemini 2.5 Pro. Specifically,
\begin{itemize}
	\item [i)] Since incorrect recognition of source polarities accounts for 35.29\% of the incorrectly solved problems, we first enhance source polarity recognition by extracting inset source images with external computer vision tools and then using the local results to correct the global recognitions. We also design specific prompts to guide Gemini 2.5 Pro in identifying the directions of labeled currents. The methodological details are provided in Section \ref{S32}.
	
	\item [ii)] There is always a nonzero probability of hallucinations in the current state-of-the-art LLMs. Manually identifying hallucinations and verifying the correctness of LLM-generated solutions is labor-intensive. To automate this process, we construct a pipeline that leverages \texttt{ngspice} for independent verification. Problems with mismatched solutions are further processed and resolved, and, if mismatches persist, they are ultimately brought to human attention. Details of this pipeline are described in Section \ref{S33}.
	
	\item [iii)] With enhanced circuit recognition and \texttt{ngspice} verification, Gemini 2.5 Pro can solve most circuit problems after several trials, as shown in Section \ref{S4}. For the remaining problems, as long as they are brought to human attention, a competent human assistant can correct the solutions either by fixing diagram recognition errors or by providing feedback to the problem-solving process. This human-in-the-loop correction method is presented in Section \ref{S33}.
\end{itemize}

\subsection{Circuit Recognition Enhancements}
\label{S32}
Correct circuit recognition and understanding is the first essential step for LLMs to accurately solve circuit analysis problems. If errors occur during the circuit recognition stage, the resulting solution is likely to be incorrect, as the problem-solving steps may not align with the actual problem. Building on the preliminary testing results presented in Section \ref{S31}, this section introduces two methods aimed at enhancing the circuit recognition capabilities of Gemini 2.5 Pro, with a particular focus on improving the identification of source polarities and current directions. 

\subsubsection{Source Detection Using External Computer Vision Tools and Correction via Inset Source Images}
\label{S321}
The core idea behind enhancing source polarity recognition lies in the observation that VLMs, such as Gemini 2.5 Pro, are more prone to hallucinations when presented with images containing excessive information, which can cause the model to lose focus. In circuit analysis scenarios, Gemini 2.5 Pro may produce incorrect source polarity recognition when given the entire circuit image---comprising all elements, connection wires, and labels. However, polarity recognition is consistently accurate when the image contains only a single source, typically provided as an inset extracted from the full diagram. Based on this observation, when sources are present in a diagram, the process of improving source polarity recognition involves three steps:

\begin{itemize}
	\item[i)] Using external computer vision tools to detect sources within a circuit diagram and generate corresponding inset images.\footnote{In our preliminary tests, Gemini 2.5 Pro was unable to reliably detect circuit elements or output their bounding box information. This limitation motivated the integration of external computer vision tools to enhance element recognition accuracy.}
	\item[ii)] Leveraging the vision capabilities of LLMs to recognize the source polarity from the inset image.
	\item[iii)] Comparing the source polarity results from global circuit recognition and local inset image recognition, and correcting the global interpretation if discrepancies are found.
\end{itemize}

In our method, we utilize two tools to detect and extract sources in circuit diagrams: a rule-based Python script built on OpenCV \citep{bradski2008learning}, and a YOLO model \citep{redmon2016you}. OpenCV is a classical computer vision library that detects lines, shapes, and corners using techniques such as the Hough transform, contour approximation, and corner detection. By combining these features, our script can identify the geometry of dependent sources, which appear as rhombus-shaped symbols in the problem set derived from \cite{svoboda2013introduction}. Another reason OpenCV is effective for detecting dependent sources is that no other components or notations in these diagrams share the rhombus shape. In contrast, independent sources are typically represented by circles---a feature that is not exclusive to them. For instance, circular shapes are also used to denote terminals of switches, among other components. This ambiguity makes it more challenging to design a rule-based script that reliably detects independent sources with high precision and recall. To overcome this limitation, we pre-trained a YOLO model, which is based on deep learning \citep{goodfellow2016deep} and driven by labeled data, to recognize and extract independent voltage and current sources. 

Alongside source recognition, the bounding box information\footnote{The bounding box is defined by the format $[x_{1}, y_{1}, x_{2}, y_{2}]$, where $(x_{1}, y_{1})$ and $(x_{2}, y_{2})$ represent the coordinates of the top-left and bottom-right corners, respectively.} of the sources is stored for two purposes. First, it enables the cropping of source inset images corresponding to the bounding boxes. Second, in cases where multiple sources are present in the original circuit diagram, these bounding boxes help the LLM distinguish between them. With this collected information, LLMs can be appropriately prompted to generate and correct circuit descriptions. Figure \ref{diagram_recognition} illustrates the overall pipeline for circuit diagram recognition, and Table \ref{io_diagram_recognition} lists the inputs and outputs for each step shown in Figure \ref{diagram_recognition}. The prompt templates for circuit recognition are provided in \ref{C1}.

Since the correctness of source polarity in the initial circuit recognition is not known in advance, the source polarity recognition step using inset images is applied to all circuit recognition problems. However, prior to performing source detection with external computer vision tools, we first verify the presence of independent or dependent sources. Detection and image cropping are conducted only when such sources are identified. In the subsequent source polarity correction step, if a mismatch is detected between the initial global diagram recognition and the local inset image recognition, the latter is assumed to be correct and is used to update the source polarity in the initial circuit description.

\begin{landscape}
		\begin{figure}
			\centering
			\includegraphics[width=1.2\textwidth]{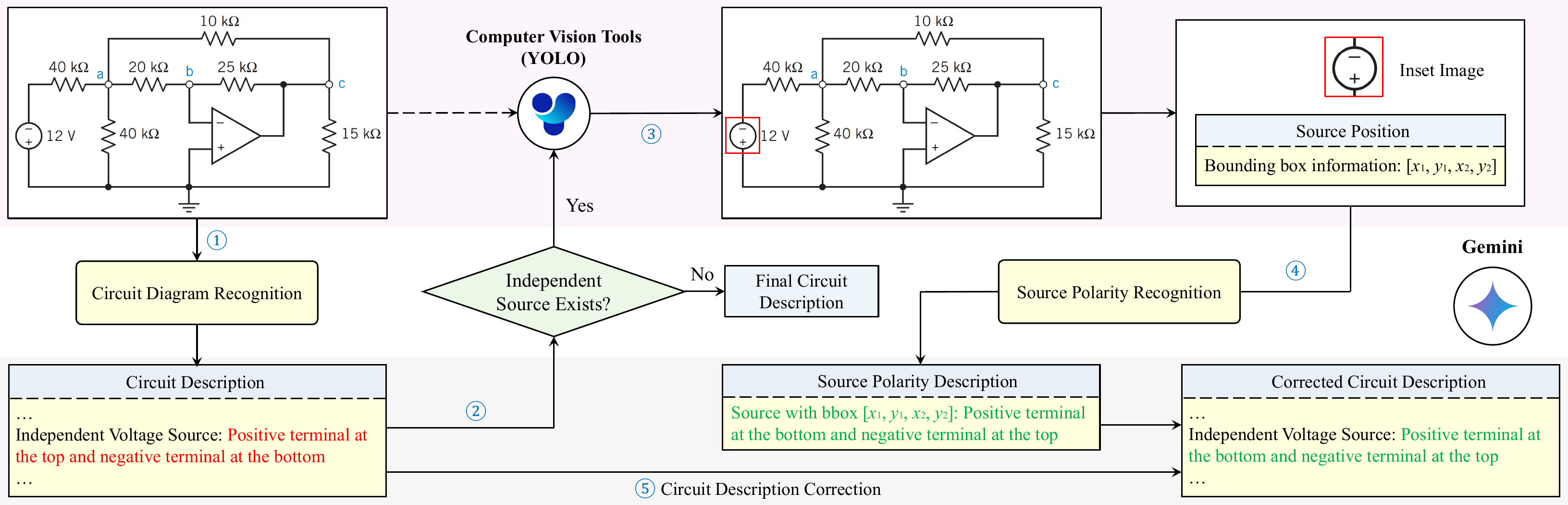}
			\caption{Source detection using external computer vision tools and correction via inset source images: 
				\raisebox{.5pt}{\textcircled{\raisebox{-.9pt}{1}}} Initial circuit diagram recognition; 
				\raisebox{.5pt}{\textcircled{\raisebox{-.9pt}{2}}} Verification of source existence; 
				\raisebox{.5pt}{\textcircled{\raisebox{-.9pt}{3}}} Source detection with external computer vision tools and bounding box extraction; 
				\raisebox{.5pt}{\textcircled{\raisebox{-.9pt}{4}}} Polarity recognition from the inset source image; 
				\raisebox{.5pt}{\textcircled{\raisebox{-.9pt}{5}}} Correction of the circuit description. 
				Note: The circuit diagram in this figure is taken from Problem 6.4-6 in \cite{svoboda2013introduction}. The source in this circuit diagram is an independent voltage source. Therefore, the computer vision tool used in step \raisebox{.5pt}{\textcircled{\raisebox{-.9pt}{3}}} is a pretrained YOLO model. For cases where dependent sources are detected, step \raisebox{.5pt}{\textcircled{\raisebox{-.9pt}{3}}} can instead employ a Python script with OpenCV to extract the bounding box information.}
			\label{diagram_recognition}
		\end{figure}
		\vspace{-0.25cm}
		\begin{table}
			\centering
			\footnotesize
			\caption{Input and output of each step in Figure \ref{diagram_recognition}}
			\label{io_diagram_recognition}
			\begin{tabular}{ccc} 
				\hline
				Step Index & Input & Output \\ 
				\hline
				\raisebox{.5pt}{\textcircled{\raisebox{-.9pt}{1}}} & Original circuit diagram ({\tt{.jpg}} or {\tt{.png}}) & Textual description of the circuit diagram (\tt{.txt}) \\ 
				\raisebox{.5pt}{\textcircled{\raisebox{-.9pt}{2}}} & Textual description of the circuit diagram ({\tt{.txt}}) & {\tt{Yes}} \text{or} {\tt{No}} \\ 
				\raisebox{.5pt}{\textcircled{\raisebox{-.9pt}{3}}} & Original circuit diagram ({\tt{.jpg}} or {\tt{.png}}) & \makecell{Inset image of the source(s) ({\tt{.jpg}} or {\tt{.png}}) \\ and corresponding bounding box information (\tt{.txt})} \\ 
				\raisebox{.5pt}{\textcircled{\raisebox{-.9pt}{4}}} & \makecell{Inset image of the source(s) ({\tt{.jpg}} or {\tt{.png}}) \\ and corresponding bounding box information (\tt{.txt})} & Textual source polarity description (\tt{.txt}) \\ 
				\raisebox{.5pt}{\textcircled{\raisebox{-.9pt}{5}}} & \makecell{Textual description of the circuit diagram (\tt{.txt}) \\ and textual source polarity description (\tt{.txt})} & Final textual description of the circuit diagram (\tt{.txt}) \\ 
				\hline
			\end{tabular}
		\end{table}
\end{landscape}

\subsubsection{Guiding Prompts for Identifying Current Directions}
\label{S322}
As demonstrated in Section \ref{S31}, the recognition of labeled current directions tends to be inaccurate when Gemini 2.5 Pro is tasked with generating circuit descriptions without specific guidance. To improve the accuracy of current direction recognition, we adopt a prompt engineering approach by introducing a specific prompting step during the initial circuit recognition phase. The content of this prompt is provided below.

\begin{tcolorbox}[breakable, enhanced, width=\textwidth, opacityfill=1., colback=verylightgray]
	This step involves recognizing the labeled current(s) in the provided circuit diagram. Please be very careful during detection.
	\begin{itemize}
	\item [1.] If there are no labeled currents in the circuit diagram, you do not need to answer the following questions.
	
	\item [2.] [BE VERY CAREFUL] For any current source or labeled currents in the circuit diagram, specify the direction of current flow by addressing the following:
	\begin{itemize}
	\item [2a)] Indicate the direction of the arrow that shows current flow (e.g., upward/downward/leftward/rightward for branch currents, and clockwise/counterclockwise for mesh currents).
	
	\item [2b)] Describe the flow path: from which node or terminal the current originates, through which component (if applicable), to which node or terminal it flows. (You may refer to terminal names using component labels.)
	\end{itemize}
	\end{itemize}
	
	NOTES:
	
	\begin{itemize}
	\item [1.] If there are two circuit diagrams labeled (a) and (b), you should provide separate recognitions for each.
	
	\item [2.] If a current starts from the negative terminal of a voltage source, clearly note that its direction is OPPOSITE to the conventional current supplied by the voltage source.
	\end{itemize}
\end{tcolorbox}

In essence, we explicitly prompt Gemini to describe the direction and flow path of labeled currents in order to reduce hallucinations in current direction recognition. This targeted recognition step is then integrated with other recognition processes to produce the complete circuit descriptions. 

Our overall prompting strategy for circuit analysis problem-solving follows a step-by-step methodology. This approach helps the LLM focus on each individual step, thereby reducing the likelihood of hallucinations. In addition to the steps involved in the circuit recognition phase, the problem-solving process itself is treated as a separate step. This step takes the textual circuit description as input and outputs the corresponding solution, including both the problem-solving steps and the final answer. The detailed prompting structure is illustrated in Figure \ref{framework}. The prompts for these steps are provided in \ref{C1} and \ref{C2}.

It is worth noting that the localized extraction and multi-stage prompt transformation strategy in this section (Section \ref{S32}) is theoretically grounded in recent findings regarding visual-language model limitations \citep{zhang2024mathverse,zhou2024visual}. As demonstrated in MathVerse \citep{zhang2024mathverse}, multimodal LLMs frequently exhibit severe visual perception hallucinations when required to resolve fine-grained diagrammatic details within complex global scenes, often misidentifying geometric relations and mathematical orientations. Furthermore, as revealed by \cite{zhou2024visual}, cross-modal interaction bottlenecks in deep transformer layers hinder effective reasoning over high-entropy visual tokens. By utilizing specialized vision tools (YOLO and OpenCV) to crop isolated component insets and generating an explicit, structured textual description of nodes, values, and polarities, our framework effectively converts dense visual circuit schematics into an intent-oriented symbolic description \citep{zhou2024visual}. This eliminates visual distractions and enables the downstream LLM to conduct symbolic and algebraic circuit derivations within its optimal linguistic reasoning domain.

\subsection{\texttt{ngspice} Verification}
\label{S33}

The methods described in Section \ref{S32} enhance the circuit recognition performance of Gemini 2.5 Pro by mitigating hallucinations related to source polarities and current directions. However, hallucinations can still occur during the problem-solving phase, even when the circuit has been correctly recognized. To address this issue, we incorporate circuit simulation using {\tt ngspice}\footnote{See \url{https://ngspice.sourceforge.io}.}, an open-source SPICE simulator for electrical and electronic circuits. The selection of {\tt ngspice} is motivated by the following features, which support the automation of the problem-solving process illustrated in Figure \ref{ngspice_flowchart}:

\begin{itemize}
	\item[i)] {\tt ngspice} accepts input for circuit simulation in the form of a \texttt{.cir} file. This file contains a textual description of the circuit formatted according to {\tt ngspice} specifications. As a result, users are not required to manually place components or connect them via a graphical user interface---an approach that is labor-intensive and unsuitable for automated problem-solving.
	
	\item[ii)] Our results demonstrate that Gemini 2.5 Pro can convert a textual circuit description into a format compliant with {\tt ngspice}. The prompt template for \texttt{.cir} file generation can be found in \ref{C3}. Leveraging this capability, along with the textual circuit description generated in the previous phase, we are able to obtain numerical solutions to circuit problems in a fully automated manner.
\end{itemize}

The numerical final answer obtained via {\tt ngspice} for each problem is compared with the corresponding solution generated by Gemini 2.5 Pro. If the LLM's solution is analytical---for example, expressed as a function of time $t$—it is first converted into numerical samples prior to comparison. The detailed comparison methodology, along with relevant hyperparameters such as tolerance thresholds, is presented in Section \ref{S4}.

The basic principles for solution verification and incorrect solution identification are outlined below:

\begin{itemize}
	\item[i)] When there are no errors in circuit recognition, problem-solving by the LLM, or {\tt .cir} file generation for {\tt ngspice} simulation\footnote{With a properly generated {\tt .cir} file, the simulation process in {\tt ngspice} will not introduce additional errors.}, the final answers obtained through both problem-solving methods will match. This consistency verifies the correctness of the solutions generated by Gemini 2.5 Pro.
	\item[ii)] When the solutions from the LLM and {\tt ngspice} do not match, the corresponding problems can be flagged for human review---such as by course instructors. This step constitutes the incorrect solution flagging process, which enhances the reliability of the circuit analysis system.
\end{itemize}

In practice, occasional hallucinations may still occur during either the {\tt .cir} file generation or the LLM-based problem-solving process. If a mismatch is detected between the Gemini 2.5 Pro and {\tt ngspice} solutions, a second trial can be initiated---re-running Gemini 2.5 Pro for either the problem-solving step or the {\tt .cir} file generation, followed by a new {\tt ngspice} simulation. If a match is obtained during these second trials, the corresponding solution is accepted as correct. This multi-trial process involving both LLM- and {\tt ngspice}-based problem solving is illustrated in Figure \ref{multi_trials}.

The underlying philosophy of our \texttt{ngspice} verification loop aligns with the tool-interactive critiquing framework formalized in CRITIC \citep{gou2024critic} and the tree-search retraction mechanisms in ToolLLM \citep{qin2024toolllm}. Because frozen foundation models are incapable of reliably verifying their own complex reasoning traces without external grounding \citep{gou2024critic}, treating the simulator as an additional external environment provides deterministic validation. When a discrepancy between the analytical solution and the numerical simulation is detected, our multi-trial mechanism initiates structured retraction and re-generation: either re-synthesizing the netlist or re-evaluating branch equations. Similar to the exploration paths used in the depth-first search-based decision tree method \citep{qin2024toolllm}, this execution-driven verification prevents cascading errors from propagating undetected into educational feedback.

However, hallucinations by the LLM during the circuit diagram recognition process can be difficult to detect. In such cases, both the LLM and {\tt ngspice} may process the same incorrectly extracted circuit information, leading to the same erroneous solution and thus passing the verification process incorrectly. In our experiments, one out of 83 problems falls into this category, as shown in Table \ref{detailed_results}. One way to address this issue is to have a human proofread the circuit information. To support this process, we are actively developing a visualization tool that converts the full textual circuit description into a circuit diagram in image format, which is more intuitive and easier for humans to verify.

\subsection{Human-in-the-Loop Corrections}
\label{S34}
To further enhance the performance of the proposed circuit problem solver, a human can be brought into the pipeline to proofread the generated circuit descriptions in order to verify the accuracy of circuit recognition and make revisions if necessary. The circuit description review can be limited to problems where the solutions produced by Gemini 2.5 Pro and \texttt{ngspice} do not match on their first attempts. Additionally, an instructor can review the Gemini-generated solutions, including both the problem-solving process and the final answers. When errors in the problem-solving process are identified, the instructor can highlight them with brief comments and allow Gemini to attempt corrections. As shown in Table \ref{detailed_results}, most of the remaining problems can be correctly solved after this human-in-the-loop correction step. It is also worth noting that the instructor only needs to review less than 20\% of the original homework problems compared to solving all problems manually, significantly reducing the workload involved in solution preparation. This human-in-the-loop correction step is also illustrated in Figures \ref{multi_trials} and \ref{framework}. The full circuit analysis problem-solving pipeline is summarized in Algorithm \ref{Alg1}.

\section{Experiments}
\label{S4}

This section presents the experimental design and results that demonstrate the effectiveness of the LLM-based circuit problem-solving pipeline introduced in Section \ref{S3}. Sections \ref{S41} and \ref{S42} describe the experimental setups and results for detecting independent sources using a YOLO model and dependent sources using a rule-based OpenCV function. Section \ref{S43} presents the results obtained from \texttt{ngspice}, and Section \ref{S44} summarizes the overall performance of the proposed problem-solving pipeline. In Section \ref{S44}, we also analyze the performance improvements contributed by each individual component to highlight their specific impact. Section \ref{S45} evaluates the performance of the proposed pipeline on circuit problems using four variations of hand-drawn diagrams. Section \ref{S46} further evaluates the generalizability of the proposed framework on a separate set of circuit problems selected from a different textbook. The computational cost and processing time are also reported for each case to demonstrate the scalability of the proposed multi-step circuit problem-solving pipeline. Section \ref{S453} presents the results of statistical significance tests, further confirming that the improvement in problem-solving accuracy achieved by the proposed pipeline over the baseline method, based on a single prompt to Gemini 2.5 Pro, is statistically significant. Finally, Section \ref{S48} analyzes the remaining incorrect solutions and identifies the primary sources of error across different problem sets and circuit diagram styles.

\begin{landscape}
		\begin{figure}
			\centering
			\includegraphics[width=1.4\textwidth]{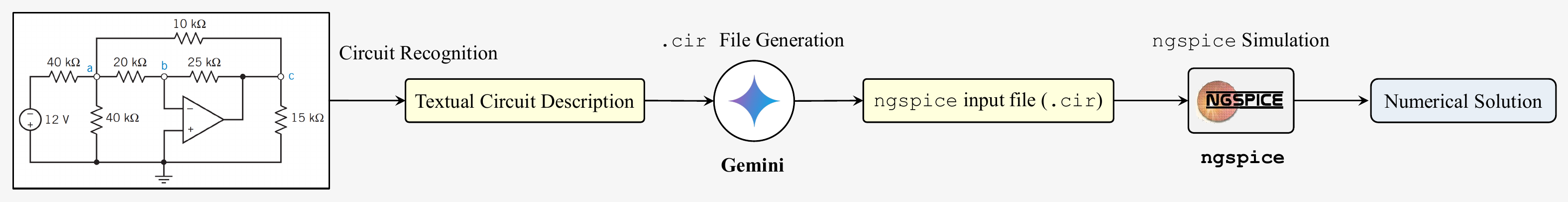}
			\caption{Problem solving pipeline of \texttt{ngspice}. Note: The circuit diagram in this figure is taken from Problem 6.4-6 in \cite{svoboda2013introduction}.}
			\label{ngspice_flowchart}
		\end{figure}
		
		\begin{figure}
			\centering
			\includegraphics[width=1.4\textwidth]{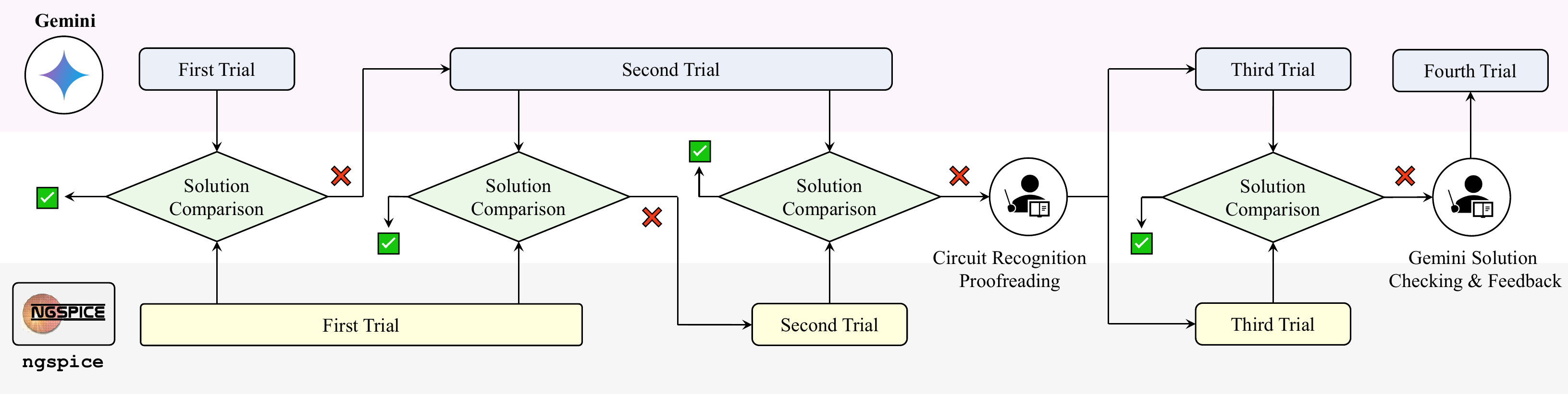}
			\caption{Multiple trials of Gemini- and \texttt{ngspice}-based problem solving processes}
			\label{multi_trials}
		\end{figure}
\end{landscape}

\begin{figure}
	\centering
	\includegraphics[width=0.69\textwidth]{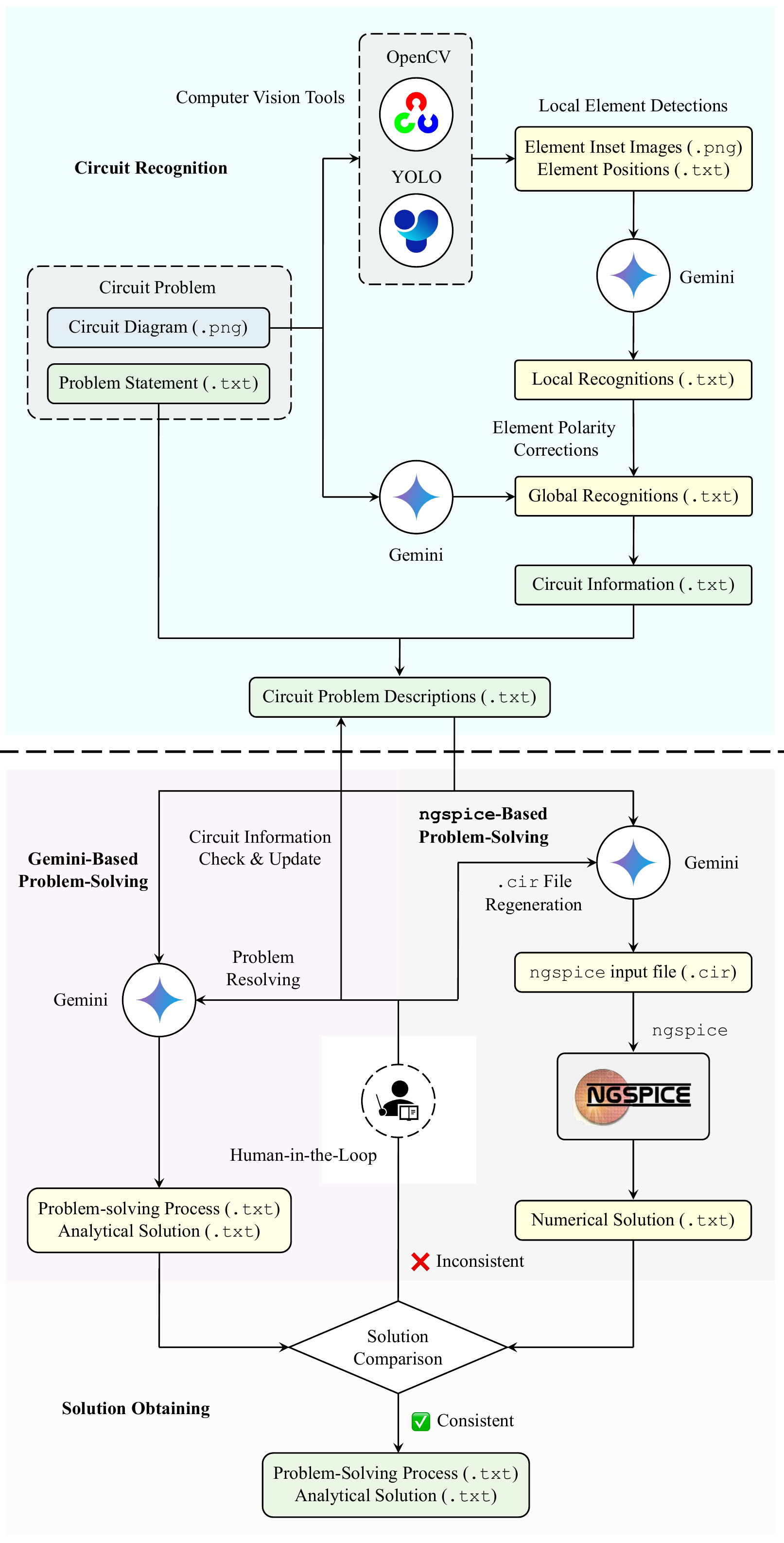}
	\caption{Framework of the proposed circuit problem solving method}
	\label{framework}
\end{figure}

\begin{breakablealgorithm}
	\caption{Circuit Analysis Problem Solving Pipeline}
	\label{Alg1}
	\begin{algorithmic}[1]
		\renewcommand{\algorithmicrequire}{\textbf{Input:}}
		\renewcommand{\algorithmicensure}{\textbf{Output:}}
		\Require A circuit analysis problem description, consisting of a textual problem statement and a circuit diagram (if available); designed prompts for Gemini 2.5 Pro; a well-trained \texttt{YOLO11} model for independent source detection; a dependent source detection function \texttt{circuit\_cv2} (see Algorithm \ref{alg:dependent_source_recognition}) implemented in \texttt{Python} using the OpenCV \texttt{cv2} package; and a comparison function \texttt{solution\_comparison} to evaluate the analytical final answer from Gemini against the numerical final answer from \texttt{ngspice}.
		\Ensure A complete solution to the given circuit analysis problem, including both the step-by-step problem-solving process and the final answer.\\
		\textit{Initialization}: Initialize the Gemini chat session \texttt{gemini\_chat} using the API key.
		\State {\texttt{\color{gray}Direct Solving for Problems Without Circuits}}
		\If{there is no associated circuit diagram}
		\State Solve the problem directly using Gemini 2.5 Pro
		
		\Return The solution generated by Gemini
		\EndIf
		\State {\texttt{\color{gray}Circuit Recognition}}
		\State Extract nodes and elements of the circuit, including types, connections, values, polarities (if applicable), and directions of labeled currents $\gets$ \texttt{gemini\_chat}(circuit diagram)
		\State circuit information\_\texttt{v1} $\gets$ Extracted circuit information
		\If{independent source recognized}
		\State Inset image(s) and \texttt{bbox} position(s) of independent source(s) $\gets$ \texttt{YOLO11}(circuit diagram)
		\ElsIf{dependent source recognized}
		\State Inset image(s) and \texttt{bbox} position(s) of dependent source(s) $\gets$ \texttt{circuit\_cv2} (circuit diagram)
		\EndIf
		\If{independent source recognized \textbf{or} dependent source recognized}
		\State Determine polarity of each source $\gets$ \texttt{gemini\_chat}(each inset source image)
		\State Circuit information\_\texttt{v2} with updated source polarities $\gets$ \texttt{gemini\_chat} (circuit information\_\texttt{v1}, polarities and \texttt{bbox} position(s) of all sources)
		\EndIf
		\State {\texttt{\color{gray}Gemini-Based Problem Solving}}
		\State Problem-solving steps and analytical final answer $\gets$ \texttt{gemini\_chat}(Circuit information\_\texttt{v2}, textual problem statement)
		\State {\texttt{\color{gray}ngspice-Based Problem Solving}}
		\State Generate \texttt{.cir} file for \texttt{ngspice} input $\gets$ \texttt{gemini\_chat}(Circuit information\_\texttt{v2}, textual problem statement)
		\State Numerical final answer $\gets$ \texttt{ngspice}(\texttt{.cir} file)
		\State {\texttt{\color{gray}Solution Comparison}}
		\State Final answer consistency $\gets$ \texttt{solution\_comparison}(analytical final answer from Gemini, numerical final answer from \texttt{ngspice})
		\State {\texttt{\color{gray}Solution Output}}
		\If{final answers match}
		
		\Return Problem-solving steps and analytical final answer generated by Gemini
		\Else
		\State Follow the procedure in Figure \ref{multi_trials} to solve further, either automatically or via human-in-the-loop
		\EndIf
	\end{algorithmic}
\end{breakablealgorithm}

\subsection{YOLO Model Training for Independent Source Detection}
\label{S41}

The YOLO model employed for detecting independent sources, including voltage and current sources, is \texttt{yolo11l}\footnote{See details at \url{https://docs.ultralytics.com/models/yolo11}.}. The training dataset is a combination of three sources available on the Roboflow platform: \raisebox{.5pt}{\textcircled{\raisebox{-.9pt}{1}}} \url{https://universe.roboflow.com/diana-eottv/circuitobjectsdetection}, \raisebox{.5pt}{\textcircled{\raisebox{-.9pt}{2}}} \url{https://universe.roboflow.com/ezpark-gcxxl/cv-circuit}, and \raisebox{.5pt}{\textcircled{\raisebox{-.9pt}{3}}} \url{https://universe.roboflow.com/deeplearning-l1bq5/circuit-diagram-eo4kn}. While datasets \raisebox{.5pt}{\textcircled{\raisebox{-.9pt}{1}}} and \raisebox{.5pt}{\textcircled{\raisebox{-.9pt}{2}}} include printed circuit diagrams, dataset \raisebox{.5pt}{\textcircled{\raisebox{-.9pt}{3}}} consists of hand-drawn circuit diagrams, enabling the detection of independent sources from hand-drawn circuit diagrams in Section \ref{S45}. The numbers of training, validation, and test samples in the datasets are presented in Table \ref{total_data}. During training, we consider 11 frequently observed classes, with the number of samples per class summarized in Table \ref{data_per_class}. Note that some images in the original datasets do not contain any of the 11 component classes listed in Table \ref{data_per_class}. These images were removed during the preprocessing stage, resulting in the instance counts shown in the combined dataset in Table \ref{data_per_class}.

\begin{table}[!h]
	\centering
	\footnotesize
	\caption{Number and percentage of images in the datasets}
	\label{total_data}
	\begin{tabular}{cccc:c}
		\hline
		Dataset & Training & Validation & Test & Total \\ 
		\hline 
		Printed (\raisebox{.5pt}{\textcircled{\raisebox{-.9pt}{1}}}+\raisebox{.5pt}{\textcircled{\raisebox{-.9pt}{2}}}) & 1607 (39.27\%) & 117 (2.86\%) & 309 (7.55\%) & 2033 (49.68\%) \\
		Hand-drawn (\raisebox{.5pt}{\textcircled{\raisebox{-.9pt}{3}}}) & 1445 (35.31\%) & 411 (10.04\%) & 203 (4.96\%) & 2059 (50.32\%) \\
		\hline
		Total & 3052 (74.58\%) & 528 (12.90\%) & 512 (12.51\%) & 4092 (100\%) \\
		\hline
	\end{tabular}
\end{table}

\begin{table}[!h]
	\centering
	\footnotesize
	\caption{Number and percentage of instances per class in the datasets}
	\label{data_per_class}
	\begin{tabular}{ccccccc:cc}
		\hline
		Class & \multicolumn{2}{c}{Training} & \multicolumn{2}{c}{Validation} & \multicolumn{2}{c}{Test} & \multicolumn{2}{c}{Total}\\
		\hline
		Dataset & \raisebox{.5pt}{\textcircled{\raisebox{-.9pt}{1}}}+\raisebox{.5pt}{\textcircled{\raisebox{-.9pt}{2}}} & \raisebox{.5pt}{\textcircled{\raisebox{-.9pt}{3}}} & \raisebox{.5pt}{\textcircled{\raisebox{-.9pt}{1}}}+\raisebox{.5pt}{\textcircled{\raisebox{-.9pt}{2}}} & \raisebox{.5pt}{\textcircled{\raisebox{-.9pt}{3}}} & \raisebox{.5pt}{\textcircled{\raisebox{-.9pt}{1}}}+\raisebox{.5pt}{\textcircled{\raisebox{-.9pt}{2}}} & \raisebox{.5pt}{\textcircled{\raisebox{-.9pt}{3}}} & \raisebox{.5pt}{\textcircled{\raisebox{-.9pt}{1}}}+\raisebox{.5pt}{\textcircled{\raisebox{-.9pt}{2}}} & \raisebox{.5pt}{\textcircled{\raisebox{-.9pt}{3}}} \\ 
		\hline
		DC Voltage Source & 906 & 517& 140 & 159 & 45 & 71 & 1091 & 747 \\ 
		Resistor & 4009 & 1240 & 775 & 351 & 235 & 142 & 5019 & 1733 \\ 
		Inductor & 795 & 717 & 111 & 195 & 50 & 95 & 956 & 1007 \\ 
		Capacitor & 993 & 725 & 206 & 191 & 98 & 94 & 1297 & 1010 \\ 
		Current Source & 562 & 528 & 58 & 150 & 9 & 60 & 629 & 738 \\ 
		Switch & 117 & 0 & 26 & 0 & 12 & 0 & 155 & 0 \\ 
		Ground & 565 & 500 & 147 & 140 & 78 & 64 & 790 & 704 \\ 
		AC Voltage Source & 145 & 115 & 44 & 34 & 20 & 13 & 209 & 162 \\ 
		Transformer & 118 & 0 & 39 & 0 & 13 & 0 & 170 & 0 \\ 
		Dep. Voltage Source & 0 & 125 & 0 & 49 & 0 & 22 & 0 & 196 \\ 
		Dep. Current Source & 0 & 145 & 0 & 33 & 0 & 23 & 0 & 201 \\ 
		\hline
		\multicolumn{9}{l}{\parbox{10cm}{* Dep. = Dependent}}
	\end{tabular}
\end{table}

Fine-tuning a YOLO model from a pre-trained backbone is more data-efficient than training from scratch, as the feature extraction layers have already been well trained. Thus, instead of training from scratch, we fine-tune the YOLO model on our combined dataset using the pre-trained \texttt{yolo11l} model\footnote{The pre-trained \texttt{yolo11l} model is available at \url{https://github.com/ultralytics/assets/releases/download/v8.3.0/yolo11l.pt}.}. The YOLO detection model is fine-tuned for 50 epochs with a batch size of 16 and an image size of 640. All other hyperparameters are kept at their default values as defined by Ultralytics.

\begin{figure}[!h]
	\centering
	\includegraphics[width=0.625\textwidth]{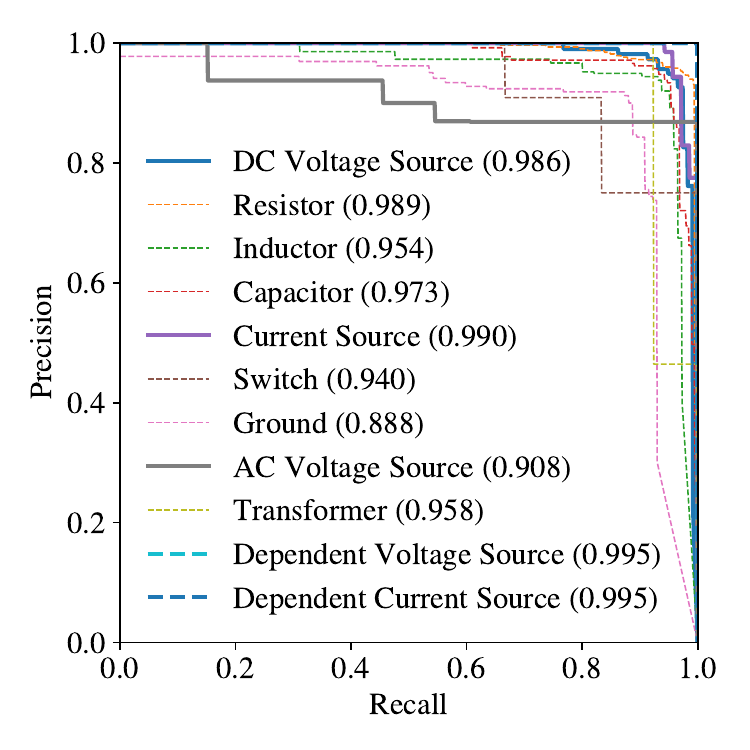}
	\caption{Precision--Recall curve of the best fine-tuned \texttt{YOLO11} model on the test dataset. Note: The numbers following the curve labels represent the mAP@0.5, i.e., the mean average precision calculated using an Intersection over Union (IoU) threshold of 0.5.}
	\label{pr_curve}
\end{figure}

We evaluate the fine-tuned model using the best-performing checkpoint on the validation set. The precision-recall curve of the resulting model, evaluated on the test dataset, is shown in Figure \ref{pr_curve}. The results indicate strong performance in detecting independent DC voltage sources, AC voltage sources, and current sources, with mAP@0.5 (mean average precision at an intersection-over-union threshold of 0.5) values of 0.986, 0.908, and 0.990, respectively.

\begin{figure}[!h]
	\centering
	\begin{subfigure}[b]{0.25\textwidth}
		\centering
		\includegraphics[width=\textwidth]{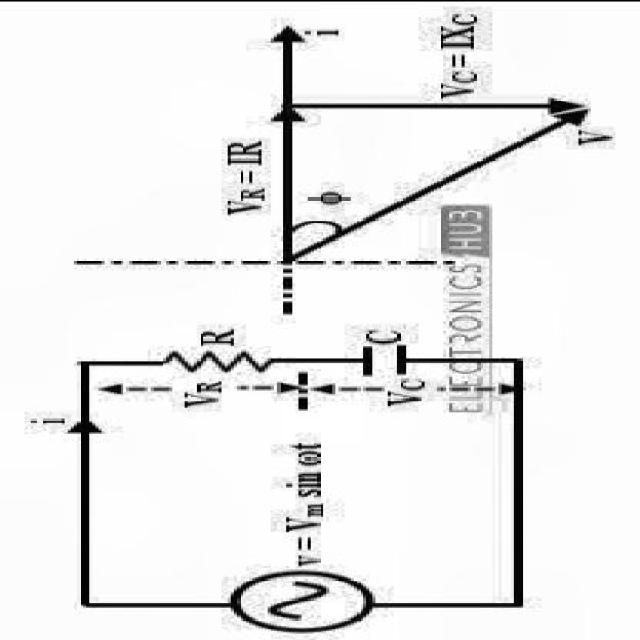}
		\caption{}
		\label{data_samples:a}
	\end{subfigure}
	\hfill
	\begin{subfigure}[b]{0.25\textwidth}
		\centering
		\includegraphics[width=\textwidth]{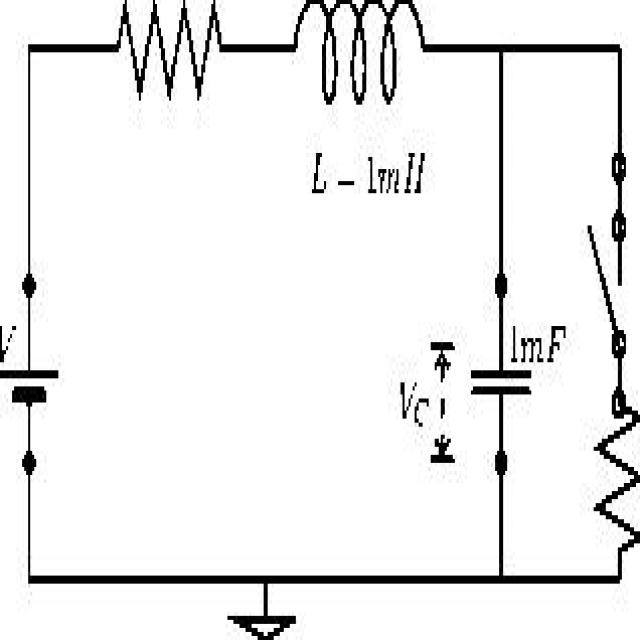}
		\caption{}
		\label{data_samples:b}
	\end{subfigure}
	\hfill
	\begin{subfigure}[b]{0.38\textwidth}
		\centering
		\includegraphics[width=\textwidth]{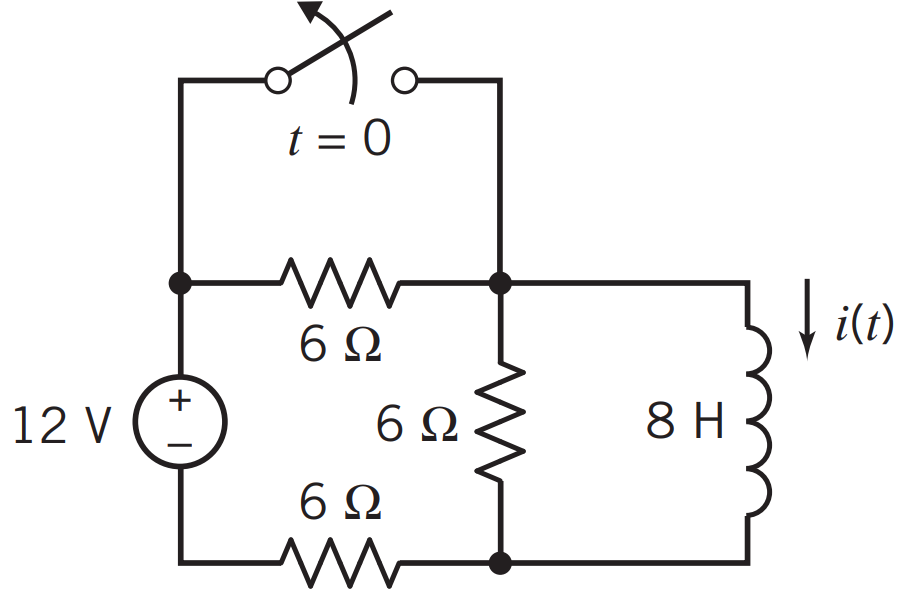}
		\caption{}
		\label{data_samples:c}
	\end{subfigure}
	
	\caption{Circuit diagram samples from the dataset used for YOLO model fine-tuning and from the problems to be solved in this paper: (a) A sample from dataset \raisebox{.5pt}{\textcircled{\raisebox{-.9pt}{1}}} \url{https://universe.roboflow.com/diana-eottv/circuitobjectsdetection}; (b) A sample from dataset \raisebox{.5pt}{\textcircled{\raisebox{-.9pt}{2}}} \url{https://universe.roboflow.com/ezpark-gcxxl/cv-circuit}; (c) The circuit diagram associated with Problem 8.3-2 in \cite{svoboda2013introduction}.}
	\label{data_samples}
\end{figure}

It is worth noting that the textbook-quality circuit diagrams in the problems addressed in this paper are of higher quality---specifically, they have better resolution and clearer labels---than the images in our training dataset. Figure \ref{data_samples} presents sample images from both the Roboflow datasets and the dataset of the target problems. Although the styles of the textbook-quality circuit diagrams differ from those in the training dataset, the YOLO detection model performs comparably or even better on the problem dataset. The confusion matrices in Table \ref{Tab_confusion1} indicate that, with a confidence threshold of 0.25, the detection model achieves perfect performance for independent voltage sources (TP = 75, FN = 0, FP = 0), meaning that all voltage sources in the textbook-quality diagrams are correctly identified without any false detections. For independent current sources, the model detects 33 out of 34 instances, with two false positives, corresponding to a recall of 97.1\% and a precision of 94.3\%. Since the cropped detections are subsequently provided to Gemini 2.5 Pro for polarity recognition (see Step 6 of the prompt template in \ref{C1}), the two false positives are effectively filtered out and therefore do not affect the final recognition results. These findings demonstrate that the proposed detection pipeline is highly reliable for textbook-quality circuit diagrams.

\begin{table}[htbp]
\centering
\footnotesize
\caption{Confusion matrices for independent source detection in textbook-quality circuit diagrams using the fine-tuned YOLO model with a prediction confidence threshold of 0.25}
\label{Tab_confusion1}
\begin{subtable}{\textwidth}
\centering
\caption{Independent Voltage Source}
\begin{tabular}{c|cc}
\hline
& Predicted Voltage Source & Predicted Background \\
\hline
Actual Voltage Source & True Positive = 75 & False Negative = 0 \\
Actual Background & False Positive = 0 & --- \\
\hline
\end{tabular}
\end{subtable}

\vspace{0.5cm}

\begin{subtable}{\textwidth}
\centering
\caption{Independent Current Source}
\begin{tabular}{c|cc}
\hline
& Predicted Current Source & Predicted Background \\
\hline
Actual Current Source & True Positive = 33 & False Negative = 1 \\
Actual Background & False Positive = 2 & --- \\
\hline
\end{tabular}
\end{subtable}
\end{table}

Note that for YOLO model fine-tuning, we consider more classes than just independent sources. This configuration is justified as follows:

\begin{itemize}
	\item[i)] Since hallucinations in LLMs and VLMs can only be reduced, not entirely eliminated, external tools like the YOLO model fine-tuned for specific tasks serve as valuable complements. In this context, the detection results for the 11 classes, shown in Figure \ref{pr_curve}, provide practical guidance on which components merit closer attention in future recognition tasks. Two noteworthy findings emerge from fine-tuning a YOLO model for circuit component detection. First, the circuit element classes considered (as listed in Table \ref{data_per_class}) reflect those commonly encountered in undergraduate circuit analysis courses. However, there is a noticeable data imbalance among these classes. For instance, resistors appear more frequently in circuits than ground symbols. Consequently, the fine-tuned \texttt{YOLO11} model exhibits comparatively lower object detection performance for ground symbols than for resistors, as shown in Figure \ref{pr_curve}. Second, the mean average precision (mAP) across classes does not strictly correlate with the number of instances per class. For example, although the switch class has the fewest instances among the 11 component types (see Table \ref{data_per_class}), it achieves a relatively high mAP@0.5 of 0.940. This discrepancy may be attributed to factors such as image variation and visual similarity between components.
	
	\item[ii)] Further results beyond Figure \ref{pr_curve} reveal that data distribution and contextual diversity play a critical role. Although the mAP@0.5 values for both the dependent voltage source and the dependent current source reach 0.995 in the test dataset (Figure \ref{pr_curve}), the detection performance in our circuit problem dataset remains poor. The reason is that almost all circuit diagrams containing dependent sources consist of only a single dependent source and no other circuit elements (i.e., 196 out of 196 cases for dependent voltage sources and 196 out of 201 cases for dependent current sources). As a result, the YOLO model tends to overfit to the pattern that dependent sources appear as standalone components occupying an entire image, rather than as interconnected elements within a complete circuit. This bias leads to poor detection performance for dependent sources in the circuit problems. This observation highlights that, for object detection models such as YOLO, the training dataset must contain sufficient contextual diversity and structural variability so that the model can learn object features within realistic circuit environments rather than memorizing isolated component patterns. To address this issue, we employ OpenCV for dependent source detection using a rule-based and data-efficient approach (see Section \ref{S42}). This method is also labor-efficient, as it avoids the need to collect and annotate extensive image datasets. Notably, this hybrid approach---combining data-driven methods (e.g., YOLO) with rule-based techniques (e.g., OpenCV)---can be generalized to other object detection tasks and has the potential to support improved recognition and understanding in VLMs.
\end{itemize}

\subsection{OpenCV-Based Dependent Source Detection}
\label{S42}

To detect the dependent sources in a circuit diagram, we leverage the characteristic rhombus-shaped contours associated with these elements. The detection and extraction process involves four main steps:

\begin{itemize}
	\item[i)] Extracting edges and identifying contours in the circuit diagram
	\item[ii)] Detecting rhombus-like convex quadrilaterals
	\item[iii)] Eliminating near-duplicate detections
	\item[iv)] Locating bounding boxes and cropping the dependent sources
\end{itemize}

\begin{figure}[!h]
	\centering
	\begin{subfigure}[b]{0.54\textwidth}
		\centering
		\includegraphics[width=\textwidth]{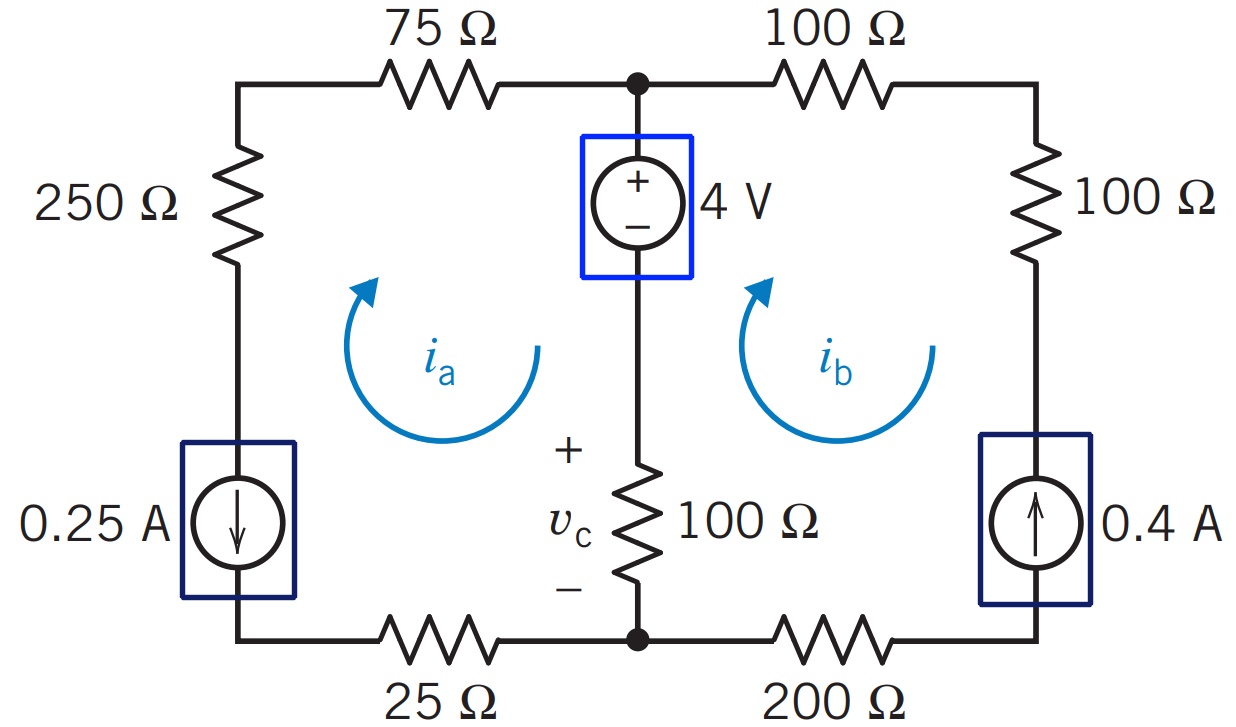}
		\caption{}
		\label{detection_samples:a}
		\vspace{0.25cm}
	\end{subfigure}
	\hfill
	\begin{subfigure}[b]{0.44\textwidth}
		\centering
		\includegraphics[width=\textwidth]{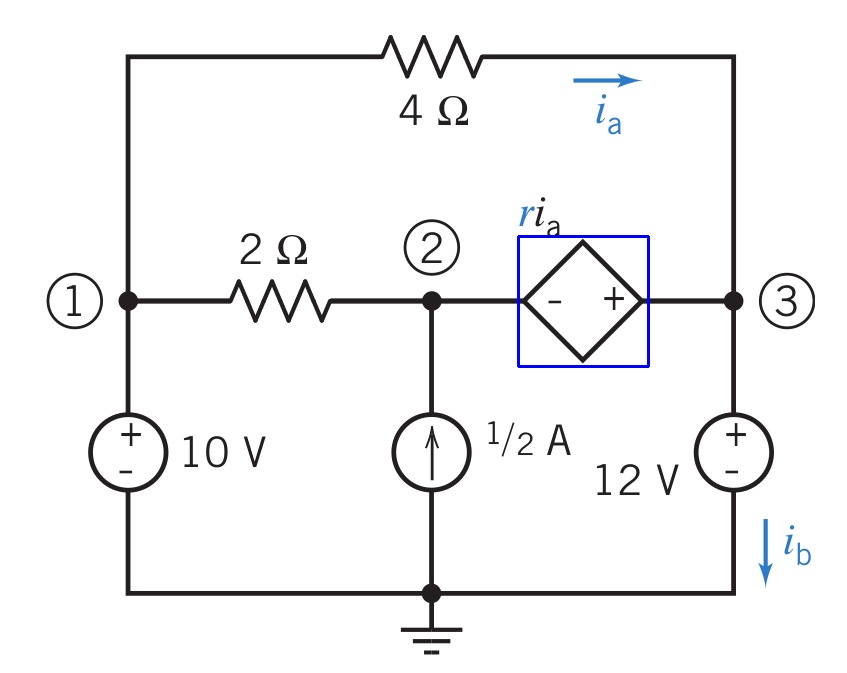}
		\caption{}
		\label{detection_samples:b}
	\end{subfigure}	
	\caption{Circuit diagram samples illustrating the detection of independent and dependent sources: (a) The circuit diagram of Problem 4.6-2 from \cite{svoboda2013introduction}, with bounding boxes marking one independent voltage source and two independent current sources in blue; (b) The circuit diagram of Problem 4.4-4 from \cite{svoboda2013introduction}, with the bounding box highlighting the dependent voltage source in blue.}
	\label{detection_samples}
\end{figure}

The pseudocode for this process using OpenCV is summarized in Algorithm \ref{alg:dependent_source_recognition}, where key functions from the Python OpenCV module (i.e., \texttt{cv2}) are highlighted. Through the detection and duplicate removal steps, dependent sources can be accurately identified. Figure \ref{detection_samples} presents examples of both independent and dependent source detections relevant to the problems to be solved.

\begin{breakablealgorithm}
	\caption{Dependent Source (Rhombus) Detection and Cropping With OpenCV}
	\label{alg:dependent_source_recognition}
	\begin{algorithmic}[1]
		\Require \texttt{circuit\_path} (path to circuit image)
		\Ensure Cropped images and bounding boxes for detected dependent sources
		
		\State {\color{gray}\texttt{Load the Image}}
		\State $image \gets$ \textbf{cv2.imread}(\texttt{circuit\_path}) \Comment{{\color{gray}Load BGR image}}
		\State $gray \gets$ \textbf{cv2.cvtColor}$(image)$ \Comment{{\color{gray}Convert to grayscale}}
		
		\State {\color{gray}\texttt{Preprocess to Obtain Edges}}
		\State $blurred \gets$ \textbf{cv2.GaussianBlur}$(gray,\ (5,5),\ 0)$ \Comment{{\color{gray}Denoise with Gaussian blur}}
		\State $edges \gets$ \textbf{cv2.Canny}$(blurred,\ 50,\ 150)$ \Comment{{\color{gray}Edge detection (low/high thresholds)}}
		
		\State {\color{gray}\texttt{Find Contours (Candidate Shapes)}}
		\State $contours \gets$ \textbf{cv2.findContours}$(edges)$
		\State $dependent\_source\_candidates \gets [\ ]$
		
		\State {\color{gray}\texttt{Detect Rhombus-like Convex Quadrilaterals}}
		\ForAll{$cnt \in contours$}
		\State $epsilon \gets 0.01 \times \textbf{cv2.arcLength} (cnt)$ \Comment{{\color{gray}Approximation tolerance}}
		\State $approx \gets$ \textbf{cv2.approxPolyDP}$(cnt,\ epsilon)$ \Comment{{\color{gray}Polygonal approximation}}
		\If{$|approx| = 4\ \land\ $ \textbf{cv2.isContourConvex}$(approx)$}
		\State $sides \gets [\ ]$
		\For{$i = 0$ to $3$}
		\State $pt1 \gets approx[i]$, $pt2 \gets approx[(i+1)\ \bmod\ 4]$
		\State $sides.\text{append}(\|pt1 - pt2\|_2)$
		\EndFor
		\State {\color{gray}\texttt{Rhombus check: all four sides nearly equal}}
		\If{$\max(sides)/\min(sides) < 1.1$}
		\State $dependent\_source\_candidates.\text{append}(approx)$
		\EndIf
		\EndIf
		\EndFor
		
		\State {\color{gray}\texttt{Remove Near-Duplicate Detections by Centroid Proximity}}
		\State $unique \gets [\ ]$
		\ForAll{$cand \in dependent\_source\_candidates$}
		\State $dup \gets \text{False}$
		\State $c\_center \gets$ centroid of $cand$ \Comment{{\color{gray}mean of vertices via \texttt{numpy}}}
		\ForAll{$u \in unique$}
		\State $u\_center \gets$ centroid of $u$
		\If{$\|c\_center - u\_center\|_2 < 10$}
		\State $dup \gets \text{True}$; \textbf{break}
		\EndIf
		\EndFor
		\If{\textbf{not} $dup$}
		\State $unique.\text{append}(cand)$
		\EndIf
		\EndFor
		
		\State $dependent\_source\_candidates \gets unique$
		
		\State $margin \gets 10$
		
		\State {\color{gray}\texttt{Find Bounding Boxes and Crop}}
		\ForAll{$poly \in dependent\_source\_candidates$}
		\State $(x,y,w,h) \gets$ \textbf{cv2.boundingRect}$(poly)$ \Comment{{\color{gray}Axis-aligned bounding box}}
		\State $crop \gets image[y - margin : y + h + margin,\ x - margin : x + w + margin]$ \Comment{{\color{gray}Clamp with \texttt{max(0,\,$\cdot$)} as needed}}
		\State $(x_1,y_1) \gets (x - margin,\ y - margin)$; $(x_2,y_2) \gets (x + w + margin,\ y + h + margin)$
		\EndFor\\
		\Return Cropped images $\{crop^{i}\}_{i=1}^{n}$ and bounding boxes $\{(x_{1}^{i},y_{1}^{i},x_{2}^{i},y_{2}^{i})\}_{i=1}^{n}$ for all detected dependent sources
	\end{algorithmic}
\end{breakablealgorithm}

\subsection{Results of \texttt{ngspice} on Circuit Problems With Textbook-Quality Diagrams}
\label{S43}

With the assistance of the external computer vision tools described in Sections \ref{S41} and \ref{S42}, Gemini 2.5 Pro can be used to generate circuit descriptions in textual format for a given problem. These descriptions can then be converted into a \texttt{.cir} file for \texttt{ngspice} simulation. Table \ref{TableA1} presents the problem categories and the corresponding \texttt{ngspice} simulation results for the 83 problems introduced in Section \ref{S31}. The categories are defined as follows:

\begin{itemize}
	\item [--] \textit{Circuit Analysis:} Problems in this category feature well-defined circuits, where the element types and values are fully specified. The required outputs are voltages, currents, or powers, expressed as numerical values or functions of time. From the perspective of \texttt{ngspice} simulation, the solutions generated by Gemini 2.5 Pro can be directly compared with those from \texttt{ngspice} to verify their correctness.
	
	\item [--] \textit{Circuit Synthesis:} Problems in this category ask for the values of one or more circuit elements, given certain voltages or currents. Since some element values are unspecified in the circuit diagram, a valid \texttt{.cir} file cannot be generated without first assigning numerical values. To verify the correctness of the solutions generated by Gemini 2.5 Pro, the following steps are taken:
	\begin{itemize}
		\item [i)] Substitute Gemini's answers into the required variables of the problem during the \texttt{.cir} file generation process.
		\item [ii)] In the \texttt{ngspice} simulation, some voltages or currents given in the problem statement are treated as unknowns at Gemini's discretion.
		\item [iii)] The resulting \texttt{.cir} file is simulated using \texttt{ngspice}, which returns the values of the assumed unknown variables.
		\item [iv)] These returned values are compared with those provided in the original problem statement. If they match, the element values solved by Gemini are considered correct.
	\end{itemize}
	For example, problem P 4.4-1 asks for the gain of a dependent source in the circuit diagram, given the values of three node voltages. In the verification process, the \texttt{.cir} file treats the dependent source gain, as solved by Gemini 2.5 Pro, as known, and solves for the three node voltages. If the node voltages returned by \texttt{ngspice} match those given in the original problem statement, the solved gain is considered correct; otherwise, it is not.
	\item [--] \textit{Network Function Analysis:} Problems in this category ask for the network functions of given circuit diagrams. Since a network function depends on angular frequency $\omega$ rather than time $t$, we use \texttt{ngspice} to generate a \texttt{.cir} file that outputs the magnitudes and phases of the network function at various values of $\omega$. The verification of the answers generated by Gemini 2.5 Pro is based on these magnitude and phase values.
	\item [--] \textit{Network Function Synthesis:} Similar to the \textit{Circuit Synthesis} category, problems in this category require determining the values of certain circuit elements to satisfy a given network function or specific magnitude and phase values of the network function at selected angular frequencies. Notably, the only problem in our set that falls into this category is P 13.2-11.
	\item [--] \textit{No Circuit Diagram:} Problems in this category do not include an associated circuit diagram, rendering them unsuitable for \texttt{ngspice} simulations. However, it is noteworthy that Gemini 2.5 Pro demonstrates strong problem-solving performance on such problems.
	\item [--] \textit{Element Type Not Defined:} The circuit elements in the diagrams for problems in this category are abstract and lack specified types. Since \texttt{ngspice} requires well-defined element types for simulation, these problems are not suitable for \texttt{ngspice} simulations. 
	\item [--] \textit{Maximum Power Transfer:} Problems in this category ask for the resistance values of certain resistors to achieve maximum power transfer. Since \texttt{ngspice} cannot directly determine whether the power achieved with the resistance calculated by Gemini is indeed maximized, these problems are not suitable for \texttt{ngspice} simulations. Notably, the only problem in our set that falls into this category is P 5.6-3.
	\item [--] \textit{Mathematical Expression Answer:} The solutions to problems in this category are purely mathematical expressions, and the element values in the circuit diagrams are unspecified. Since \texttt{ngspice} requires concrete element values to perform simulations, these problems are not suitable for \texttt{ngspice} simulations.
	\item [--] \textit{No Need for \texttt{ngspice} Simulation:} Problem P 10.4-1 requires determining the phasor representations of a resistance, a capacitance, two inductances, and a voltage, making an \texttt{ngspice} simulation unnecessary. 
\end{itemize}

We observe that, for problems without circuit diagrams and not requiring \texttt{ngspice} simulation, Gemini 2.5 Pro demonstrates relatively strong performance, as these problems primarily involve reasoning rather than the interpretation of circuit diagrams. This suggests that, with the support of \texttt{ngspice}, a human instructor or teaching assistant would only need to verify the results of Gemini 2.5 Pro for $\frac{2+1+4}{83} \approx 8.43\%$ of the problems---specifically, those categorized as ``Element Type Not Defined,'' ``Maximum Power Transfer,'' and ``Mathematical Expression Answer'' in Table \ref{TableA1}. This significantly reduces the human workload. Furthermore, in our experiments, Gemini 2.5 Pro correctly solved all 17 problems in the categories that are not suitable for \texttt{ngspice} simulation.

\begin{table}[h!]
	\centering
	\fontsize{9pt}{10pt}\selectfont
	\caption{Categories of problems and results for \texttt{ngspice} simulations}
	\label{TableA1}
	\begin{tabular}{cc}
		\hline
		Category (Number of Problems) & Problem Index(es) \\  
		\hline
		Circuit Analysis (56) &  \makecell[l]{\color{darkgreen}P 2.4-6, {\color{orange}P 2.5-1,} P 2.5-2, P 3.2-6, P 3.4-4, P 3.6-1,\\ \color{darkgreen}P 3.6-3, P 4.2-2, P 4.2-5, P 4.3-3, P 4.4-6, {\color{orange}P 4.6-2,} \\ \color{darkgreen}P 4.7-1, P 4.7-2, P 4.7-6, {\color{red}{P 5.2-1,}} P 5.2-2, P 5.2-3,\\ \color{darkgreen}P 5.2-5, P 5.3-4, P 5.3-6, P 5.3-8, P 5.4-5, P 5.5-5,\\ \color{darkgreen}P 6.3-2, P 6.3-3, P 6.4-1, P 6.4-2, P 6.4-6, P 7.2-4, \\\color{darkgreen}P 7.4-1, P 7.5-5, P 7.7-2, P 7.8-2, P 8.3-1, {\color{brown}P 8.3-2,} \\\color{darkgreen}P 8.3-4, P 8.3-5, P 8.3-10, {\color{blue}P 8.4-1,} {\color{orange}P 8.7-3,} P 8.7-5, \\\color{darkgreen}P 9.4-1, P 9.4-2, P 9.5-1, {\color{blue}P 9.6-1,} P 9.7-2, {\color{brown}P 9.8-2},\\ {\color{brown}P 10.4-6,} \color{darkgreen}P 10.6-1, {\color{red}P 10.6-4,} P 10.7-1, P 10.7-6, \\\color{darkgreen}P 10.8-1, P 10.8-6, P 10.10-4}\\
		\hline
		Circuit Synthesis (4) & \makecell[l]{\color{darkgreen}P 4.4-1, P 4.5-2, P 7.4-3, P 8.3-9} \\
		\hline
		Network Function Analysis (5) & \makecell[l]{\color{darkgreen}P 13.2-1, P 13.2-2, P 13.2-3, P 13.2-8, P 13.2-9} \\
		\hline
		Network Function Synthesis (1) & \makecell[l]{\color{darkgreen}P 13.2-11} \\
		\hline\hline
		No Circuit Diagram (9) & \makecell[l]{\color{darkgray}P 1.5-2, P 1.5-3, P 2.4-7, P 7.2-2, P 7.3-2, P 7.5-4, \\\color{darkgray}P 7.6-2, P 10.3-2, P 10.3-3} \\
		\hline
		Element Type Not Defined (2) & \makecell[l]{\color{darkgray}P 3.2-1, P 3.2-2} \\
		\hline
		Maximum Power Transfer (1) & \makecell[l]{\color{darkgray}P 5.6-3} \\
		\hline 
		Mathematical Expression Answer (4) & \makecell[l]{\color{darkgray}P 6.4-4, P 6.4-5, P 9.2-1, P 9.3-3} \\
		\hline 
		No Need for \texttt{ngspice} Simulation (1) & \makecell[l]{\color{darkgray}P 10.4-1} \\
		\hline
		\multicolumn{2}{l}{\parbox{12cm}{* The problem indices in this table are taken from \cite{svoboda2013introduction}.}}
	\end{tabular}
\end{table}

The text color of the problem indices in Table \ref{TableA1} indicates the results of the \texttt{ngspice} simulations when the problem diagrams are of textbook quality. The correctness of the solutions is manually determined by comparing the final answers obtained from \texttt{ngspice} simulations with the corresponding reference answers.

\begin{itemize}
	\item [--] Problems with {\color{darkgreen}green} indices were correctly solved by \texttt{ngspice} in the first trial.
	\item [--] Problems with {\color{blue}blue} indices indicate that the original circuit diagram recognitions from Gemini 2.5 Pro were inaccurate, resulting in incorrect \texttt{.cir} files for \texttt{ngspice} simulations. However, once the recognition issues were corrected manually, the \texttt{.cir} files were generated correctly on the first attempt, and the \texttt{ngspice} results were accurate.
	\item [--] For problems with {\color{brown}brown} indices, \texttt{ngspice} encountered issues when executing the Gemini-generated \texttt{.cir} file in the first trial. However, when Gemini was used to re-generate the \texttt{.cir} file for a second trial, the \texttt{ngspice} simulation ran successfully and produced correct results.
	\item [--] For problems with {\color{orange}orange} indices, \texttt{ngspice} ran successfully but produced incorrect results with the Gemini-generated \texttt{.cir} file in the first trial. When using the Gemini-regenerated \texttt{.cir} file, \texttt{ngspice} produced correct results in the second trial. This category differs from the {\color{blue}blue} one in that the circuit diagrams are correctly recognized for problems with {\color{orange}orange} indices but not for those with {\color{blue}blue} indices. Additionally, the difference between the {\color{orange}orange} and {\color{brown}brown} categories is that, in the former, \texttt{ngspice} ran successfully in the first trial, although the result was incorrect. In contrast, for problems with {\color{brown}brown} indices, \texttt{ngspice} did not produce any results due to implementation bugs.
	\item [--] For problems with {\color{red}red} indices, \texttt{ngspice} returned incorrect results in both the first and second trials.
	\item [--] Problems with {\color{darkgray}gray} indices make them unsuitable for \texttt{ngspice} simulations for various reasons, as detailed in the table.
\end{itemize}

Since \texttt{ngspice} is a rule-based simulation tool, any errors in its results stem from inaccuracies in the \texttt{.cir} files generated by Gemini 2.5 Pro. Therefore, if a second trial is necessary, we regenerate the \texttt{.cir} file based on the original problem statement and circuit description, and then execute the newly generated file using \texttt{ngspice}.

Table \ref{TableA1} shows that, out of 66 problems suitable for solution by \texttt{ngspice}, 56 were correctly solved on the first attempt, corresponding to a success rate of 84.85\%. In three cases, \texttt{ngspice} encountered implementation issues during the first trial and failed to produce results; however, correct results were obtained after regenerating the corresponding \texttt{.cir} files. Since such implementation issues are easily identifiable, the effective problem-solving success rate for \texttt{ngspice} can be considered $\frac{59}{66} \approx 89.39\%$ upon the first \textit{successful} trial. As a verification approach, some of the remaining errors can be identified and corrected by comparing the outputs of \texttt{ngspice} and Gemini 2.5 Pro, as will be discussed in Section \ref{S44}.

\subsection{Results of the Proposed Problem-Solving Method With \texttt{ngspice} Verification on Circuit Problems with Textbook-Quality Diagrams}
\label{S44}

Using the complete problem-solving pipeline illustrated in Figure \ref{framework}, all 83 problems introduced in Section \ref{S31} can be solved. In this section, we first describe an automated sub-pipeline that compares \texttt{ngspice}-simulated and Gemini-generated answers. We then present and analyze the results of the overall problem-solving framework on the circuit problem dataset consisting of textbook-quality diagrams.

\subsubsection{Automated Comparison Sub-pipeline Between \texttt{ngspice}-Simulated and Gemini-Generated Answers}
\label{S441}

As shown in Algorithm \ref{Alg1} and Figure \ref{framework}, the simulated outputs produced by \texttt{ngspice} can be used to verify the solutions generated by Gemini 2.5 Pro. To further automate the proposed framework, we developed a sub-pipeline that automatically compares the final answers produced by \texttt{ngspice} and Gemini 2.5 Pro. This sub-pipeline consists of the following steps.

\begin{itemize}
	\item[1.] \textit{Extract the numerical values of the required variables from the \texttt{ngspice} output}
	
	The output of the \texttt{.cir} file simulation contains the numerical values of the required variables, formatted as:
	\begin{small}
	\begin{verbatim}
		----------------------------------------------------
		Index [time/frequency] [required variable(s)]
		
		[The values are printed here line by line]
		----------------------------------------------------
	\end{verbatim}
	\end{small}
	\vspace{-0.4cm}
	Whether the second column uses \texttt{time} or \texttt{frequency} depends on whether the outputs are functions of time\footnote{To maintain a unified LLM prompt format for \texttt{.cir} file generation, we configure \texttt{ngspice} to output the required circuit variables as numerical values with respect to time, regardless of whether the circuit operates in steady state or exhibits time-varying behavior. For steady-state circuits, all variables are constant functions of time $t$.} (i.e., ``the Circuit Analysis'' and ``Circuit Synthesis'' categories in Table \ref{TableA1}) or frequency (i.e., the ``Network Function Analysis'' and ``Network Function Synthesis'' categories in Table \ref{TableA1}). If multiple variables are required, each variable appears in a separate column. For problems that require a network function, the output includes the numerical values of both the magnitude and phase of the network function. An example of the \texttt{ngspice} output is provided in \ref{SB32}.
	
	Given the clear and consistent format, we can reliably apply regular expressions to extract the simulated numerical values of each required variable and store them in a \texttt{json} file.
	
	\item[2.] \textit{Convert the Gemini-generated textual solutions to numerical solutions}
	
	The solutions generated by Gemini are expressed in natural language. To enable direct comparison with the \texttt{ngspice}-simulated results, these textual outputs must be converted into numerical values defined on the same time or frequency steps. To achieve this, we instruct Gemini 2.5 Pro to produce a Python script that can be executed directly to compute the numerical values corresponding to the Gemini-generated solution. The inputs to the prompt include:
	\begin{itemize}
		\item the Gemini-generated solution,
		\item the summarized final answer that directly and concisely addresses the problem, and
		\item the time or frequency range used in the \texttt{ngspice} simulation.
	\end{itemize}
	
	The prompt template for this conversion is provided in \ref{C4}. Using the information above, we can generate numerical values of the Gemini-derived solution \textit{over the same time or frequency interval} as the \texttt{ngspice} output. These values are further processed using linear interpolation to obtain numerical values that align \textit{exactly with the time or frequency steps} of the \texttt{ngspice} simulation. Similar to the processing of \texttt{ngspice} results in Step 1, the final numerical outputs can also be stored in a \texttt{json} file. 
	
	\item [3.] \textit{Compare \texttt{ngspice}-simulated and Gemini-generated solutions}
	
	Given two \texttt{json} files containing the numerical solutions produced by \texttt{ngspice} and Gemini 2.5 Pro at the same time or frequency steps, a Python script can be used to compare the two results for consistency. In this process, we first compute the differences between the two numerical solutions and then verify whether all differences are sufficiently small. However, for certain transient circuits, \texttt{ngspice} simulations may require time to reach a steady state. In such cases, discrepancies during the transient phase are permissible, provided that the steady-state values ultimately agree. Therefore, failure of the global check described above does not necessarily indicate a mismatch. To account for this, when the global check fails, we perform an additional local check that considers only the final 5\% of data points in the time or frequency domain. If this local check passes, we still consider the results from \texttt{ngspice} and Gemini 2.5 Pro to be consistent. Our results demonstrate that this procedure successfully classifies matched and mismatched solutions with 100\% accuracy across the 66 circuit problems evaluated.
\end{itemize}

\subsubsection{Result Summary of the Proposed Circuit Problem-Solving Framework on Problems with Textbook-Quality Diagrams}

Table \ref{detailed_results} summarizes the results of the proposed Gemini-based circuit problem-solving pipeline and the \texttt{ngspice} simulations on the circuit problem dataset consisting of textbook-quality diagrams. In the second Gemini trial, the model temperature was set to 0.2 to slightly increase response randomness, with the aim of potentially generating alternative correct solutions. When calculating the correctness ratio, a solution is considered correct as long as its final answer is accurate. We have manually verified that the overall problem-solving processes are valid whenever the final answers are correct.

Across three respective trials, the cumulative correct ratios improve as follows: 85.54\% $\to$ 91.57\% $\to$ 93.98\% for Gemini 2.5 Pro, and 89.39\% $\to$ 93.94\% $\to$ 95.45\% for the \texttt{ngspice} simulations. With human-in-the-loop corrections, the cumulative correctness reaches 97.59\%. Compared to the baseline correct ratio of 79.52\% when using Gemini 2.5 Pro alone, our pipeline improves performance by 18.07 percentage points (a relative improvement of 22.72\%).

Two problems, P 8.4-1 and P 10.6-4, were solved incorrectly even after the full pipeline. For Problem 8.4-1, the initial circuit recognition was incorrect, leading both Gemini and \texttt{ngspice} to solve a misinterpreted circuit. As a result, they produced matching but incorrect solutions. This issue was not flagged by the \texttt{ngspice} verification mechanism introduced in this paper, underscoring the critical importance of accurate circuit diagram recognition for correct solution generation. Additionally, we emphasize that this error---caused by incorrect circuit recognition and not detected by the automated verification loop---is still counted as an incorrect solution in the reported accuracy, since all reported accuracies are determined through manual comparison between the framework-generated solutions and the ground-truth solutions. The failure of Problem 10.6-4 stems from its complexity. This problem requires computing the values of three phasor mesh currents. After formulating the KVL equations for the three meshes, the following system of equations must be solved:
$$
\left[
\begin{array}{ccc}
	65 - j 35 & 0 & -25 + j 50 \\
	0 & 32 - j 34 & j 50 \\
	-25 + j 50 & j 50 & 25 - j 60
\end{array}
\right] 
\left[
\begin{array}{l}
	\mathbf{I}_1 \\\mathbf{I}_2 \\\mathbf{I}_3
\end{array}
\right] 
= \left[
\begin{array}{c} 48 \angle 75^{\circ} \\-48 \angle 75^{\circ} \\0 \end{array}
\right]
$$
This level of mathematical analysis presents a significant challenge for Gemini 2.5 Pro, indicating that the mathematical reasoning capabilities of current state-of-the-art LLMs still require further advancement.

The results in Table \ref{detailed_results} also provide ablation studies of the key components in the proposed framework. We summarize the main observations as follows.

\begin{itemize}
	\item[i)] In the first Gemini trial, 12 problems were solved incorrectly, and \texttt{ngspice} successfully flagged 11 of these cases by producing different final answers. For problems with incorrectly recognized circuit diagrams, \texttt{ngspice} and Gemini do not necessarily converge to the same (incorrect) solutions, likely because vision hallucinations in VLMs are largely random and may render a circuit harder or even unsolvable. This high flagging ratio demonstrates the effectiveness of the proposed \texttt{ngspice}-based verification method.
	
	\item[ii)] Among the 11 problems that Gemini 2.5 Pro solved incorrectly in the first trial, five were solved correctly simply by rerunning Gemini. This highlights the importance of verifying solutions with an external simulator such as \texttt{ngspice} and identifying Gemini's incorrect outcomes, since a large portion ($\frac{5}{11}\approx45.45\%$) can be corrected with minimal intervention.
	
	\item[iii)] Similarly, three out of the 12 problems incorrectly solved by \texttt{ngspice} in its first trial were solved correctly in the second trial. From this perspective, Gemini 2.5 Pro and \texttt{ngspice} effectively serve as cross-verification mechanisms for each other.
	
	\item[iv)] Manual proofreading of circuit recognition is performed only for problems in which the solutions generated by Gemini 2.5 Pro and \texttt{ngspice} do not match on their first attempts. Thus, the human proofreader is required to review only 12 circuit diagrams and examine 5 + 17 = 22 complete Gemini-generated solutions\footnote{The 22 Gemini-generated solutions requiring manual revision include 17 problems that are unsuitable for \texttt{ngspice} simulation (as indicated in Table \ref{TableA1}) and 5 problems where the solutions from \texttt{ngspice} and Gemini do not match after three attempts from each tool (as shown in Table \ref{detailed_results}). It is worth noting that Gemini 2.5 Pro accurately solved all 17 problems unsuitable for \texttt{ngspice} simulation, thereby minimizing the effort needed by the human instructor to correct those solutions.}, corresponding to 15.66\% and 26.51\% of the 83 problems, respectively. This demonstrates that the 
\end{itemize}
	
\begin{landscape}
	\begin{table}[h!]
		\centering
		\footnotesize
		\caption{Results of the proposed Gemini-based circuit problem-solving pipeline and \texttt{ngspice} simulations}
		\label{detailed_results}
		\scalebox{0.9}{
			\begin{tabular}{p{2.5cm}p{3.5cm}p{2.5cm}p{2.25cm}p{2.75cm}p{2.75cm}p{2.75cm}}
				\hline
				\multicolumn{2}{c}{\multirow{2}{*}{\# Trial}} & \makecell{1st Gemini trial} & \multicolumn{2}{c}{\makecell{2nd Gemini trial (\texttt{temperature=0.2})}} & \makecell{3rd Gemini trial} & \multirow{2}{*}{\makecell{4th Gemini trial w/ \\human in the loop}}\\ \cline{3-6}
				& & \multicolumn{2}{c}{\makecell{1st \texttt{ngspice} trial}} & \makecell{2nd \texttt{ngspice} trial} & \makecell{3rd \texttt{ngspice} trial} & \\
				\hline
				\multirow{2.1}{*}{\makecell{Cumulative\\Correct Ratio}} & \makecell{Gemini} & \makecell{71/83 (85.54\%)} & \multicolumn{2}{c}{\makecell{76/83 (91.57\%)}} & \makecell{77/83 (93.98\%)} & \makecell{81/83 (97.59\%)} \\ \cline{2-7}
				& \makecell{\texttt{ngspice}} & \multicolumn{2}{c}{\makecell{59/66 (89.39\%)}} & \makecell{62/66 (93.94\%)} & \makecell{63/66 (95.45\%)} & \makecell{---} \\ 
				\hline
				\multicolumn{2}{c}{\makecell{Circuit Recognition Proofreading}} & \makecell{No} & \makecell{No} & \makecell{No} & \makecell{Yes} & \makecell{Yes} \\
				\hline
				\multirow{17}{*}{\makecell{Reason of \\Incorrectness}} 
				& \makecell{Incorrect recognitions\\for source polarities} & \makecell{None} & \makecell{None} & \makecell{None} & \makecell{None} & \makecell{None} \\ \cline{2-7}
				& \makecell{Hallucinations of\\current directions} & \makecell{P 3.4-4$^{\color{darkgreen}\dagger}$ \\ P 4.6-2$^{\color{orange}\dagger}$ \\ P 4.7-2$^{\color{darkgreen}\dagger}$ \\ \textit{P 5.2-1}$^{\color{red}\dagger}$ \\ P 5.3-6$^{\color{darkgreen}\dagger}$ \\ P 5.5-5$^{\color{darkgreen}\dagger}$ \\ P 6.3-2$^{\color{darkgreen}\dagger}$ \\ P 8.3-4$^{\color{darkgreen}\dagger}$ \\ P 10.8-1$^{\color{darkgreen}\dagger}$} & \makecell{P 3.4-4$^{\color{darkgreen}\dagger}$ \\ P 4.6-2$^{\color{orange}\dagger}$ \\ \textit{P 5.2-1}$^{\color{red}\dagger}$ \\ P 5.5-5$^{\color{darkgreen}\dagger}$} & \makecell{P 3.4-4$^{\color{darkgreen}\dagger}$ \\ P 4.6-2$^{\color{orange}\dagger}$ \\ \textit{P 5.2-1}$^{\color{red}\dagger}$ \\ P 5.5-5$^{\color{darkgreen}\dagger}$} & \makecell{\color{darkgray}{P 3.4-4}$^{\color{darkgreen}\dagger}$ \\ \color{darkgray}{P 4.6-2}$^{\color{orange}\dagger}$ \\ \textit{P 5.2-1}$^{\color{red}\dagger}$ \\ \color{darkgray}P 5.5-5$^{\color{darkgreen}\dagger}$} & \makecell{None} \\ \cline{2-7}
				& \makecell{Hallucinations of\\element connections} & \makecell{\fbox{P 2.5-2} \\ \textit{P 5.2-1}$^{\color{red}\dagger}$ \\ {P 8.4-1} \\ P 9.6-1$^{\color{blue}\dagger}$} & \makecell{\textit{P 5.2-1}$^{\color{red}\dagger}$ \\ \color{darkgray}{P 8.4-1} \\ P 9.6-1$^{\color{blue}\dagger}$} & \makecell{\textit{P 5.2-1}$^{\color{red}\dagger}$ \\ \color{darkgray}{P 8.4-1} \\ P 9.6-1} & \makecell{\textit{P 5.2-1}$^{\color{red}\dagger}$ \\ \color{darkgray}{P 8.4-1}} & \makecell{\color{darkgray}{P 8.4-1}} \\ \cline{2-7}
				& \makecell{Hallucinations of\\circuit analyses} & \makecell{P 10.6-4$^{\color{red}\dagger}$} & \makecell{P 10.6-4$^{\color{red}\dagger}$} & \makecell{P 10.6-4$^{\color{red}\dagger}$} & \makecell{\color{darkgray}P 10.6-4$^{\color{red}\dagger}$} & \makecell{P 10.6-4} \\ 
				\hline
				\multicolumn{2}{c}{\makecell{Correct for Gemini trials\\ but incorrect for \texttt{ngspice} trials}} & \makecell{P 2.5-1$^{\color{orange}\dagger}$ \\ P 8.7-3$^{\color{orange}\dagger}$} & \makecell{P 2.5-1$^{\color{orange}\dagger}$ \\ P 8.7-3$^{\color{orange}\dagger}$} & \makecell{None} & \makecell{None} & \makecell{None} \\
				\hline
				\multicolumn{7}{l}{\parbox{21.75cm}{$^{\dagger}$ Problem indices marked with a dagger ($\dagger$) indicate discrepancies between the solutions provided by Gemini 2.5 Pro and \texttt{ngspice}. These problems are highlighted because either the Gemini solution, the \texttt{ngspice} solution, or both are incorrect. The colors of the daggers match those used in Table \ref{TableA1} and indicate the outcomes of the \texttt{ngspice} simulations. \\
						* Problem indices in italic font (e.g., \textit{P 5.2-1}) indicate that there are multiple reasons for their incorrectness. \\
						** Problem \fbox{P 2.5-2}, highlighted with a box, involves hallucinated element connections during the circuit recognition step. However, the textual description states that ``a current source and a voltage source are connected in series with a resistor,'' which helps mitigate the recognition errors. As a result, both the Gemini and \texttt{ngspice} solutions are correct.\\
						*** Problem indices in {\color{darkgray}gray} indicate cases that were not tested at the corresponding step. This is either because both Gemini and \texttt{ngspice} produced matching but incorrect final solutions (e.g., {\color{darkgray}P 8.4-1}), or because no circuit recognition errors were identified by human reviewers during proofreading (e.g., {\color{darkgray}P 10.6-4}).}}
		\end{tabular}}
	\end{table}
\end{landscape}

\begin{itemize}	
	\item [] proposed framework can substantially reduce the time and effort required from instructors when preparing reference solutions. Moreover, among the two human-in-the-loop steps, providing solution feedback in the fourth Gemini trial yields a larger marginal improvement in accuracy than proofreading the circuit diagrams. This observation indicates that, within our pipeline, Gemini 2.5 Pro faces greater challenges in circuit analysis than in circuit-diagram recognition. With these limitations identified, future work may focus on enhancing the fundamental reasoning capability of LLMs through targeted fine-tuning.
\end{itemize}

Table \ref{tab:ablation} summarizes the marginal gain of each module in terms of problem-solving accuracy, where the values are derived from Tables \ref{table:1} and \ref{detailed_results}. Each module---namely, i) YOLO model and OpenCV-based recognition enhancement, ii) \texttt{ngspice}-driven verification, and iii) human-in-the-loop feedback---leads to an additional 5 correctly solved problems, combining to improve the overall problem-solving accuracy from 79.52\% to 97.59\%. These ablation results also highlight why physical simulation alone (i.e., direct \texttt{ngspice} integration without localized vision preprocessing) is insufficient. Without YOLO and OpenCV, source-polarity misidentifications---the primary cause of baseline failures, accounting for 35.29\% of the errors, as shown in Table \ref{table:1}---propagate directly into the synthesized netlists. Consequently, because both the analytical reasoning and the simulation inherit the same visual misinterpretations, \texttt{ngspice} evaluates an erroneous circuit, potentially resulting in false-positive consensus. Upstream localized preprocessing is therefore essential for establishing more faithful topological grounding before simulation-based verification.

\begin{table}[!t]
	\centering
	\footnotesize
	\caption{The marginal gain of each module in terms of problem-solving accuracy}
	\label{tab:ablation}
	\begin{tabular}{ccc}
		\hline
		Method & Accuracy & Marginal Gain  \\
		\hline
		Gemini baseline & 66/83 (79.52\%) & --- \\
		\makecell{+ YOLO model and OpenCV-based \\recognition enhancement}  & 71/83 (85.54\%) & 5/83 (6.02\%) \\
		+ \texttt{ngspice}-driven verification  & 76/83 (91.57\%) & 5/83 (6.02\%) \\
		+ Human-in-the-loop feedback & 81/83 (97.59\%) & 5/83 (6.02\%)  \\
		\hline
	\end{tabular}
\end{table}

Note that the solutions generated by Gemini 2.5 Pro can be easily converted into different formats for various educational purposes. For example, a concise reference solution containing only the essential steps can be distributed to students as course material after each homework deadline, while more detailed solutions and high-level summaries can be prepared as problem-specific context for an AI-enabled online tutor \citep{chen2025wip}. An example of circuit problem solving, including detailed intermediate steps from both \texttt{ngspice} and Gemini, is provided in \ref{SB}.

\subsection{Experiments and Results on Circuit Problems With Hand-Drawn Diagrams}
\label{S45}
Sections \ref{S43} and \ref{S44} showed that the proposed pipeline can improve circuit-problem-solving accuracy from 79.52\% to 97.59\% on a dataset with textbook-quality circuit diagrams. In this section, we detail the experiments and results obtained with various types of hand-drawn circuit diagrams to demonstrate the generalizability and robustness of the designed pipeline. Additionally, we report the time and cost of implementing our framework to solve circuit problems.

\subsubsection{Hand-Drawn Circuit Diagram Datasets}
\label{S451}
As summarized in Table \ref{TableA1}, among the 83 circuit problems, 66 are suitable for both Gemini 2.5 Pro and \texttt{ngspice}. To evaluate the capability of the full proposed pipeline on circuit problems with hand-drawn diagrams, these 66 problems are selected as the test set in this section.\footnote{Some problems (e.g., P5.2-1) contain multiple circuit diagrams; in such cases, the diagrams are treated as a single combined diagram.} The four variations of the corresponding hand-drawn circuit diagram datasets are introduced below.

\begin{figure}
\centering
\begin{subfigure}{\textwidth}
\centering
\includegraphics[width=0.55\textwidth]{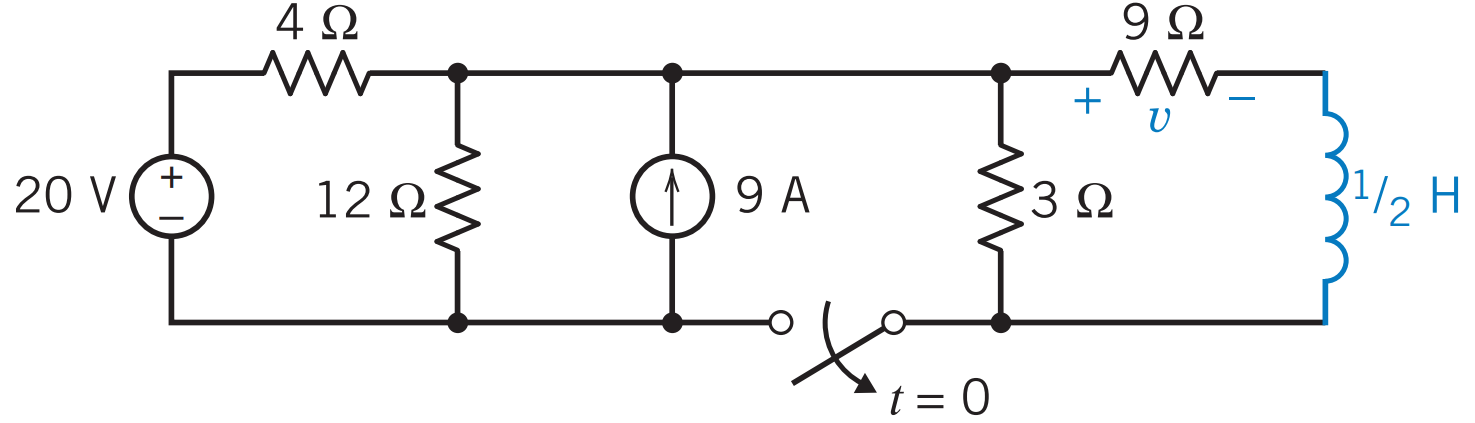}
\caption{Example textbook-quality circuit diagram}
\label{fig:handdrawn_examples_a}
\end{subfigure}
\vspace{-0.75em}

\begin{subfigure}{\textwidth}
\centering
\includegraphics[width=0.55\textwidth]{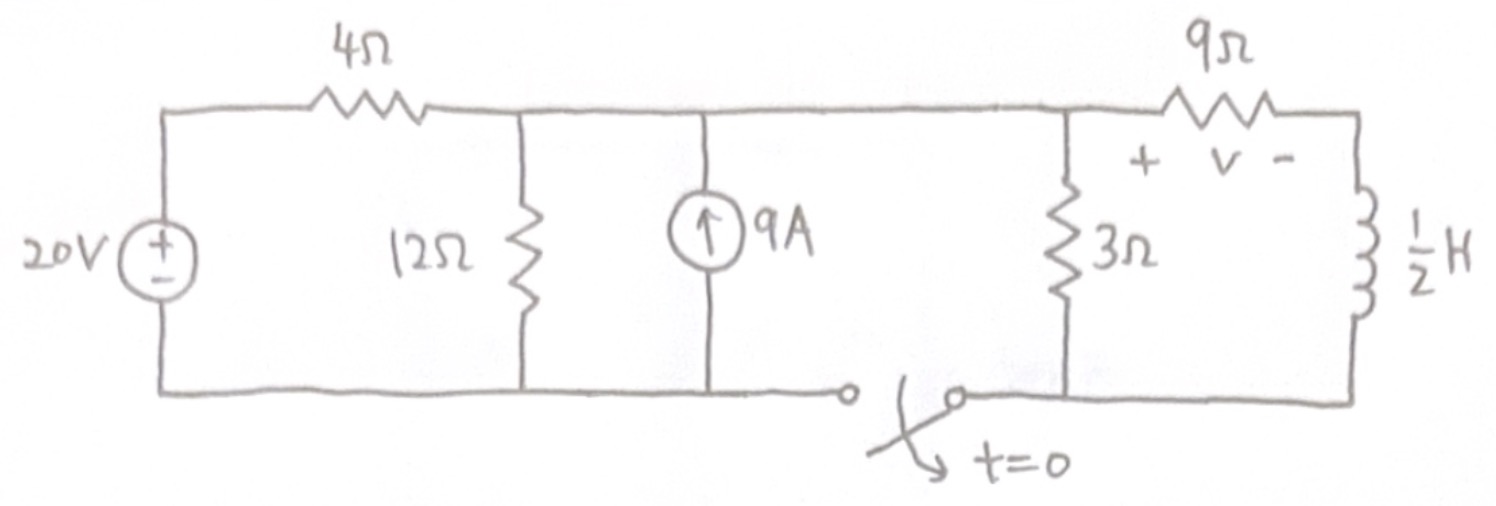}
\caption{Example hand-drawn circuit diagram of type i)}
\label{fig:handdrawn_examples_b}
\end{subfigure}
\vspace{-0.75em}

\begin{subfigure}{\textwidth}
\centering
\includegraphics[width=0.55\textwidth]{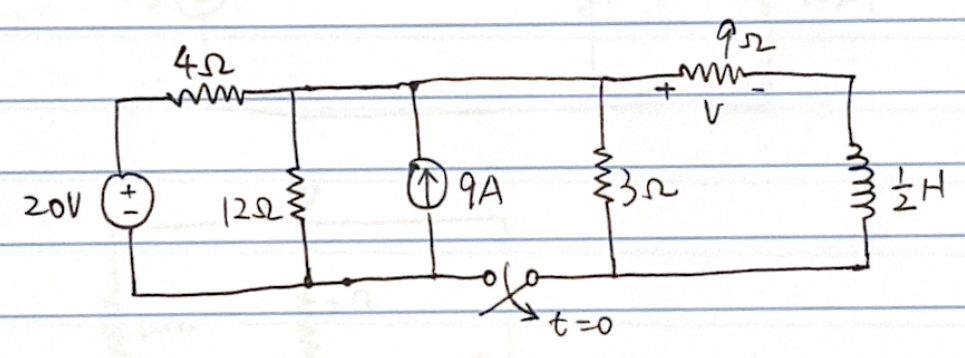}
\caption{Example hand-drawn circuit diagram of type ii)}
\label{fig:handdrawn_examples_c}
\end{subfigure}
\vspace{-0.75em}

\begin{subfigure}{\textwidth}
\centering
\includegraphics[width=0.55\textwidth]{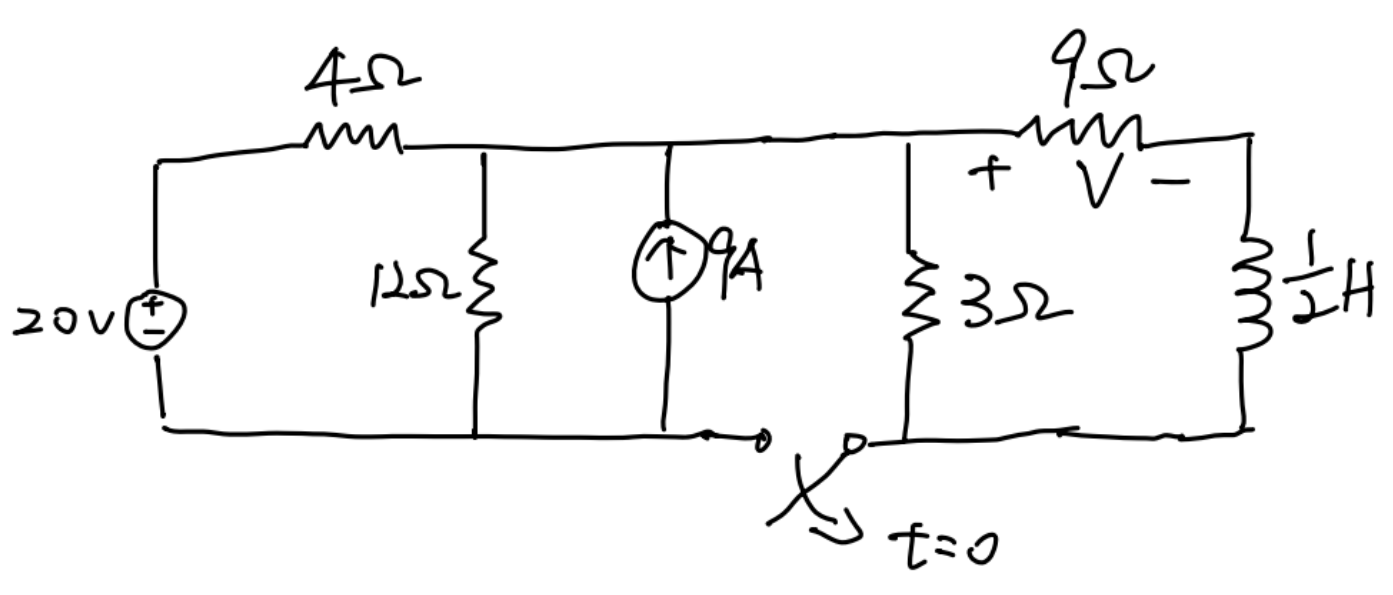}
\caption{Example hand-drawn circuit diagram of type iii)}
\label{fig:handdrawn_examples_d}
\end{subfigure}
\vspace{-0.75em}

\begin{subfigure}{\textwidth}
\centering
\includegraphics[width=0.55\textwidth]{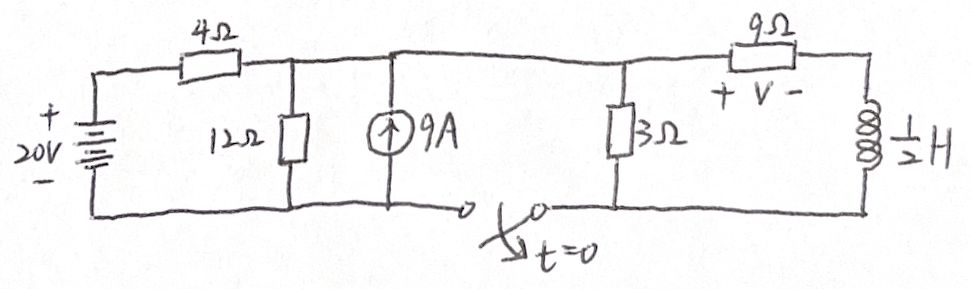}
\caption{Example hand-drawn circuit diagram of type iv)}
\label{fig:handdrawn_examples_e}
\end{subfigure}
\caption{Examples of textbook-quality and various hand-drawn circuit diagrams for Problem 8.3-10 in \cite{svoboda2013introduction}. Subfigure (e), corresponding to type iv), adopts IEC standard symbols for the 20 V DC voltage source, the resistors, and the $\tfrac{1}{2}$ H inductor.}
\label{fig:handdrawn_examples}
\end{figure}

\begin{itemize}
\item[i)] \textit{Scanned hand-drawn circuit diagrams on white paper:} Circuit diagrams were drawn on white paper and subsequently scanned to produce digital images suitable for input to Gemini 2.5 Pro.
\item[ii)] \textit{Scanned hand-drawn circuit diagrams on lined paper:} Unlike type i), the circuit diagrams were drawn on lined paper and then scanned to generate digital images.
\item[iii)] \textit{Screenshots of hand-drawn circuit diagrams from a digital notebook:} In this variation, circuit diagrams were hand-drawn using a digital notebook (e.g., an iPad with Apple Pencil) and captured as screenshots to produce digital circuit images.
\item[iv)] \textit{Scanned hand-drawn circuit diagrams on white paper with IEC symbol standards:} The textbook \cite{svoboda2013introduction} and the diagrams in types i)--ii) adopt IEEE symbol standards. In this variation, the symbols of elements such as resistors, inductors, and DC voltage sources were converted from IEEE to IEC standards, as illustrated in Figure \ref{fig:handdrawn_examples}(e). Similar to type i), the diagrams were drawn on white paper and then scanned to generate digital images.
\end{itemize}

The four types of hand-drawn circuit diagrams were created by different authors of this paper, each exhibiting distinct drawing and handwriting styles, thereby enhancing the stylistic diversity of the dataset. Figure \ref{fig:handdrawn_examples} presents representative examples of hand-drawn circuit diagrams for Problem 8.3-10 in \cite{svoboda2013introduction}.

\subsubsection{Results With Hand-Drawn Circuit Diagrams}
\label{S452}

With the four types of hand-drawn circuit diagram variations, we apply the proposed pipeline to solve circuit problems in each dataset. The results presented in this section include not only the problem-solving accuracy, characterized by the consistency between the reference final answers and those derived using the proposed pipeline, but also the time and cost of the problem-solving process to demonstrate the scalability of the method. As baselines, Table \ref{Tab3} reports the accuracies, times, and costs obtained by directly leveraging Gemini 2.5 Pro for circuit problem solving without any assistance. The results for textbook-style circuit diagrams are also included as baselines. Note that the times and costs reported in the tables of this section, including Tables \ref{Tab3}--\ref{Tab2}, account only for the processing time of Gemini and \texttt{ngspice}, as well as the detection time of the YOLO model and OpenCV, where applicable. The time required for human review and commenting in the human-in-the-loop process is estimated separately. The reported accuracy values are presented in the format of ``number of correct circuit recognitions or problem solutions / total number of circuit diagrams or problems in the corresponding step (percentage)''. The time and cost values are reported in the form of ``mean $\pm$ standard deviation''.

\begin{table}[!t]
	\centering
	\scriptsize
	\caption{Results of the original Gemini 2.5 Pro}
	\label{Tab3}
	\begin{tabular}{cccc}
		\hline
		Circuit diagram type & Accuracy & Time (s) & LLM cost (\textcent) \\
		\hline
		Textbook-style & 46/66 (69.70\%) & $102.50\pm74.47$ & $14.11\pm10.21$ \\
		Scanned hand-drawn (white paper) & 47/66 (71.21\%) & $74.95\pm54.66$ & $11.03\pm8.15$ \\
		Scanned hand-drawn (lined paper) & 41/66 (62.12\%) & $84.80\pm52.85$ & $11.70\pm7.46$ \\
		Screenshot of hand-drawn (digital notebook) &  40/66 (60.61\%) & $94.38\pm70.78$ & $13.38\pm11.13$ \\
		\makecell{Scanned hand-drawn \\(IEC symbol standard) (white paper)} & 45/66 (68.18\%) & $83.29\pm62.54$ & $11.87\pm9.02$ \\
		\hline
		Total & 219/330 (66.36\%) & $87.98\pm64.34$ & $12.42\pm9.36$ \\
		\hline
	\end{tabular}
\end{table}

Table \ref{Tab1} presents the results of Gemini 2.5 Pro in circuit problem solving using the proposed pipeline with different types of circuit diagrams. The results are reported in detail for the circuit recognition stage and for the first through fourth trials. For the circuit recognition stage, the reported times and costs represent the aggregated values of all five steps detailed in Figure \ref{diagram_recognition} and Table \ref{io_diagram_recognition}, including Gemini processing as well as YOLO and OpenCV detection. The accuracy is determined based on human review of the generated circuit textual description (\texttt{.txt}) files. In this process, a circuit recognition result is considered inaccurate if any minor inconsistency appears in the corresponding description. The average totals are computed as aggregated values across the different stages or trials.

The circuit recognition accuracies reported in Table \ref{Tab1} indicate inferior performance on hand-drawn circuit diagrams compared to textbook-style diagrams. This outcome is expected, as hand-drawn diagrams are typically noisier and exhibit greater stylistic variability, which can introduce random errors in the extracted textual circuit descriptions. However, most problems with erroneous circuit recognition yield inconsistent final answers between Gemini 2.5 Pro and \texttt{ngspice}. This discrepancy is reflected in the larger number of problems requiring the third trial for the hand-drawn circuit diagrams, in which human intervention is introduced to proofread the circuit diagrams for problems whose solutions from Gemini 2.5 Pro and \texttt{ngspice} differ in the first trial. With this circuit proofreading and correction step, followed by a human-in-the-loop feedback process in the fourth trial, the final cumulative accuracies for the four types of hand-drawn circuit diagrams reach 89.39\%---92.42\%, approaching the 96.97\% accuracy achieved for textbook-style diagrams despite their substantially lower circuit recognition accuracies. The detailed causes of the remaining incorrect solutions are analyzed in Section \ref{S48}.

Table \ref{Tab2} presents the results of \texttt{ngspice} in circuit problem solving using different types of circuit diagrams. For cases involving hand-drawn circuit diagrams, the cumulative accuracies are lower than those for textbook-style diagrams. This discrepancy is primarily attributed to the lower circuit recognition accuracy for hand-drawn diagrams, as shown in Table \ref{Tab1}.  However, it is important to emphasize that the role of \texttt{ngspice} in our pipeline is to 

\begin{landscape}
\begin{table}
\centering
\scriptsize
\caption{Results of Gemini 2.5 Pro in circuit problem solving using different types of circuit diagrams}
\label{Tab1}
\begin{tabular}{cccccccc}
\hline
Circuit diagram type & & Circuit recognition & 1st trial & 2nd trial & 3rd trial & 4th trial & Average total \\
\hline
\multirow{3}{*}{\makecell{Textbook-style}} & Accuracy & 64/66 (96.97\%) & 54/66 (81.82\%) & 5/11 (45.45\%) & 1/2 (50.00\%) & 4/5 (80.00\%) & 64/66 (96.97\%) \\
& Time (s) & $63.10\pm18.25$ & $78.88\pm43.20$ & $100.55\pm40.32$ & $106.39\pm15.19$ & --- & $165.77\pm91.07$ \\
& LLM cost (\textcent) & $7.98\pm2.60$ & $11.67\pm6.48$ & $14.51\pm5.74$ & $15.16\pm2.54$ & --- & $23.08\pm13.23$ \\
\hline
\multirow{3}{*}{\makecell{Scanned hand-drawn\\(white paper)}} & Accuracy & 56/66 (84.85\%) & 50/66 (75.76\%) & 11/18 (61.11\%) & 3/6 (50.00\%) & 11/12 (91.67\%) & 60/66 (90.91\%) \\
& Time (s) & $64.46\pm17.56$ & $81.26\pm49.67$ & $105.53\pm51.72$ & $112.43\pm44.14$ & --- & $184.72\pm119.88$ \\
& LLM cost (\textcent) & $8.17\pm2.49$ & $11.91\pm7.18$ & $14.99\pm7.21$ & $15.37\pm6.22$ & --- & $25.56\pm17.01$ \\
\hline
\multirow{3}{*}{\makecell{Scanned hand-drawn\\(lined paper)}} & Accuracy & 53/66 (80.30\%) & 50/66 (75.76\%) & 10/17 (58.82\%) & 5/9 (55.56\%) & 6/8 (75.00\%) & 59/66 (89.39\%) \\
& Time (s) & $65.78\pm25.22$ & $73.52\pm38.08$ & $85.62\pm29.84$ & $85.42\pm29.48$ & --- & $173.00\pm94.58$ \\
& LLM cost (\textcent) & $7.92\pm3.26$ & $10.62\pm5.65$ & $12.16\pm4.54$ & $11.93\pm4.35$ & --- & $23.30\pm13.66$ \\
\hline
\multirow{3}{*}{\makecell{Screenshot of hand-drawn\\(digital notebook)}} & Accuracy & 58/66 (87.88\%) & 52/66 (78.79\%) & 7/14 (50.00\%) & 3/4 (75.00\%) & 10/10 (100.00\%) & 60/66 (90.91\%) \\
& Time (s) & $68.38\pm50.44$ & $80.40\pm55.13$ & $103.82\pm73.40$ & $105.40\pm37.75$ & --- & $177.19\pm133.46$ \\
& LLM cost (\textcent) & $8.88\pm8.27$ & $11.44\pm7.65$ & $14.31\pm9.50$ & $14.37\pm4.91$ & --- & $24.23\pm18.69$ \\
\hline
\multirow{3}{*}{\makecell{Scanned hand-drawn\\(IEC symbol standard)\\(white paper)}} & Accuracy & 57/66 (86.36\%) & 51/66 (77.27\%) & 8/20 (40.00\%) & 5/8 (62.50\%) & 10/12 (83.33\%) & 61/66 (92.42\%) \\
& Time (s) & $61.80\pm17.41$ & $73.14\pm39.59$ & $103.89\pm46.23$ & $137.84\pm99.74$ & --- & $183.13\pm140.08$ \\
& LLM cost (\textcent) & $7.52\pm2.21$ & $10.69\pm5.73$ & $14.72\pm6.39$ & $20.46\pm15.15$ & --- & $25.16\pm19.99$ \\
\hline
\multirow{3}{*}{Total} & Accuracy & 288/330 (87.27\%) & 257/330 (77.88\%) & 39/80 (48.75\%) & 17/29 (58.62\%) & 41/47 (87.23\%) & 304/330 (92.12\%) \\
& Time (s) & $64.70\pm28.81$ & $77.44\pm45.71$ & $99.93\pm50.15$ & $111.77\pm65.24$ & --- & $176.76\pm117.72$ \\
& LLM cost (\textcent) & $8.09\pm4.42$ & $11.27\pm6.61$ & $14.15\pm6.87$ & $15.85\pm9.83$ & --- & $24.26\pm16.76$ \\
\hline
\end{tabular}
\end{table}
\end{landscape}

\begin{landscape}
	\begin{table}
		\centering
		\footnotesize
		\caption{Results of \texttt{ngspice} in circuit problem solving using different types of circuit diagrams}
		\label{Tab2}
		\begin{tabular}{cccccc}
			\hline
			Circuit diagram type & & 1st trial & 2nd trial & 3rd trial & Average total \\
			\hline
			\multirow{3}{*}{\makecell{Textbook-style}} & Accuracy & 59/66 (89.39\%) & 8/11 (72.73\%) & 1/2 (50.00\%) & 63/66 (95.45\%) \\
			& Time (s) & $108.64\pm94.06$ & $120.49\pm64.84$ & $113.89\pm25.01$ & $136.31\pm132.33$ \\
			& LLM cost (\textcent) & $16.06\pm13.28$ & $17.13\pm9.19$ & $16.31\pm3.49$ & $19.99\pm18.78$ \\
			\hline
			\multirow{3}{*}{\makecell{Scanned hand-drawn\\(white paper)}} & Accuracy & 53/66 (80.30\%) & 13/18 (72.22\%) & 2/6 (33.33\%) & 54/66 (81.82\%) \\
			& Time (s) & $113.25\pm95.37$ & $119.35\pm41.67$ & $131.36\pm70.04$ & $157.75\pm139.04$ \\
			& LLM cost (\textcent) & $17.04\pm13.86$ & $17.31\pm5.76$ & $19.24\pm9.88$ & $23.50\pm20.21$ \\
			\hline
			\multirow{3}{*}{\makecell{Scanned hand-drawn\\(lined paper)}} & Accuracy & 51/66 (77.27\%) & 11/17 (56.87\%) & 6/9 (66.67\%) & 55/66 (83.33\%) \\
			& Time (s) & $108.18\pm79.77$ & $116.99\pm56.87$ & $120.30\pm49.25$ & $154.72\pm135.30$ \\
			& LLM cost (\textcent) & $15.70\pm11.25$ & $16.47\pm8.06$ & $17.60\pm6.84$ & $22.34\pm19.23$ \\
			\hline
			\multirow{3}{*}{\makecell{Screenshot of hand-drawn\\(digital notebook)}} & Accuracy & 53/66 (80.30\%) & 8/14 (57.14\%) & 0/4 (0.00\%) & 51/66 (77.27\%) \\
			& Time (s) & $117.67\pm82.66$ & $127.77\pm65.14$ & $196.46\pm67.24$ & $156.68\pm132.89$ \\
			& LLM cost (\textcent) & $16.70\pm11.13$ & $17.79\pm8.44$ & $27.46\pm9.58$ & $22.14\pm18.23$ \\
			\hline
			\multirow{3}{*}{\makecell{Scanned hand-drawn\\(IEC symbol standard)\\(white paper)}} & Accuracy & 53/66 (80.30\%) & 11/20 (55.50\%) & 6/8 (75.00\%) & 58/66 (87.88\%) \\
			& Time (s) & $106.32\pm70.12$ & $106.05\pm29.09$ & $104.36\pm31.13$ & $151.10\pm110.49$ \\
			& LLM cost (\textcent) & $15.73\pm10.17$ & $14.99\pm4.01$ & $15.35\pm4.26$ & $22.14\pm15.85$ \\
			\hline
			\multirow{3}{*}{Total} & Accuracy & 269/330 (80.91\%) & 51/80 (63.75\%) & 15/29 (51.72\%) & 281/330 (85.15\%) \\
			& Time (s) & $110.81\pm85.02$ & $117.27\pm52.15$ & $129.66\pm60.04$ & $151.31\pm130.63$ \\
			& LLM cost (\textcent) & $16.25\pm12.03$ & $16.63\pm7.19$ & $18.80\pm8.30$ & $22.02\pm18.55$ \\
			\hline
		\end{tabular}
	\end{table}
\end{landscape}

\noindent
verify the correctness of Gemini-generated solutions and to identify incorrect ones. The high final problem-solving accuracies achieved for both textbook-style and hand-drawn diagrams demonstrate that \texttt{ngspice} effectively fulfills this role.

For the time and cost of the proposed pipeline\footnote{With human-in-the-loop solution feedback in the fourth trial, online LLM platforms can be leveraged to interactively incorporate human feedback into the problem-solving process. The time and cost for each interaction are typically less than 3 minutes and 15 cents, respectively. In our experiments, we used Google AI Studio (\url{https://aistudio.google.com}) for all fourth trials. Since these time and cost estimates are obtained from interactive usage rather than being strictly measured through controlled time and token counts (as in Table \ref{Tab1}, which are computed via our Python scripts and the Gemini API), they are not directly comparable to the reported values. Therefore, these estimates are not included in Table \ref{Tab1}.}, although there are variations across different problems and circuit diagram types, the average cumulative time required to solve a single problem is $176.76 + 151.31 = 328.07$ seconds (i.e., 5.47 minutes), while the average cost is $24.26 + 22.02 = 46.28$ cents. These values are higher than those obtained using a single LLM prompt, which has an average time of 87.98 seconds (i.e., 1.47 minutes) and an average cost of 12.42 cents, as reported in Table \ref{Tab3}. However, the proposed pipeline achieves significantly higher problem-solving accuracy, as demonstrated by the statistical significance analysis in Section \ref{S453}. Additionally, Table \ref{tab:manual_proofread} reports the number of problems requiring manual circuit diagram proofreading in the third trial and solution feedback in the fourth trial. On average, 14.8 problems require circuit diagram proofreading and 9.4 require solution feedback. In our experiments, each proofreading and feedback operation takes approximately 2 and 3 minutes, respectively, when performed by an expert with a solid background in circuit analysis and familiarity with the tasks. In contrast, the same expert typically requires around 10 minutes to solve a circuit problem of average difficulty with handwritten solutions. If the solution is required to be neatly typed and include sufficient intermediate steps for educational purposes, the time increases to 15 minutes or longer. These workload estimates are intended as an approximate comparison under the assumption that the human user is already proficient in circuit analysis and primarily needs to inspect and correct the intermediate outputs generated by the proposed pipeline. They are based on our experimental experience rather than a controlled human-subject timing study, and the actual workload may vary depending on user expertise, problem difficulty, the required level of solution detail, and the specific educational setting. We also note that this comparison concerns active human workload rather than total wall-clock time, since most stages of the proposed pipeline are automated and do not require continuous human involvement. Therefore, under these assumptions, the proposed pipeline not only reduces human effort in solving individual circuit problems, but also provides faster generation of detailed, pedagogically useful solutions.

\begin{table}[!t]
	\centering
	\scriptsize
	\caption{Number of problems requiring manual circuit diagram proofreading in the third trial and solution feedback in the fourth trial}
	\label{tab:manual_proofread}
	\begin{tabular}{ccc}
		\hline
		Circuit Diagram Type & Recognition Proofreading & Solution Feedback \\
		\hline
		\makecell[l]{Textbook-style} & 12 & 5  \\
		\makecell[l]{Scanned hand-drawn (white paper)} & 16& 12 \\
		\makecell[l]{Scanned hand-drawn (lined paper)} & 16 & 8  \\
		\makecell[l]{Scanned hand-drawn (digital notebook)} & 14 & 10  \\
		\makecell[l]{Scanned hand-drawn (IEC symbol standard)} & 15 & 12 \\
		\hline
		Average & 14.8 & 9.4 \\
		\hline
	\end{tabular}
\end{table}

As further explanation regarding the types and scopes of human-in-the-loop interventions in Table \ref{tab:manual_proofread}, we note that circuit recognition proofreading applies to problems where the analytical result generated by Gemini 2.5 Pro mismatches the numerical solution obtained from the \texttt{ngspice} simulation in their \textit{first} trials. The solution feedback intervention is reserved for problems where mismatches persist after the \textit{third} trial of Gemini 2.5 Pro and the corresponding automated \texttt{ngspice} verification loop. The ``3rd trial'' and ``4th trial'' columns in Table \ref{Tab1} show that $\frac{17}{5}=3.4$ and $\frac{41}{5}=8.2$ additional problems are correctly solved on average after the circuit recognition proofreading and solution feedback steps, respectively; these account for $\frac{3.4}{66} \approx 5.15\%$ and $\frac{8.2}{66} \approx 12.42\%$ of the 66 problems that contain circuit diagrams. Accordingly, the reported human workload should be interpreted in the context of these targeted interventions rather than as continuous human supervision throughout the entire solution process.

To better understand the contribution of individual components within the pipeline, we further examine the detection accuracy of the YOLO model, despite the strong overall problem-solving performance. Table \ref{Tab_confusion2} presents the confusion matrices for independent source detection across the four types of hand-drawn circuit diagrams, using the fine-tuned YOLO model with a prediction confidence threshold of 0.25. For voltage source detection in type iv) hand-drawn diagrams that follow IEC symbol standards for DC voltage sources, only 48 out of 75 voltage sources are detected, resulting in a recall of 64\%. This outcome is expected, as the IEC-style DC voltage symbol has a shape similar to that of the ground symbol, and comparatively fewer IEC-style samples were included in the YOLO training process. In addition, the recalls for current-source detection in types ii) and iii) are both $\frac{22}{34} \approx 64.71\%$. This suggests that variations in drawing style substantially affect detection performance, particularly when there is a mismatch between the styles represented in the training data and those in the prediction data. It is important to note, however, that source polarity recognition is achieved through the combined efforts of Gemini 2.5 Pro and the YOLO model, with YOLO primarily serving as an assistant in the source polarity rectification process. Therefore, the reduced recall in source detection for hand-drawn circuit diagrams has a smaller impact on the overall circuit recognition performance than the detection results alone might suggest, as the polarities of most missed sources were correctly identified by Gemini in the earlier stages.

\begin{table}[htbp]
\centering
\footnotesize
\caption{Confusion matrices for independent source detection in the four types of hand-drawn circuit diagrams using the fine-tuned YOLO model with a prediction confidence threshold of 0.25}
\label{Tab_confusion2}

\begin{subtable}{\textwidth}
\centering
\caption{Independent Voltage Source}
\begin{tabular}{c|*{8}{C{0.045\textwidth}}}
\hline
& \multicolumn{4}{c}{Predicted Voltage Source} & \multicolumn{4}{c}{Predicted Background} \\
\hline
Circuit Type & i) & ii) & iii) & iv) & i) & ii) & iii) & iv) \\
\hline
Actual Voltage Source & 75 & 75 & 74 & 48 & 0 & 0 & 1 & 27 \\
Actual Background & 0 & 0 & 5 & 2 & \multicolumn{4}{c}{------} \\
\hline
\end{tabular}
\end{subtable}

\vspace{0.5cm}

\begin{subtable}{\textwidth}
\centering
\caption{Independent Current Source}
\begin{tabular}{c|*{8}{C{0.045\textwidth}}}
\hline
& \multicolumn{4}{c}{Predicted Current Source} & \multicolumn{4}{c}{Predicted Background} \\
\hline
Circuit Type & i) & ii) & iii) & iv) & i) & ii) & iii) & iv) \\
\hline
Actual Current Source & 33 & 22 & 22 & 33 & 1 & 12 & 12 & 1 \\
Actual Background & 0 & 1 & 0 & 0 & \multicolumn{4}{c}{------} \\
\hline
\end{tabular}
\end{subtable}
\end{table}

\subsection{Experiments and Results on Circuit Problems from a Different Textbook}
\label{S46}
The experiments in Sections \ref{S43}--\ref{S45} use the same set of circuit problems from a single textbook \citep{svoboda2013introduction}, with Sections \ref{S43} and \ref{S44} focusing on textbook-style circuit diagrams and Section \ref{S45} examining four handwritten variations of the same underlying problems. These experiments therefore primarily evaluate robustness to changes in circuit-diagram representation rather than generalization to new problems. To address problem-level generalization, this section extends the evaluation to a separate set of circuit problems selected from a different textbook \citep{irwin2015basic}.

Specifically, we selected 43 problems from \cite{irwin2015basic} that cover circuit-analysis topics similar to those considered in Sections \ref{S43}--\ref{S45}. Although the topics are comparable, the newly selected problems differ from the original set in their circuit configurations, component values, problem formulations, and diagram representations, thereby providing an independent evaluation set with larger problem-level diversity. All selected problems include circuit diagrams and are suitable for solving with both Gemini and \texttt{ngspice}. The indices of the selected problems are provided in Table \ref{TabA1}. Beyond the intrinsic differences in the individual problems and circuit diagrams, several broader differences between the two problem sets are summarized below; these differences also constitute the principal dimensions of generalization examined in this section.

\begin{figure}
	\centering
	\begin{subfigure}{\textwidth}
		\centering
		\includegraphics[width=0.3\textwidth]{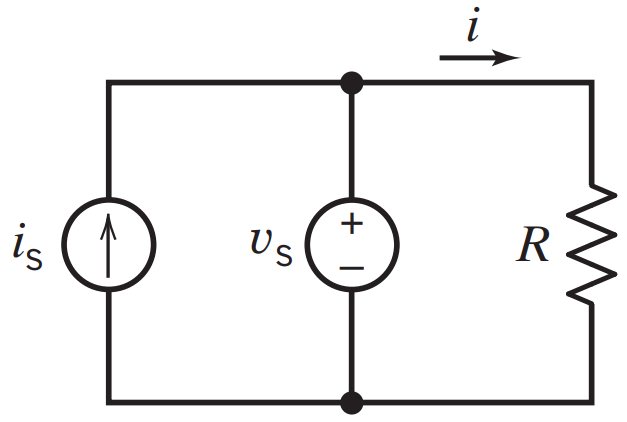}
		\caption{Circuit-diagram style in \cite{svoboda2013introduction}, featuring current-direction arrows placed adjacent to the corresponding wires.}
	\end{subfigure}
	\vspace{-0.75em}
	
	\begin{subfigure}{\textwidth}
		\centering
		\includegraphics[width=0.835\textwidth]{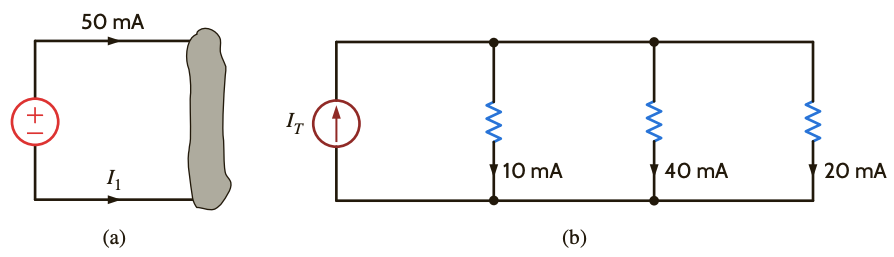}
		\caption{Circuit-diagram style in \cite{irwin2015basic}, featuring current-direction arrows embedded in the wires, more varied use of color for circuit elements, and irregular filled shapes representing unknown circuit elements.}
	\end{subfigure}
	\caption{Examples illustrating differences in circuit-diagram styles between \cite{svoboda2013introduction} and \cite{irwin2015basic}.}
	\label{Fig:DiagramDiff}
\end{figure}

\begin{itemize}
\item [i)] \textit{Different circuit-diagram styles:} As illustrated in Figure \ref{Fig:DiagramDiff}, the circuit diagrams in \cite{irwin2015basic} differ visually from those in \cite{svoboda2013introduction}. For example, current-direction arrows are embedded directly in the wires rather than placed adjacent to them, circuit elements are depicted using a more varied range of colors, and unknown circuit elements may be represented by irregular filled shapes. These differences introduce additional variation in the visual representation of otherwise similar circuit concepts.
\item [ii)] \textit{Different circuit-diagram complexity:} As summarized in Table \ref{Tab:DiagramDiff}, the selected problems from \cite{irwin2015basic} contain slightly more circuit elements per diagram on average than those from \cite{svoboda2013introduction}, with an average of 6.09 elements per diagram compared with 5.70. This difference provides an additional test of whether the proposed framework generalizes to circuit diagrams with somewhat greater structural complexity.
\end{itemize}

\begin{table}[!t]
	\centering
	\footnotesize
	\caption{Number of circuit elements per diagram for the selected problems from the two textbooks. Values are reported as mean $\pm$ standard deviation.}
	\label{Tab:DiagramDiff}
	\begin{tabular}{ccc}
		\hline
		Textbook & \# Selected Problems & Elements per Diagram \\
		\hline
		\cite{svoboda2013introduction} & 66 & $5.70 \pm 1.92$ \\
		\cite{irwin2015basic} & 43 & $6.09 \pm 1.96$ \\
		\hline
	\end{tabular}
\end{table}

The results on the newly selected problems from \cite{irwin2015basic} are summarized in Table \ref{Tab:ResultExtended}. We first note that the problems in \cite{irwin2015basic} appear to be more challenging for Gemini 2.5 Pro than those in \cite{svoboda2013introduction}, as the accuracy obtained by directly prompting Gemini 2.5 Pro decreases from 69.70\% to 58.14\%. In terms of the performance of the proposed framework, despite the additional variation in circuit-diagram style and complexity introduced by the new textbook, the framework achieves a final Gemini accuracy of 83.72\%, compared with 96.97\% on the textbook-style problems from \cite{svoboda2013introduction}. The circuit-recognition accuracy is 72.09\% on the diagrams from \cite{irwin2015basic}, compared with 96.97\% on the textbook-style diagrams from \cite{svoboda2013introduction}. The lower recognition and final problem-solving accuracies are associated with the larger variation in circuit-diagram representations and complexity across the two textbooks. These variations can introduce additional challenges at the recognition stage and subsequently affect problem-solving performance, as analyzed in Section \ref{S48}.

It is noteworthy that, under the setting with additional variation in circuit-diagram style and complexity, the final accuracy of 83.72\% achieved by the proposed framework is significantly higher than the 58.14\% accuracy obtained when Gemini 2.5 Pro is directly prompted to solve the same problems. As will be shown in the statistical analysis in Section \ref{S453}, this improvement is significant at the level of $\alpha=0.001$. These results support the cross-textbook generalizability of the proposed framework in terms of improving problem-solving accuracy, while also suggesting that the prompts may require further refinement to achieve the best performance on a new problem set from a different source.

\begin{table}[!h]
	\centering
	\footnotesize
	\caption{Results on circuit problems from \cite{irwin2015basic}}
	\label{Tab:ResultExtended}
	\begin{tabular}{cccc}
		\hline
		& Metric & Gemini & \texttt{ngspice} \\
		\hline
		\multirow{3}{*}{\makecell{Original\\Gemini}} & Accuracy & 25/43 (58.14\%) & \\
		& Time (s) & $87.95\pm80.35$ & --- \\
		& LLM cost (\textcent) & $10.08\pm7.38$ & \\
		\hline
		\hline
		\multirow{3}{*}{\makecell{Circuit\\recognition}} & Accuracy & 31/43 (72.09\%) & \\
		& Time (s) & $61.49\pm23.39$ & --- \\
		& LLM cost (\textcent) & $8.57\pm3.02$ & \\
		\hline
		\multirow{3}{*}{\makecell{1st trial}} & Accuracy & 27/43 (62.79\%) & 31/43 (72.09\%) \\
		& Time (s) & $62.82\pm39.21$ & $77.95\pm27.93$ \\
		& LLM cost (\textcent) & $9.79\pm6.36$ & $12.56\pm4.33$ \\
		\hline
		\multirow{3}{*}{\makecell{2nd trial}} & Accuracy & 5/11 (45.45\%) & 6/11 (54.55\%) \\
		& Time (s) & $70.37\pm60.57$ & $76.32\pm30.83$ \\
		& LLM cost (\textcent) & $11.47\pm10.11$ & $13.38\pm5.02$ \\
		\hline
		\multirow{3}{*}{\makecell{3rd trial}} & Accuracy & 2/6 (33.33\%) & 6/6 (100.00\%) \\
		& Time (s) & $96.46\pm67.52$ & $97.30\pm39.48$ \\
		& LLM cost (\textcent) & $15.55\pm11.24$ & $16.66\pm6.28$ \\
		\hline
		\multirow{3}{*}{\makecell{4th trial}} & Accuracy & 6/6 (100.00\%) &  \\
		& Time (s) & \multirow{2}{*}{---}  & --- \\
		& LLM cost (\textcent) &  & \\
		\hline
		\multirow{3}{*}{\makecell{Average\\total}} & Accuracy & 36/43 (83.72\%) & 35/43 (81.40\%) \\
		& Time (s) & $155.47\pm111.47$ & $111.05\pm83.03$ \\
		& LLM cost (\textcent) & $23.26\pm17.56$ & $18.31\pm13.79$ \\
		\hline
	\end{tabular}
\end{table}

\subsection{Statistical Significance Analysis}
\label{S453}
We conducted a statistical significance test to evaluate whether the improvement in problem-solving accuracy achieved by our proposed pipeline over the baseline method is statistically significant. Specifically, we applied the exact binomial test \citep{conover1999practical} to the discordant outcomes, that is, problems solved correctly by one method but incorrectly by the other.\footnote{Because both the baseline method and the proposed pipeline are evaluated on the same set of $N=66$ problems with binary outcomes (correct or incorrect), the comparison involves paired nominal data. In this setting, McNemar’s test \citep{conover1999practical} is the standard method, as it is based on the discordant pairs where the two methods disagree. However, because the number of discordant pairs may be small, we adopt the exact binomial formulation of McNemar’s test to obtain valid inference without relying on large-sample approximations.} Let $n_{01}$ denote the number of problems for which the baseline method is correct but the proposed pipeline is incorrect, and let $n_{10}$ denote the number of problems for which the baseline method is incorrect but the proposed pipeline is correct. Under the null hypothesis that the two methods have equivalent problem-solving capability, each method is equally likely to be correct on a discordant pair. Therefore, the number of problems, $X$, for which the baseline method is correct but the proposed pipeline is incorrect follows the binomial distribution
$$
X\sim \mathrm{Binomial}(n_{01} + n_{10}, 0.5).
$$
To assess statistical significance, we compute the two-tailed $p$-value as
$$
p\text{-value} = \sum_{x:P(X=x)\le P(X=n_{01})}P(X=x),
$$
which represents the probability, under the null hypothesis, of observing a discordant split at least as extreme as the one obtained.

\begin{table}[htbp]
    \centering
    \scriptsize
    \caption{Significance test results for the proposed method and the baseline method (directly prompting Gemini 2.5 Pro to solve the circuit problems). The total number of problems is $N=66$ for each of the five cases based on \cite{svoboda2013introduction} and $N=43$ for the case based on \cite{irwin2015basic}. $n_{01}$ represents the number of problems for which the baseline method is correct but the proposed method is incorrect. $n_{10}$ represents the number of problems for which the baseline method is incorrect but the proposed method is correct. The 95\% CI column reports the Clopper-Pearson exact confidence intervals for the probability that the proposed method is correct when the two methods yield discordant outcomes.}
    \label{tab:mcnemar_results}
    \begin{tabular}{ccccccc}
        \hline
        \multirow{2}{*}{Circuit diagram type} & \multicolumn{2}{c}{Accuracy} & \multirow{2}{*}{\textbf{$n_{01}$}} & \multirow{2}{*}{\textbf{$n_{10}$}} & \multirow{2}{*}{\makecell{$p$-value}} & \multirow{2}{*}{95\% CI} \\ \cline{2-3}
        & Baseline & Framework & & & & \\
        \hline
		\makecell[l]{Textbook-style} & 69.70\% & 96.97\% & 1 & 19 & $0.00004^{***}$ & [0.7513, 0.9987] \\
        \makecell[l]{Hand-drawn (white paper)} & 71.21\% & 90.91\% & 3 & 16 &$0.00443^{**}$ & [0.6042, 0.9662] \\
        \makecell[l]{Hand-drawn (lined paper)} & 62.12\% & 89.39\% & 4 & 22 & $0.00053^{***}$ & [0.6513, 0.9564] \\
        \makecell[l]{Hand-drawn (digital notebook)} & 60.61\% & 90.91\% & 1 & 21 & $0.00001^{***}$ & [0.7716, 0.9988] \\
        \makecell[l]{Hand-drawn (IEC symbols)} & 68.18\% & 92.42\% & 0 & 16 & $0.00003^{***}$ & [0.7941, 1.0000] \\
        \hline
        \makecell[l]{\cite{irwin2015basic}} & 58.14\% & 83.72\% & 0 & 11 & $0.00010^{***}$ & [0.7151, 1.0000] \\
        \hline
        \multicolumn{7}{l}{** and *** denote statistical significances at the $\alpha=0.005$ and $\alpha=0.001$ levels, respectively.}
    \end{tabular}
\end{table}

Table \ref{tab:mcnemar_results} summarizes the exact binomial test results for the six evaluation cases, including the five circuit diagram variations based on \cite{svoboda2013introduction} and the additional case based on \cite{irwin2015basic}. Across all cases, we observe that $n_{10} \gg n_{01}$, indicating that the proposed pipeline correctly solves many problems that the baseline method fails to solve, while only rarely introducing new errors. Accordingly, the resulting $p$-values are all below $0.005$, with five of the six cases yielding $p$-values below $0.001$. Thus, the null hypothesis is rejected at the $\alpha=0.005$ significance level in all six cases and at the more stringent $\alpha=0.001$ level in five cases, providing strong statistical evidence that the proposed pipeline outperforms the baseline method. Moreover, the consistently significant improvements across different hand-drawn styles, symbol standards, and problems drawn from an additional textbook further demonstrate the robustness and generalizability of the proposed framework across variations in both circuit-diagram representation and problem source.

Furthermore, we computed the 95\% Clopper-Pearson exact confidence intervals (CIs) \citep{clopper1934use} for the probability that our pipeline yields the correct answer when the two methods produce discordant outcomes. As shown in Table \ref{tab:mcnemar_results}, the lower bounds of the 95\% confidence intervals are greater than 0.5 in all six cases, ranging from 0.6042 to 0.7941. Because all of these intervals exclude 0.5 (the probability under the null hypothesis), they provide a conservative measure of uncertainty and further support the conclusion that the proposed pipeline has a statistically significant performance advantage over the baseline method, even when accounting for the relatively small numbers of discordant outcomes.

\subsection{Error Analysis of Incorrect Solutions}
\label{S48}

In this section, we systematically examine the failure modes observed in the evaluation results reported in Tables \ref{Tab1} and \ref{Tab:ResultExtended}. To this end, we categorize the causes of incorrect solutions into five primary error types:

\begin{itemize}
	\item [--] \textit{Circuit Recognition:} Errors arising during the circuit recognition stage that lead Gemini 2.5 Pro and \texttt{ngspice} to produce internally consistent but incorrect solutions. Examples include misreading current directions or misinterpreting branch connections.
	\item [--] \textit{Curve Recognition:} Errors in extracting quantitative information from plotted curves.
	\item [--] \textit{Circuit Understanding:} Errors in interpreting the circuit during the problem-solving process, even when the circuit itself has been correctly recognized. For example, a mesh current may be correctly recognized as counterclockwise, while the direction of the current in a particular branch associated with that mesh may subsequently be misinterpreted during problem solving.
	\item [--] \textit{Reasoning:} Errors in formulating circuit equations or performing logical derivations, such as incorrectly treating the voltage at the negative terminal of a voltage source as zero.
	\item [--] \textit{Arithmetic:} Algebraic, matrix inversion, or numerical computation errors that occur despite the circuit equations having been correctly formulated.
\end{itemize}

\begin{table}
	\centering
	\footnotesize
	\caption{Numbers and ratios of incorrect problems by problem type and error type}
	\label{Tab:IncorrectReasons}
	\begin{tabular}{cccc}
		\hline
		Problem Type & Error Type & Count & Ratio \\
		\hline
		\multirow{2}{*}{\makecell{Problems from \cite{svoboda2013introduction}\\with textbook-style diagrams}} 
		& Circuit Recognition & 1 & 50.00\% \\
		& Reasoning & 1 & 50.00\% \\
		\hline
		\multirow{5}{*}{\makecell{Problems from \cite{svoboda2013introduction}\\with hand-drawn diagrams}} 
		& Circuit Recognition & 11 & 45.83\% \\
		& Curve Recognition & 2 & 8.33\% \\
		& Circuit Understanding & 6 & 25.00\% \\
		& Reasoning & 4 & 16.67\% \\
		& Arithmetic & 1 & 4.17\% \\
		\hline\hline
		\makecell{Problems from \cite{irwin2015basic}\\with textbook-style diagrams} 
		& Circuit Recognition & 7 & 100.00\% \\
		\hline
	\end{tabular}
\end{table}

Table \ref{Tab:IncorrectReasons} summarizes the numbers and relative proportions of these error categories across different problem sources and circuit diagram styles. The results show that circuit recognition is the dominant source of errors across the evaluated problem sets. Of the two incorrect solutions for the textbook-style diagrams from \cite{svoboda2013introduction}, one is caused by a circuit recognition error and the other by a reasoning error. In contrast, the hand-drawn diagrams exhibit a broader range of failure modes, with circuit recognition accounting for the largest proportion of errors (45.83\%), followed by circuit understanding (25.00\%) and reasoning (16.67\%). This indicates that the visual variations introduced by hand-drawn diagrams not only increase the likelihood of recognition errors but also create additional challenges in correctly interpreting the recognized circuits during subsequent problem solving. For the problems from \cite{irwin2015basic}, all seven incorrect solutions are attributable to circuit recognition, suggesting that the reduced performance on this dataset is primarily associated with difficulties in interpreting the different circuit-diagram representations rather than with downstream reasoning or numerical computation. 

Future directions for further improving circuit recognition quality are discussed in Section \ref{S52}.

\section{Discussion}
This section discusses several extensions and limitations of the proposed circuit problem-solving framework, including its potential for parametric analysis and design optimization, the limitations of the verification loop due to circuit recognition hallucinations along with possible mitigation strategies, and the quality of intermediate reasoning and its educational implications, as well as the selection of the LLM backbone and the cross-model generalizability of the proposed framework.

\subsection{Parametric Analysis and Design Optimization}
The focus of this paper is primarily on circuit problem solving, i.e., solving for unknown quantities in a given circuit with specified topology and component values. As shown in Table \ref{TableA1}, among the 83 problems considered, only 5 are circuit synthesis problems, in which certain circuit behaviors are prescribed and the task is to determine the parameters of some components. For these problems, the \texttt{ngspice} verification procedure treats the component parameters solved by Gemini 2.5 Pro as known quantities and simulates the circuit to check whether the resulting outputs match the prescribed behaviors. Notably, our pipeline correctly solves all 5 of these problems, demonstrating its preliminary capability in circuit synthesis and design-oriented tasks. It should be noted that all 5 circuit synthesis problems considered here have predefined circuit structures, and their relatively simple topologies also enable Gemini 2.5 Pro to recognize the circuits correctly. Nevertheless, the strong performance on these problems provides useful insight into how the proposed framework can be naturally extended to more general parametric analysis and design optimization tasks, particularly in scenarios where the circuit structure is fixed and only a subset of parameters needs to be determined or optimized. 

In such settings, the prescribed circuit structure can be converted into both a textual description for the LLM and a \texttt{.cir} file for \texttt{ngspice} in which target component values are represented as tunable parameters. Building on this representation, a closed-loop workflow can be established, where the LLM proposes candidate parameter values, \texttt{ngspice} evaluates the resulting circuit performance, and the simulation feedback is used to iteratively refine the parameter selection. This feedback-driven process enables systematic exploration of the parameter space and supports tasks such as sensitivity analysis and constraint-driven design. Furthermore, for more complex design optimization tasks, the LLM can be integrated as a high-level optimizer \citep{yang2023large}, generating successive candidate solutions based on prior simulation outcomes, while \texttt{ngspice} provides objective performance evaluations. This iterative LLM-simulation loop allows the framework to move beyond single-pass problem solving toward goal-oriented circuit design, thereby extending its applicability to a broader class of engineering design problems.

From an engineering education perspective, by iteratively varying component parameters and invoking \texttt{ngspice} simulations, the system can generate performance-parameter curves (e.g., power transfer versus resistance), enabling students to visualize circuit sensitivities and deepen conceptual understanding. Furthermore, by leveraging simulation feedback in a closed-loop manner, the framework can guide learners in adjusting parameters to meet specified performance targets, thereby fostering design-oriented thinking and bridging the gap between analysis and practical engineering design.

\subsection{Limitations of the Verification Loop Due to Circuit Recognition Hallucinations and Potential Mitigation Strategy}
\label{S52}
While the \texttt{ngspice} verification loop serves as an effective safeguard against problem-solving hallucinations by validating the consistency between the Gemini-derived solution and circuit simulation results, it does not fully eliminate errors arising from incorrect circuit recognition. In particular, if Gemini 2.5 Pro misinterprets a hand-drawn circuit diagram during the recognition stage, the same erroneous circuit representation will be used both for generating the solution and for constructing the \texttt{.cir} file for \texttt{ngspice}. In this scenario, both the subsequent Gemini-based problem solving and the \texttt{ngspice} simulation will faithfully analyze the same incorrect circuit. As a result, the verification loop may still confirm internally consistent but fundamentally incorrect solutions, allowing such errors to pass undetected. This explains why circuit recognition constitutes the dominant source of the remaining errors reported in Table \ref{Tab:IncorrectReasons} of Section \ref{S48}.

To mitigate this limitation, future work could incorporate an explicit circuit visualization and human-in-the-loop proofreading stage. For example, the textual circuit representations can be automatically converted into a netlist format by an LLM, which can then be further translated into a schematic representation using tools such as the \texttt{CircuiTikZ} package\footnote{See details at \url{https://ctan.dcc.uchile.cl/graphics/pgf/contrib/circuitikz/doc/circuitikzmanual.pdf}.} in \LaTeX, enabling direct visual comparison with the original diagram. This approach provides an intuitive and efficient mechanism for identifying structural discrepancies, particularly in cases involving ambiguous or noisy hand-drawn inputs. When discrepancies are identified, a human proofreader can either revise the netlist directly or describe the discrepancies in natural language to an LLM, enabling interactive correction of the generated circuit representation. More broadly, integrating such visualization tools into the existing pipeline could establish a bidirectional verification framework, in which both textual (circuit description-level) and visual (diagram-level) consistency are enforced. Since the accuracies reported in Section \ref{S4} treat unflagged failures caused by incorrect circuit recognition as incorrect outcomes, the proposed integration of \texttt{CircuiTikZ} visualization and the corresponding circuit recognition verification mechanism is intended to further improve problem-solving accuracy beyond the results currently reported in this paper.

\subsection{Quality of Intermediate Reasoning and Educational Implications}
The evaluation in this work is primarily based on final-answer correctness, which provides a clear, objective, and reproducible criterion for performance comparison across different configurations. However, the correctness of the final answer does not necessarily guarantee that all intermediate steps are reasonable and pedagogically sound. Indeed, for related tasks such as automated homework grading \citep{sun2026circuit}, simple evaluation metrics may overlook errors in the intermediate reasoning process. To verify the quality of the underlying reasoning in the results produced by our framework, we conducted a comprehensive manual audit of the step-by-step reasoning processes. This qualitative review confirmed that the intermediate reasoning steps---including node identification, the formulation of nodal and mesh equations, transient-state transitions, and algebraic manipulations---are logically consistent and mathematically valid. In our experiments, the framework does not arrive at correct final answers through faulty or self-canceling derivations.

However, we note that manually proofreading the intermediate reasoning process is labor-intensive. A more comprehensive and automated evaluation framework that explicitly incorporates intermediate reasoning quality would further enhance the reliability, scalability, and educational value of the system while minimizing human effort. Potential directions include step-level verification to ensure the validity of each derivation, consistency checks across multiple reasoning paths, and rubric-based assessments of explanation quality in terms of clarity, completeness, and instructional usefulness. When necessary, incorporating human-in-the-loop evaluation could provide expert judgment on reasoning quality, particularly in complex or ambiguous cases flagged by the automated system \citep{sun2026circuit}. These extensions would enable the framework to move beyond outcome-oriented evaluation toward a more holistic assessment paradigm aligned with the goals of engineering education. 

\subsection{LLM Backbone Selection and Cross-Model Generalizability}
The proposed circuit problem-solving framework is fundamentally model-agnostic in its architectural design. The core pipeline---comprising decoupled visual component detection, structured symbolic netlist synthesis, verification through deterministic physical simulation, and interactive human-in-the-loop refinement---operates independently of any specific proprietary foundation model. The decision to base the primary evaluation on Gemini 2.5 Pro stems from the hierarchical dependencies inherent in end-to-end circuit solving. Accurate upstream circuit-topology recognition is a critical prerequisite: if an LLM fails to correctly parse basic schematic connectivity or component values, these errors can propagate through subsequent analytical derivation and simulation-verification stages. Prior benchmarking studies have highlighted the substantial challenges that contemporary multimodal models face in circuit interpretation. For example, \cite{skelic2025circuit} reported that GPT-4o achieved only 48.04\% accuracy on a benchmark of undergraduate circuit problems, with many failures associated with visual diagram interpretation. \cite{akbari2026circuitsense} conducted a comprehensive evaluation of eight MLLMs, including Gemini 2.5 Pro, GPT-4o, and Claude Sonnet 4, using CircuitSense, a benchmark designed to assess circuit understanding and mathematical model extraction from technical diagrams. Their results showed that Gemini 2.5 Pro achieved the strongest overall performance among the evaluated models, tying or leading the perception subtasks---including component detection, connection identification, and function classification---and outperforming the other models across all reported circuit-analysis subcategories. These findings provide independent empirical support for our selection of Gemini 2.5 Pro as the backbone model for this investigation. It achieved 79.52\% zero-shot accuracy on textbook-style circuit problems and demonstrated sufficiently reliable visual diagram parsing and analog-circuit reasoning capabilities to support the development of the proposed framework. Although specific perception interventions, such as YOLO-based source-polarity correction, were designed to address localized visual hallucinations observed in Gemini 2.5 Pro, the underlying principles---using specialized computer vision tools to reduce visual ambiguity and coupling LLM-generated analytical solutions with closed-loop physical simulation---are broadly applicable across multimodal LLMs. As the schematic-interpretation capabilities of vision-language models continue to improve, the modular architecture presented in this work can readily incorporate alternative or emerging LLMs as the underlying backbone.

\section{Conclusion}
This work presents a comprehensive enhancement framework that substantially improves the reliability of Gemini 2.5 Pro---one of the state-of-the-art LLMs---in solving undergraduate circuit analysis problems, a domain where multimodal recognition, precise mathematical reasoning, and domain-specific knowledge pose significant challenges for current models. By systematically evaluating Gemini 2.5 Pro and characterizing its dominant failure modes, we identify two key bottlenecks: hallucinations in circuit-element recognition, particularly in source polarity identification, and hallucinations in analytical reasoning, such as incorrect current directions. The proposed pipeline addresses these issues through two complementary strategies: a fine-tuned YOLO-based detector combined with OpenCV preprocessing to eliminate polarity-recognition errors, and an iterative \texttt{ngspice}-based verification loop to detect and correct reasoning inconsistencies.

Extensive experiments across 83 circuit problems demonstrate the effectiveness of this hybrid approach. The framework elevates Gemini 2.5 Pro's problem-solving accuracy from the original 79.52\% to 97.59\% when applied to textbook-quality circuit diagrams, with \texttt{ngspice} successfully flagging the vast majority of incorrect LLM outputs through inconsistencies between the LLM-generated and simulation-based solutions. Ablation results further show that many reasoning errors can be corrected with minimal reruns, underscoring the value of a simulation-driven feedback loop. Moreover, the required human oversight is limited and strategically targeted, reducing instructor workload to a small fraction of the total problem set.

Additional experimental results further demonstrate that the proposed multi-step pipeline significantly enhances circuit problem-solving performance across a wide range of hand-drawn circuit diagram types. While Gemini 2.5 Pro degrades substantially on diverse hand-drawn circuit diagrams, the proposed framework maintains high accuracy (89.39\%---92.42\%), approaching textbook-quality performance. Moreover, statistical significance analysis based on the exact binomial test confirms that these performance gains are statistically significant across all six evaluation cases ($p<0.005$), with five of the six cases reaching $p<0.001$. Evaluation on an additional set of 43 circuit problems from a different textbook further demonstrates the cross-textbook generalizability of the proposed framework, improving problem-solving accuracy from 58.14\% with direct Gemini prompting to 83.72\%. The corresponding error analysis further reveals that circuit recognition remains the dominant source of residual errors across the evaluated problem sets, with all seven incorrect solutions on the additional textbook dataset attributable to circuit recognition. Additionally, although the enhanced pipeline incurs moderately higher time and cost compared to single-prompt LLM usage, it remains substantially more efficient than manual problem solving while achieving significantly higher accuracy. Overall, these results highlight the practical applicability, scalability, and generalizability of the proposed framework for real-world circuit analysis tasks in engineering practice.

Importantly, the LLM-generated solutions can be transformed into diverse educational materials, including concise reference answers and detailed pedagogical explanations. This directly supports scalable applications such as automated feedback, problem recommendation, and AI-assisted tutoring. More broadly, this work offers a principled blueprint for integrating LLMs with external engineering tools to detect and mitigate hallucinations---a strategy that is readily transferable to other engineering domains with well-established simulation or verification pipelines. By augmenting LLM-based problem-solving systems with external perception and verification modules, and enabling the generation of high-quality, structured domain-specific data, this study lays the foundation for broader integration of trustworthy AI systems into engineering education and practice.

Future work will focus on further improving circuit recognition and verification, for example, by integrating circuit visualization for bidirectional diagram--description consistency checking and developing more automated methods for evaluating the quality of intermediate reasoning. We also plan to extend the framework toward parametric analysis and circuit design optimization and to investigate its generalizability across alternative multimodal LLM backbones.

\section{Declaration of Generative AI and AI-assisted Technologies in the Manuscript Preparation Process}
During the preparation of this work, the authors used ChatGPT to polish the language. After using this tool, the authors reviewed and edited the content as needed and take full responsibility for the content of the submitted article.

\section*{Acknowledgments}
The authors appreciate the support provided by the School of Electrical and Computer Engineering and the College of Engineering at the Georgia Institute of Technology. 

\appendix
\section{Details of the Circuit Problem Dataset}
\label{SA}

Table \ref{TabA1} presents the problem indices and the number of data instances in each topic for the dataset introduced in Section \ref{S31}. 

\begin{table}[!h]
	\centering
	\scriptsize
	\caption{Problem/exercise indices and number of data instances in each topic}
	\label{TabA1}
	\begin{tabular}{cccc}
		\hline \noalign{\vskip 3pt}
		Topic & \# Data & Problem Indices* & Exercise Indices** \\ 
		\noalign{\vskip 2pt} \hline \noalign{\vskip 2pt}
		Electric Circuit Variables & 2/0 & \makecell[l]{P 1.5-2, P 1.5-3} & --- \\ \vspace{0.1cm}
		Circuit Elements & 4/2 & \makecell[l]{P 2.4-6, P 2.4-7, P 2.5-1, P 2.5-2} & \makecell[l]{E 2.1, E 2.2} \\ \vspace{0.1cm}
		Resistive Circuits & 6/5 & \makecell[l]{P 3.2-1, P 3.2-2, P 3.2-6, P 3.4-4, \\P 3.6-1, P 3.6-3} & \makecell[l]{E 2.4, E 2.5, E 2.10,\\ E 2.17, E 2.21} \\ \vspace{0.1cm}
		\makecell{Methods of Analysis of\\ Resistive Circuits} & 10/9 & \makecell[l]{P 4.2-2, P 4.2-5, P 4.3-3, P 4.4-1,\\ P 4.4-6, P 4.5-2, P 4.6-2, P 4.7-1, \\ P 4.7-2, P 4.7-6} & \makecell[l]{E 2.6, E 2.9, E 2.27, \\ E 2.31, E 3.4, E 3.7, \\E 3.9, E 3.14, E 3.23} \\ \vspace{0.1cm}
		\makecell{Circuit Theorems} & 10/5 & \makecell[l]{P 5.2-1, P 5.2-2, P 5.2-3, P 5.2-5,\\ P 5.3-4, P 5.3-6, P 5.3-8, P 5.4-5, \\P 5.5-5, P 5.6-3} & \makecell[l]{E 5.1, E 5.5, E 5.10,\\E 5.12, E 5.15} \\ \vspace{0.1cm}
		The Operational Amplifier & 7/3 & \makecell[l]{P 6.3-2, P 6.3-3, P 6.4-1, P 6.4-2,\\ P 6.4-4, P 6.4-5, P 6.4-6} & \makecell[l]{E 4.1, E 4.4, E 4.6} \\ \vspace{0.1cm}
		Energy Storage Elements & 10/1 & \makecell[l]{P 7.2-2, P 7.2-4, P 7.3-2, P 7.4-1,\\ P 7.4-3, P 7.5-4, P 7.5-5, P 7.6-2, \\P 7.7-2, P 7.8-2} & \makecell[l]{E 6.10} \\ \vspace{0.1cm}
		\makecell{The Complete Response of\\ \textit{RL} and \textit{RC} Circuits} & 9/6 & \makecell[l]{P 8.3-1, P 8.3-2, P 8.3-4, P 8.3-5, \\P 8.3-9, P 8.3-10, P 8.4-1, P 8.7-3,\\ P 8.7-5} & \makecell[l]{E 7.1, E 7.2, E 7.3,\\E 7.4, E 7.6, E 7.9} \\ \vspace{0.1cm}
		\makecell{The Complete Response of\\ Circuits with Two Energy\\
			Storage Elements} & 8/3 & \makecell[l]{P 9.2-1, P 9.3-3, P 9.4-1, P 9.4-2,\\P 9.5-1, P 9.6-1, P 9.7-2, P 9.8-2} & \makecell[l]{E 7.14, E 7.16, E 7.18} \\ \vspace{0.1cm}
		\makecell{Sinusoidal Steady-State \\Analysis} & 11/6 & \makecell[l]{P 10.3-2, P 10.3-3, P 10.4-1,\\ P 10.4-6, P 10.6-1, P 10.6-4,\\ P 10.7-1, P 10.7-6, P 10.8-1,\\ P 10.8-6, P 10.10-4} & \makecell[l]{E 8.8, E 8.9, E 8.14,\\E 8.15, E 8.17, E 8.22} \\
		\makecell{Frequency Response\\(Network Function)} & 6/3 & \makecell[l]{P 13.2-1, P 13.2-2, P 13.2-3,\\P 13.2-8, P 13.2-9, P 13.2-11} & \makecell[l]{E 12.21, E 12.22,\\ E 12.23} \\ 
		\noalign{\vskip 0.5pt} \hline \noalign{\vskip 2pt}
		\color{darkgray}{\makecell{Frequency Response\\(Bode Plot)}} & {\color{darkgray}3/0} & \makecell[l]{\color{darkgray}P 13.3-1, P 13.3-2, P 13.3-3} & ---\\ 
		\noalign{\vskip 2pt} \hline \noalign{\vskip 2pt}
		\makecell{Total Number of Problems} & --- & 83 {\color{darkgray}(86)} & 43 \\
		\noalign{\vskip 1pt} \hline \noalign{\vskip 2pt}
		\multicolumn{4}{l}{\parbox{13cm}{* The problem indices are taken from \cite{svoboda2013introduction}.\\
				** The exercise indices are taken from \cite{irwin2015basic}.\\
				*** The problem indices in {\color{darkgray}gray} indicate the topic of the Bode plot, which is not considered in this paper.}}
	\end{tabular}
\end{table}

During the testing process, we observed that Gemini 2.5 Pro encounters difficulties in identifying the switch status and correctly interpreting certain labeled currents in some circuits. To address these issues, we made minimal modifications to selected problem statements. The details are summarized below.

\begin{itemize}
	\item[i)] For circuits containing operational amplifiers, if a labeled current $i_{\rm o}$ and a labeled voltage $v_{\rm o}$ appear in the diagram but are not associated with the same element, Gemini 2.5 Pro almost always misinterprets $i_{\rm o}$ as the current through the element whose voltage is labeled as $v_{\rm o}$. To prevent this misinterpretation, we add a clarifying sentence to the problem description explicitly indicating the branch with which $i_{\rm o}$ is associated. Three problems fall into this category in Table \ref{TabA1}, namely P 6.3-2, P 6.3-3, and P 6.4-2. As an example, we present the original and adapted statements of Problem 6.3-3 below. The newly added sentence is highlighted using \textit{italic} font.
	
	[Original Problem Statement]
	
	Find $v_{\rm{o}}$ and $i_{\rm{o}}$ for the circuit of Figure P 6.3-2.
	
	[Adapted Problem Statement] 
	
	Find $v_{\rm{o}}$ and $i_{\rm{o}}$ for the circuit of Figure P 6.3-2. 
	
	\textit{Note: The current $i_{\rm o}$ flows from the inverting input terminal of the op-amp, through a resistor, to the output terminal of the op-amp.}
	
	\begin{figure}[!h]
		\centering
		\includegraphics[width=0.45\linewidth]{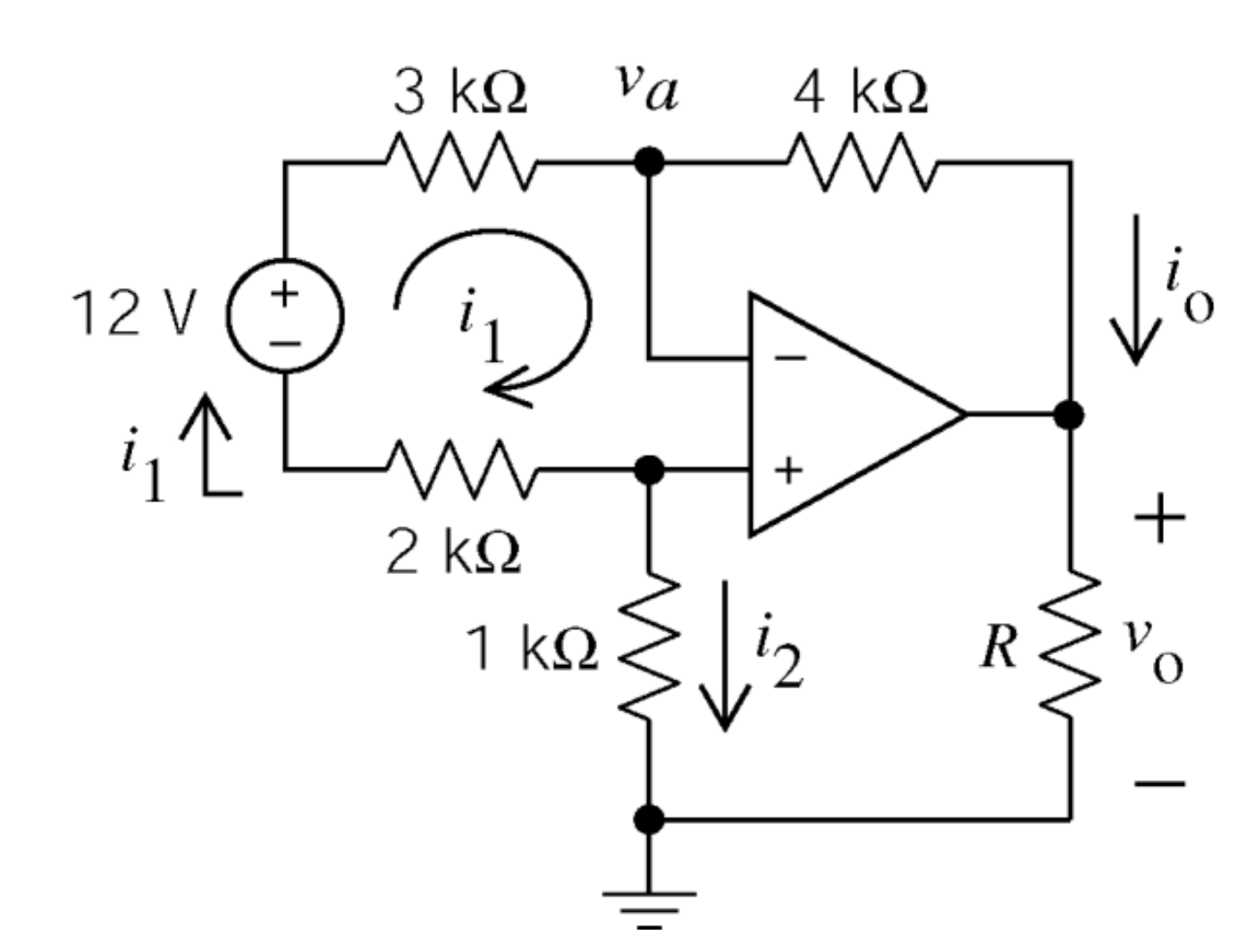}
		\captionsetup{labelformat=empty}
		\caption{Figure P 6.3-2}
	\end{figure}
	
	\item [ii)] Gemini 2.5 Pro is not reliable in recognizing switch types or interpreting switch status changes when a circuit diagram contains switches. Therefore, if the original problem statement does not specify the switch behavior, we add this information to support accurate problem solving and instruct Gemini 2.5 Pro to prioritize the textual problem description when inconsistencies arise. One problem in Table \ref{TabA1}, namely P 9.6-1, falls into this category. We present the original and adapted statements below, with the original and adapted sentences highlighted in \textit{italic} font.
	
	[Original Problem Statement]
	
	A communication system from a space station uses short pulses to control a robot operating in space. The transmitter circuit is modeled in Figure P 9.6-1. Find the output voltage $v_{\rm{c}}(t)$ for $t > 0$. \textit{Assume steady-state conditions at $t = 0^{-}$.}
	
	[Adapted Problem Statement] 
	
	A communication system from a space station uses short pulses to control a robot operating in space. The transmitter circuit is modeled in Figure P 9.6-1. Find the output voltage $v_{\rm{c}}(t)$ for $t > 0$. \textit{The switch has been open for a long time before closing at $t = 0$.}
	
	\begin{figure}[!h]
		\centering
		\includegraphics[width=0.525\linewidth]{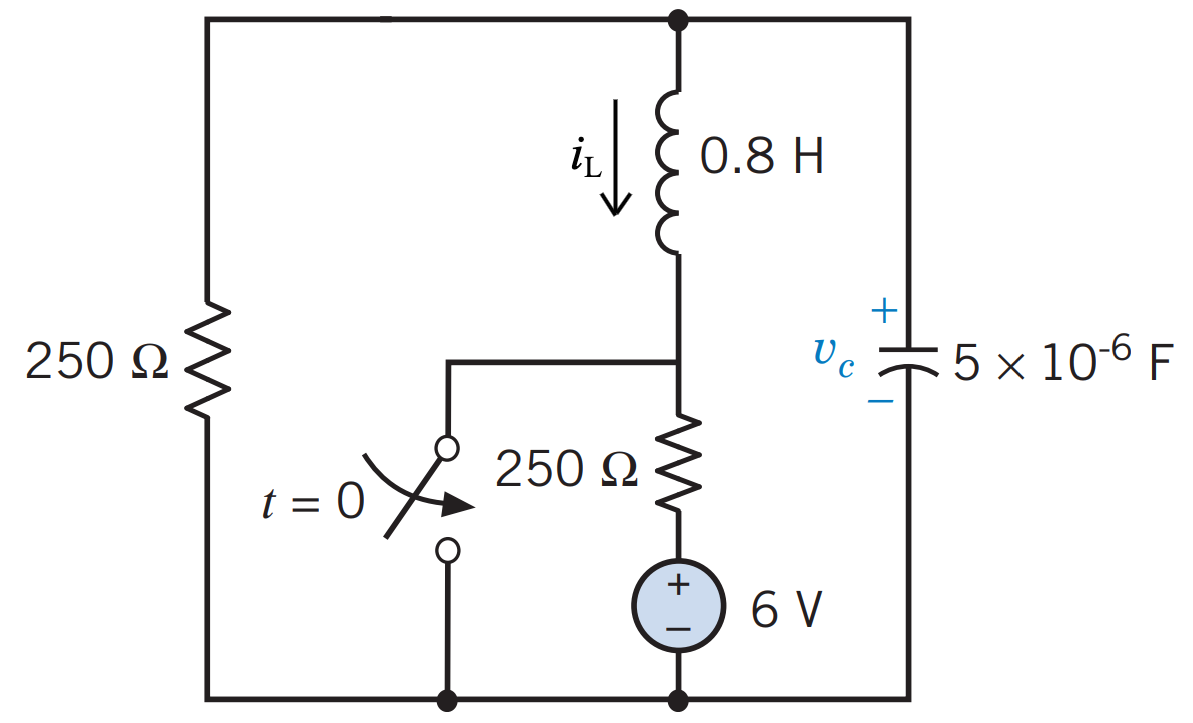}
		\captionsetup{labelformat=empty}
		\caption{Figure P 9.6-1}
	\end{figure}
	
\end{itemize}

Note that these two types of issues can be easily identified, and the corresponding problem statements can therefore be adjusted in advance. As a result, before applying the proposed circuit problem-solving pipeline, instructors or teaching assistants can simply make these minimal adaptations to the problem statements to improve the overall solution accuracy.

\section{An Example of Circuit Problem Solving}
\label{SB}
This section presents an example of circuit recognition, \texttt{ngspice}-related results, and a reference solution generated by Gemini 2.5 Pro. The detailed intermediate steps and the reference solution help clarify the outcomes of our proposed framework. Moreover, it is demonstrated that the Gemini-generated solution contains sufficient information to be adapted into various formats for different educational purposes.

\subsection{Problem Description}
We select Problem 8.3-5 from \cite{svoboda2013introduction} as an example. The circuit diagram associated with this problem is relatively complex, consisting of seven elements and five nodes. The circuit includes a variety of components: one independent voltage source, one operational amplifier, three resistors, one capacitor, and a single-pole single-throw switch that opens at $t=0$. Due to the number and diversity of elements, this problem serves as a relatively challenging case for Gemini 2.5 Pro to generate a complete solution, including the problem-solving process and a corresponding \texttt{.cir} file for \texttt{ngspice} simulation. The problem description is provided below.

\textbf{P 8.3-5} The circuit shown in Figure P 8.3-5 is at steady state before the switch opens at time $t = 0$. Determine the voltage, $v_{\rm o}(t)$, for $t > 0$.

\begin{figure}[!h]
	\centering
	\includegraphics[width=0.5\linewidth]{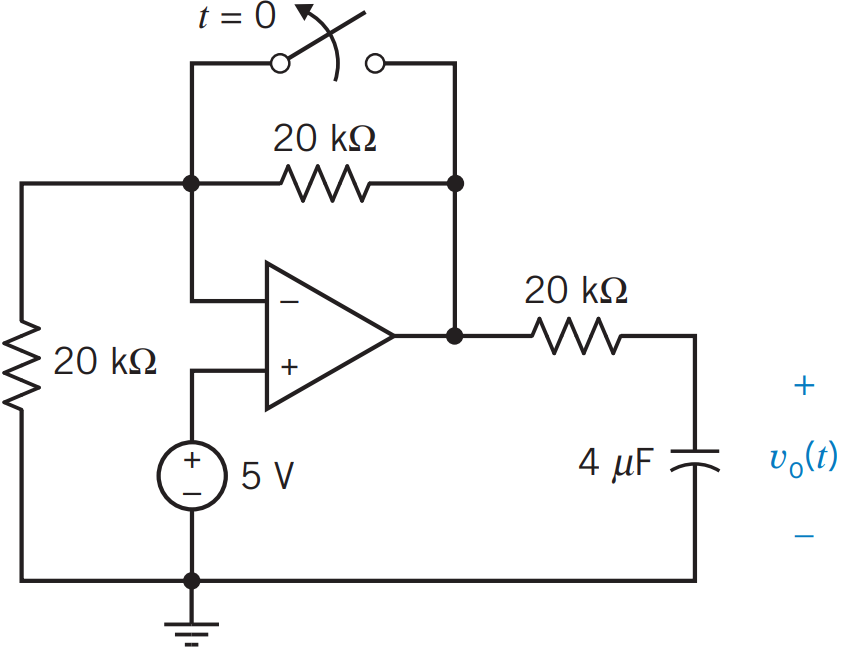}
	\captionsetup{labelformat=empty}
	\caption{Figure P 8.3-5}
\end{figure}

\subsection{Circuit Diagram Recognition Result}
Using the full circuit diagram recognition pipeline, the circuit in Figure 8.3-5 is described as follows\footnote{The original content, generated by Gemini 2.5 Pro, was formatted in Markdown. We adapted this format to improve organization and readability in the printed PDF version.}. The description includes all circuit components, wire connections, component labels, and the voltage $v_{\rm o}$, thereby capturing the complete information presented in the original circuit diagram.

\begin{tcolorbox}[breakable, enhanced, width=\textwidth, opacityfill=1., colback=white]
	Here is a detailed description of the circuit's components, their values, and their connections.
	
	\begin{itemize}
		\item \textbf{Independent Voltage Source (V\_s):}
		\begin{itemize}
			\item Value: 5 V.
			\item Polarity: The positive terminal is on top, and the negative terminal is on the bottom.
			\item Connections: The positive terminal is connected to the non-inverting ($+$) input of the op-amp. The negative terminal is connected to the ground node.
		\end{itemize}
		
		\item \textbf{Resistor R1:}
		\begin{itemize}
			\item Value: 20 k$\Omega$.
			\item Connections: One terminal is connected to the inverting ($-$) input of the op-amp. The other terminal is connected to the ground node.
		\end{itemize}
		
		\item \textbf{Resistor R2:}
		\begin{itemize}
			\item Value: 20 k$\Omega$.
			\item Connections: It is connected in the feedback path between the op-amp's output terminal and its inverting ($-$) input terminal. It is in parallel with the switch S.
		\end{itemize}
		
		\item \textbf{Resistor R3:}
		\begin{itemize}
			\item Value: 20 k$\Omega$.
			\item Connections: One terminal is connected to the output of the op-amp. The other terminal is connected to the positive terminal of the capacitor C.
		\end{itemize}
		
		\item \textbf{Capacitor C:}
		\begin{itemize}
			\item Value: 4 $\mu$F.
			\item Labeled Voltage: The voltage across the capacitor is labeled as $v_o(t)$.
			\item Polarity for $v_o(t)$: The positive reference is at the terminal connected to R3, and the negative reference is at the terminal connected to ground.
			\item Connections: The positive terminal is connected to resistor R3. The negative terminal is connected to the ground node.
		\end{itemize}
		
		\item \textbf{Operational Amplifier (Op-amp):}
		\begin{itemize}
			\item Connections:
			\begin{itemize}
				\item The non-inverting ($+$) input is connected to the positive terminal of the 5 V voltage source.
				\item The inverting ($-$) input is connected to one terminal of resistor R1, one terminal of resistor R2, and one terminal of the switch S.
				\item The output terminal is connected to the other terminal of resistor R2, the other terminal of the switch S, and one terminal of resistor R3.
			\end{itemize}
		\end{itemize}
		
		\item \textbf{Switch S:}
		\begin{itemize}
			\item Type: Single-pole single-throw (SPST).
			\item Time-varying behavior: The switch opens at $t = 0$.
			\item Connections: It is connected in parallel with resistor R2. One terminal is connected to the inverting ($-$) input of the op-amp, and the other terminal is connected to the output of the op-amp.
		\end{itemize}
	\end{itemize}
\end{tcolorbox}

\subsection{\texttt{ngspice}-Related Results}
This section presents the \texttt{ngspice}-related results for Problem 8.3-5 from \cite{svoboda2013introduction}, including the \texttt{.cir} file generated by Gemini 2.5 Pro and the corresponding \texttt{ngspice} simulation output.

\subsubsection{\texttt{.cir} File for \texttt{ngspice} Simulation}
The content of the \texttt{.cir} file used for the \texttt{ngspice} simulation is provided below. This code simulates the circuit over the time interval $[0.5\times10^{-3}\ \mathrm{s}, 0.5\ \mathrm{s}]$. The output is presented in two formats: the numerical values of \texttt{vout} ($v_{\rm{o}}$) using the command \texttt{print vout}, and the corresponding waveform using \texttt{plot vout}. The simulation results generated from the \texttt{.cir} file are presented in \ref{SB32}.

\begin{tcolorbox}[breakable, enhanced, width=\textwidth, opacityfill=1., colback=white]
	\fontsize{9pt}{10pt}\selectfont
	\begin{verbatim}
		* P 8.3-5: Op-amp circuit with a switch
		
		.PARAM pi = 3.141592653589793
		
		* Circuit components
		* Node 1: Op-amp non-inverting input
		* Node 2: Op-amp inverting input
		* Node 3: Op-amp output
		* Node 4: Capacitor positive terminal (output voltage vo_t)
		* Node 5: Switch control voltage
		* Node 0: Ground
		
		* Independent Voltage Source (V_s)
		Vs 1 0 5
		
		* Resistors
		R1 2 0 20e3
		R2 3 2 20e3
		R3 3 4 20e3
		
		* Capacitor C
		C1 4 0 4e-6
		
		* Operational Amplifier (Op-amp)
		Xopamp 1 2 3 OPAMP_IDEAL
		
		* Switch S (in parallel with R2)
		* The switch opens at t = 0.
		S1 3 2 5 0 SW1
		
		************************************
		* Switch (Closed to Open at t = 0) *
		************************************
		* Control voltage for switch (opens just after t=0)
		Vctrl 5 0 PULSE(5 0 1e-9 1e-9 1e-9 1e6 1e6)
		
		* Switch model
		.model SW1 SW(RON=1e-3 ROFF=1e6 Vt=2.5 Vh=0)
		************************************
		
		************************************
		*           Ideal Op-Amp           *
		************************************
		.subckt OPAMP_IDEAL p n o
		* p: non-inverting input
		* n: inverting input
		* o: output
		Bop o 0 V = {1e6 * (v(p) - v(n))}
		.ends
		************************************
		
		***********************************************************************
		*   Standard Simulation Control Template (DO NOT ADD OTHER COMMANDS)  *
		***********************************************************************
		.control
		tran 0.5e-3 0.5
		let vout = v(4)
		print vout
		plot vout
		.endc
		***********************************************************************
		
		.end
	\end{verbatim}
\end{tcolorbox}

\subsubsection{Simulation Result of \texttt{ngspice}}
\label{SB32}
\begin{figure}[!h]
	\centering
	\includegraphics[width=0.7\textwidth]{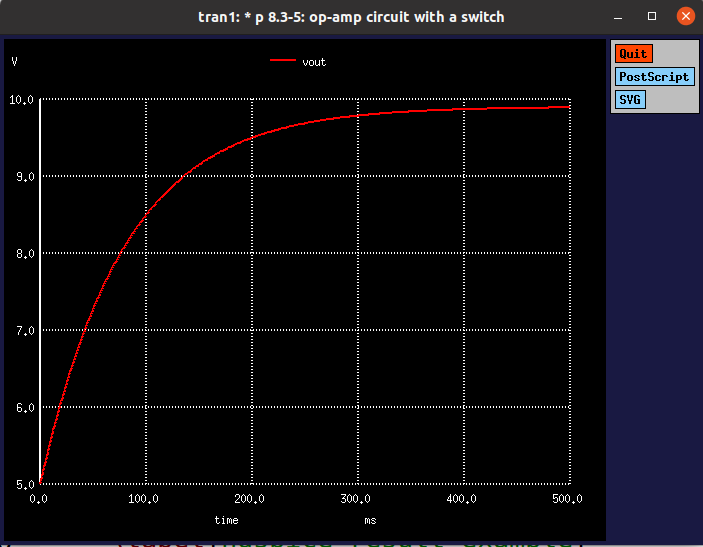}
	\caption{Simulation result from \texttt{ngspice} for Problem 8.3-5 in \cite{svoboda2013introduction}. The plot shows the output voltage ($v_{\rm{o}}$) over time, consistent with the expected analytical solution ($v(t) = 10 - 5e^{-12.5t}\ \mathrm{V})$.}
	\label{ngspice_result_example}
\end{figure}

Since the textual output of the \texttt{print vout} command is long, we include only the first and last several values of the variable \texttt{vout} below. In addition, the \texttt{plot vout} command produces a figure that can be obtained by running \texttt{ngspice P8.3-5.cir} in the terminal, assuming \texttt{ngspice} is properly installed. The screenshot of this figure is shown in Figure \ref{ngspice_result_example}, where the curve agrees with the correct analytical solution to Problem 8.3-5, namely $v(t)=10 - 5e^{-12.5t}\ \mathrm{V}$.

\begin{tcolorbox}[breakable, enhanced, width=\textwidth, opacityfill=1., colback=white]
	\fontsize{9pt}{10pt}\selectfont
	\begin{verbatim}
		
		Note: No compatibility mode selected!
		
		
		Circuit: * p 8.3-5: op-amp circuit with a switch
		
		Doing analysis at TEMP = 27.000000 and TNOM = 27.000000
		
		Using SPARSE 1.3 as Direct Linear Solver
		
		Initial Transient Solution
		--------------------------
		
		Node                                   Voltage
		----                                   -------
		1                                            5
		2                                            5
		3                                            5
		4                                            5
		5                                            5
		b.xopamp.bop#branch                -0.00025025
		vctrl#branch                                 0
		vs#branch                                    0
		
		
		No. of Data Rows : 1045
		* p 8.3-5: op-amp circuit with a switch
		Transient Analysis  Sun Jul 20 22:12:02  2025
		-----------------------------------------------------------------------
		Index   time            vout            
		-----------------------------------------------------------------------
		0	0.000000e+00	4.999995e+00	
		1	1.000000e-11	4.999995e+00	
		2	2.000000e-11	4.999995e+00	
		3	4.000000e-11	4.999995e+00	
		4	8.000000e-11	4.999995e+00	
		5	1.600000e-10	4.999995e+00	
		6	3.200000e-10	4.999995e+00	
		7	6.400000e-10	4.999995e+00	
		8	1.000000e-09	4.999995e+00	
		9	1.064000e-09	4.999995e+00	
		10	1.192000e-09	4.999995e+00
		
		... 
		
		Index   time            vout            
		-----------------------------------------------------------------------
		1041	4.987352e-01	9.892328e+00	
		1042	4.992352e-01	9.892387e+00	
		1043	4.997352e-01	9.892447e+00	
		1044	5.000000e-01	9.892478e+00	
		
		Warning: command 'plot' is not available during batch simulation, 
		ignored! You may use Gnuplot instead.
		
		Note: Simulation executed from .control section 
	\end{verbatim}
\end{tcolorbox}

\subsection{Gemini 2.5 Pro Generated Solution}

This section presents the step-by-step solution to Problem 8.3-5, generated by Gemini 2.5 Pro.\footnote{The original content, generated by Gemini 2.5 Pro, was formatted in Markdown. We adapted this format to improve organization and readability in the printed PDF version.}

\begin{tcolorbox}[breakable, enhanced, width=\textwidth, opacityfill=1., colback=white]
	Here is a step-by-step solution to the circuit analysis problem.
	
	The problem asks for the voltage $v_o(t)$ for $t > 0$. The switch in the circuit opens at $t = 0$. We will solve this by analyzing the circuit in two time intervals: $t < 0$ to find the initial condition, and $t > 0$ to find the transient response.
	
	\subsection*{\textbf{Part 1: Analysis for $t < 0$ (Finding the Initial Condition)}}
	
	For $t < 0$, the problem states that the circuit is in a steady state and the switch is closed.
	
	\textbf{1. Circuit Configuration for $t < 0$:}
	\begin{itemize}
		\item \textbf{Switch Status:} The switch S is closed.
		\item \textbf{Circuit Structure:} The closed switch is in parallel with the 20 k$\Omega$ feedback resistor (R2). A closed switch acts as a short circuit (zero resistance), so it effectively shorts out resistor R2. This creates a direct connection from the op-amp's output terminal to its inverting ($-$) input terminal. This configuration is known as a voltage follower or buffer.
		\item \textbf{Capacitor State:} Since the circuit is in DC steady state, the capacitor C acts as an open circuit. This means no DC current can flow through it.
	\end{itemize}
	
	\textbf{2. Calculating the Initial Voltage $v_o(0^-)$:}
	\begin{itemize}
		\item We assume an ideal op-amp. For an ideal op-amp, the voltage at the non-inverting input ($v_+$) is equal to the voltage at the inverting input ($v_-$): $v_+ = v_-$.
		\item The non-inverting ($+$) input is connected directly to the positive terminal of the 5 V DC source. The negative terminal of the source is connected to ground. Therefore, $v_+ = 5\text{ V}$.
		\item Due to the ideal op-amp property, $v_- = v_+ = 5\text{ V}$.
		\item In the voltage follower configuration for $t < 0$, the output voltage of the op-amp, $v_{\text{out}}$, is equal to the inverting input voltage:
		$$v_{\text{out}}(t) = v_-(t) = 5\text{ V}\quad\text{for } t < 0.$$
		\item The voltage $v_o(t)$ is the voltage across the capacitor C. The capacitor is connected to the op-amp output via resistor R3.
		\item Since the capacitor acts as an open circuit in steady state, the current flowing through it is zero. Consequently, the current flowing from the op-amp output, through resistor R3, to the positive terminal of the capacitor is also zero.
		\item With zero current through R3, there is no voltage drop across it (Ohm's Law: $V = I \cdot R = 0 \cdot 20\text{ k}\Omega = 0\text{ V}$).
		\item Therefore, the voltage at the positive terminal of the capacitor is the same as the op-amp's output voltage:
		$$v_o(0^-) = v_{\text{out}}(0^-) = 5\text{ V}.$$
	\end{itemize}
	
	The voltage across a capacitor cannot change instantaneously. Thus, the voltage across the capacitor just after the switch opens is the same as the voltage just before it opens:
	$$v_o(0^+) = v_o(0^-) = 5\text{ V}.$$
	
	\subsection*{\textbf{Part 2: Analysis for $t > 0$ (Finding the Transient Response)}}
	
	At $t = 0$, the switch opens. We now analyze the new circuit configuration for $t > 0$.
	
	\textbf{1. Circuit Configuration for $t > 0$:}
	\begin{itemize}
		\item \textbf{Switch Status:} The switch S is open.
		\item \textbf{Circuit Structure:} With the switch open, the 20 k$\Omega$ feedback resistor (R2) is now part of the circuit. The op-amp is now configured as a non-inverting amplifier. The feedback network consists of resistor R2 (from output to inverting input) and resistor R1 (from inverting input to ground).
	\end{itemize}
	
	\textbf{2. Analyzing the Op-Amp Circuit:}
	\begin{itemize}
		\item The non-inverting input voltage remains $v_+ = 5\text{ V}$.
		\item For an ideal op-amp, $v_- = v_+ = 5\text{ V}$.
		\item The output of the op-amp, $v_{\text{out}}$, is determined by the non-inverting amplifier gain formula:
		$$
		v_{\text{out}} = v_+ \left(1 + \frac{R2}{R1}\right)
		= 5\text{ V} \left(1 + \frac{20\text{ k}\Omega}{20\text{ k}\Omega}\right)
		= 5\text{ V} \cdot 2
		= 10\text{ V}.
		$$
		\item So, for all $t > 0$, the output of the op-amp is a constant $v_{\text{out}}(t) = 10\text{ V}$.
	\end{itemize}
	
	\textbf{3. Analyzing the RC Circuit:}
	\begin{itemize}
		\item For $t > 0$, the capacitor C is charged through resistor R3 by the op-amp's output voltage, $v_{\text{out}} = 10\text{ V}$. This forms a simple first-order RC circuit.
		\item The circuit consists of a voltage source $v_{\text{out}} = 10\text{ V}$, a resistor $R = R3 = 20\text{ k}\Omega$, and the capacitor $C = 4\ \mu\text{F}$. The voltage across the capacitor is $v_o(t)$.
	\end{itemize}
	
	\textbf{4. Finding the Final Value and Time Constant:}
	\begin{itemize}
		\item \textbf{Final Value ($v_o(\infty)$):} As $t \to \infty$, the circuit will reach a new steady state. The capacitor will be fully charged and will again act as an open circuit. No current will flow through R3. Therefore, the voltage across the capacitor will be equal to the source voltage driving it:
		$$
		v_o(\infty) = v_{\text{out}} = 10\text{ V}.
		$$
		\item \textbf{Time Constant ($\tau$):} The time constant of this RC circuit is given by the product of the resistance and capacitance:
		$$
		\tau = R \cdot C = R3 \cdot C
		= (20 \times 10^3\ \Omega)(4 \times 10^{-6}\ \text{F})
		= 80 \times 10^{-3}\ \text{s}
		= 0.08\ \text{s}.
		$$
	\end{itemize}
	
	\textbf{5. Formulating the Complete Solution:}
	
	The general expression for the voltage across a capacitor in a first-order RC circuit is:
	$$
	v_o(t) = v_o(\infty) + \big[v_o(0^+) - v_o(\infty)\big]\ e^{-t/\tau}.
	$$
	
	Substituting the values we found:
	$$
	v_o(\infty) = 10\text{ V},\qquad
	v_o(0^+) = 5\text{ V},\qquad
	\tau = 0.08\text{ s}.
	$$
	
	$$
	v_o(t) = 10 + \big[5 - 10\big]\ e^{-t/0.08}
	= 10 - 5\ e^{-t/0.08}\ \text{V}.
	$$
	
	The term in the exponent can be simplified: $1/0.08 = 12.5$.\\
	So, the final expression for $v_o(t)$ for $t > 0$ is:
	$$
	v_o(t) = 10 - 5e^{-12.5t}\ \text{V}.
	$$
	
	\subsection*{\textbf{Final Answer}}
	
	The voltage $v_o(t)$ for $t > 0$ is given by the expression: $v_o(t) = 10 - 5e^{-12.5t}\ \text{V}$.
\end{tcolorbox}

The Gemini-generated solution includes all the key steps in the problem-solving process. Its detailed explanation makes it well-suited for conversion into various formats for educational purposes.
\begin{itemize}
	\item [1.] Before ($t = 0$), the closed switch forces the op-amp into a voltage-follower configuration, giving the capacitor an initial voltage of 5 V.
	\item [2.] At ($t = 0$), the switch opens and the op-amp becomes a non-inverting amplifier with a constant output of 10 V.
	\item [3.] For ($t > 0$), the capacitor charges toward 10 V through a 20 k$\Omega$ resistor, forming a first-order RC circuit.
	\item [4.] The time constant of this RC circuit is calculated as 0.08 s.
	\item [5.] Using the standard first-order charging formula, the capacitor voltage is obtained as $v_{\rm o}(t)=10 - 5e^{-12.5t}$.
\end{itemize}

\section{Prompt Templates}
\label{SC}
This section presents the prompt templates used with Gemini 2.5 Pro throughout the circuit problem-solving workflow. The prompts fall into four distinct categories, each serving a specific purpose: circuit recognition, problem solving, \texttt{.cir} file generation, and Python script generation for converting a Gemini-generated solution into a numerical one.

For tasks involving multiple steps, multiple rounds of interaction with Gemini 2.5 Pro are required. In such cases, the Gemini API's multi-turn conversation feature\footnote{See details at \url{https://ai.google.dev/gemini-api/docs/text-generation\#multi-turn-conversations}.}, also known as chat mode, is used in the implementation. Additionally, while the actual prompts were implemented in plain text format, the presentations in this section have been adapted to improve organization and readability in the printed PDF version.

\subsection{Prompt Templates for Circuit Recognition}
\label{C1}
This section presents the prompt templates used for circuit diagram recognition in Section \ref{S32}. To achieve highly accurate circuit descriptions, a multi-step prompting method is employed. The prompts are described as follows:

\begin{itemize}
	\item [1.] \textit{Circuit Diagram Input:} In this step, the circuit diagram to be recognized is input into Gemini 2.5 Pro.
	\item [2.] \textit{Recognition of Circuit Components and Nodes:} As the basic recognition step, the components and nodes in the circuit diagram are identified.
	\item [3.] \textit{Identification of Current Direction:} As shown in Section \ref{S31}, current direction recognition is prone to hallucinations. Therefore, a dedicated prompting step is used to focus on current direction identification. This corresponds to the content in Section \ref{S322}.
	\item [4.] \textit{Check for Independent Source Existence:} Before applying the fine-tuned YOLO model to detect the positions of independent sources, their existence is first checked. Detection is carried out only if such sources are present.
	\item [5.] \textit{Check for Dependent Source Existence:} Similarly, before using OpenCV to detect the positions of dependent sources, their existence is verified. The detection step is executed only if such sources are found.
	\item [6.] \textit{Recognition of Inset Images:} This step involves recognizing the source polarities from cropped images of circuit sources, as described in Section \ref{S321}.
	\item [7.] \textit{Summarization of Circuit Recognition:} The final step summarizes the circuit recognition results and outputs the complete textual circuit description for use in subsequent problem-solving tasks and \texttt{ngspice} simulation. If the source polarity identified through inset image recognition differs from the initial recognition from the full circuit diagram, the result from the inset image is considered correct.
\end{itemize}

The detailed prompt templates are presented below.

\begin{tcolorbox}[title={Step 1: Circuit Diagram Input}, breakable, enhanced, width=\textwidth, opacityfill=1., colback=verylightgray]
	You are a circuit analysis expert. Let's solve a circuit analysis problem step by step, following my prompts. The attached circuit diagram will be used throughout the entire question-and-answer process.\\
	
	Please reply ``Yes'' if you are ready to answer questions about the attached circuit diagram. You do not need to do any calculations or analyses at this step. 
\end{tcolorbox}

\begin{tcolorbox}[title={Step 2: Recognition of Circuit Components and Nodes}, breakable, enhanced, width=\textwidth, opacityfill=1., colback=verylightgray]
	This step involves recognizing the components and nodes in the provided circuit. 
	
	\begin{itemize}
		\item [1.] List all the components in the circuit along with their associated values, such as the voltage or current sources, the resistance, capacitance, and inductance of the components, and the controlling variables for controlled sources. 
		
		\item [2.] List all the nodes in the circuit diagram. Specify how each node is connected to the terminals of the components.
	\end{itemize}
	
	Very Important Notes:  
	
	\begin{itemize}
		\item [1.] For the detected voltage/current sources, you should clearly indicate whether the sources are independent or dependent. 
		
		\item [2.] If the voltage across a component is labeled in a circuit diagram, regardless of whether it is positive or negative, it equals the voltage at the positive node minus the voltage at the negative node. 
		
		\item [3.] If a switch is detected, then it must be a single-pole single-throw (SPST) type and have only two states: open or closed. If you detect a three-terminal switch, your detection is incorrect and the recognition process must be redone. The detected SPST switch should be treated as a solid component, meaning its two terminals should be regarded as distinct nodes. Note that you do not need to identify the switch's state (i.e., open or closed) at this stage, as this information will be provided in the problem statement later.
	\end{itemize}
\end{tcolorbox}

\textit{Note for Step 2:} In our problem sets, all switches are single-pole single-throw (SPST). For problems involving other types of switches, such as single-pole double-throw (SPDT), the switch type, connection relationships, status, and switching behavior should be explicitly included in the textual problem statement. This is because we found that Gemini 2.5 does not reliably recognize switch-related information for these more complex types. 

\begin{tcolorbox}[title={Step 3: Identification of Current Direction}, breakable, enhanced, width=\textwidth, opacityfill=1., colback=verylightgray]
	This step involves recognizing the labeled current(s) in the provided circuit. Please be very careful when doing detections. 
	
	\begin{itemize}
		\item [1.] If there are not labeled currents in the provided circuit diagram, then you do not need to answer the following question. 
		
		\item [2.] [BE VERY CAREFUL] For a current source or any currents labeled in the circuit diagram, specify the direction of current flow, such as 
		
		\begin{itemize}
			\item [2a)] the direction of the arrow that indicates the current direction (upward/downward/leftward/rightward for branch currents and clockwise/anti-clockwise for mesh currents)
			
			\item [2b)] it flows from which node/terminal, and through which component (if applicable), to which node/terminal (You may use component names to describe the terminals)
		\end{itemize}
	\end{itemize}
	
	NOTES:
	\begin{itemize}
		\item [1.] If there are two circuit diagrams labeled (a) and (b), you should make separate recognitions. 
		
		\item [2.] If a current is starting from the negative terminal of a voltage source, you should clearly note that its direction is OPPOSITE to the current supplied by the voltage source. 
	\end{itemize}
\end{tcolorbox}

\begin{tcolorbox}[title={Step 4: Check for Independent Source Existence}, breakable, enhanced, width=\textwidth, opacityfill=1., colback=verylightgray]
	Based on the component list you generated, please respond ``Yes'' if there is at least one INDEPENDENT source in the given circuit. Otherwise, respond ``No''. \\
	
	Note that you should only respond ``Yes'' or ``No'' without any other words. 
\end{tcolorbox}

\begin{tcolorbox}[title={Step 5: Check for Dependent Source Existence}, breakable, enhanced, width=\textwidth, opacityfill=1., colback=verylightgray]
	Based on the component list you generated, please respond ``Yes'' if there is at least one DEPENDENT source in the given circuit. Otherwise, respond ``No''. \\
	
	Note that you should only respond ``Yes'' or ``No'' without any other words. 
\end{tcolorbox}

\begin{tcolorbox}[title={Step 6: Recognition of Inset Images}, breakable, enhanced, width=\textwidth, opacityfill=1., colback=verylightgray]
	An external tool detected \{NUM\_COMPONENTS\} \{TYPE\_COM-PONENT\}s from the provided circuit diagram. One of the \{TYPE\_COMPONENT\} has the xyxy bounding box information of \{BBOX\_XYXY\}. The inset of this \{TYPE\_COMPONENT\}, which is from the main circuit diagram, is attached. 
	
	\begin{itemize}
		\item [1.] Please double check if the inset as provided is truly a(n) \{TYPE\_COMPONENT\}. If not, you may ignore the second request below. 
		
		\item [2.] Please recognize the \{RECOGNITION\_CONTENT\}? Note that your response should be based on the inset circuit element provided specifically in this chat. 
	\end{itemize}
\end{tcolorbox}

\textit{Notes for Step 6:}
\begin{itemize}
	\item[i)] Depending on the detection scenario, the placeholders \{NUM\_COMPO-NENTS\}, \{TYPE\_COMPONENT\}, \{RECOGNITION\_CONTENT\}, and \{BBOX\_XYXY\} should be replaced with specific information. Notably, the available options for \{TYPE\_COMPONENT\} and \{RECOGNITION\_CONTENT\} are:
	\begin{itemize}
		\item \{TYPE\_COMPONENT\}: [``independent voltage source'', ``independent current source'', ``dependent source'']
		\item \{RECOGNITION\_CONTENT\}: [``positive and negative terminals of this independent voltage source'', ``current direction through this independent current source'', ``polarity/direction of this dependent source'']
	\end{itemize}
	\item[ii)] The prompt template above is designed for cases where multiple sources are detected. It can be easily adapted for scenarios involving a single detected source.
\end{itemize}

\begin{tcolorbox}[title={Step 7: Summarization of Circuit Recognition}, breakable, enhanced, width=\textwidth, opacityfill=1., colback=verylightgray]
	\begin{itemize}
		\item [1.] Check if the polarities/directions of sources one-by-one in the main and inset circuit recognitions are different. If there is any discrepancy, you should treat the INSET RECOGNITION RESULT(S) (i.e., your LAST RESPONSE) as the correct ones, re-recognize the main circuit diagram, and revise the main circuit recognitions according to the summary of the inset circuit recognitions. Make sure to answer the following questions according to the revised circuit recognitions.
		
		\item [2.] Based on your component and node detections in the chat history, specify each component's value and its direct connections with other components. If the component terminals have polarities---such as the positive and negative terminals of a voltage source---clearly indicate them. For a current source or any currents labelled in the circuit diagram, specify the direction of current flow, such as 
		\begin{itemize}
			\item [2a)] the direction of the arrow that indicates the current direction (upward/downward/leftward/rightward for branch currents and clockwise/anti-clockwise for mesh currents)
			
			\item [2b)] it flows from which node/terminal, and through which component (if applicable), to which node/terminal (You may use component names to describe the terminals)
		\end{itemize}
		\item [3.] If any time-varying values are provided in the form of a function curve in the initial figure, you should identify those values and express the corresponding variables as functions of time.\\ 
	\end{itemize}
	
	NOTES:
	
	\begin{itemize}
		\item [1.] If a current is starting from the negative terminal of a voltage source, you should clearly note that its direction is OPPOSITE to the current supplied by the voltage source. 
		
		\item [2.] Your output should consist only of the complete circuit information, as instructed in point 2 above, although you must implicitly address point 1 during the reasoning process.
	\end{itemize}
\end{tcolorbox}

\subsection{Prompt Templates for Solving Circuit Problems Based on Recognized Circuits}
\label{C2}
With the recognized circuits in textual format, we can solve the corresponding circuit problems using Gemini 2.5 Pro. The detailed prompting steps are summarized below:

\begin{itemize}
	\item [1.] \textit{Circuit Problem Solving:} This step involves applying Gemini 2.5 Pro to perform the analysis and computation based on the recognized circuit description. The prompt includes specific instructions designed to enhance problem-solving performance.
	\item [2.] \textit{Solution Summarization to Final Answer:} In this step, the solution is summarized and formatted to clearly and concisely extract the final answer. The final answer is then compared with the \texttt{ngspice} simulation result for verification.
\end{itemize}

The detailed prompt templates are presented below.

\begin{tcolorbox}[title={Step 1: Circuit Problem Solving}, breakable, enhanced, width=\textwidth, opacityfill=1., colback=verylightgray]
	You are an expert in circuit analysis tasked with solving a circuit analysis problem. The problem includes a problem statement, a text-based description of the circuit, and an attached circuit diagram. In the case of any discrepancy between these sources, prioritize the problem statement first, followed by the text description. Your response should include both the problem-solving process and the final answer. \\
	
	[Problem Statement]
	
	\{PROBLEM\_STATEMENT\}\\
	
	[Circuit Information]
	
	\{CIRCUIT\_INFORMATION\}\\
	
	VERY IMPORTANT NOTES:
	
	\begin{itemize}
		\item [1.] If the circuit diagram includes a switch, 
		
		\begin{itemize}
			\item [1a)] You should restate or rephrase the switch status change(s) described in the problem statement and follow them when solving the problem. 
			
			\item [1b)] [Very Important and Be Careful] Under different switch states, you should first clearly describe the circuit's structure and connections before proceeding with the analysis.
		\end{itemize}
		
		\item [2.] If a current is described in the problem statement or indicated in the circuit diagram or circuit information, you should first restate or rephrase the current and adhere to it when solving the problem. In the event of any discrepancy between the current described in the problem statement and your previous interpretation, the description in the problem statement should be considered correct. 
		
		\item [3.] When describing/defining a current, 
		
		\begin{itemize}
			\item [3a)] You should always describe or define it by a sentence like ``the current flows from node/terminal X, through component Z (if applicable, Z MUST be included), to node/terminal Y'', where ``X'', ``Y'' can be replaced with node/terminal names, and ``Z'' can be replaced with (a) component name(s). The circuit problem-solving process should be based on such descriptions. 
			
			\item [3b)] [Very Important] The definitions of a node and a terminal differ. A node is defined as a junction where two or more circuit elements are connected, whereas a terminal refers to a part of a specific circuit element. When describing the direction of current, it is crucial to distinguish whether the current is flowing from or to a particular node or terminal, especially when calculating powers. 
			
			\item [3c)] [Very Important] Whenever you mention a current, you should always restate its definition (from which node/terminal, through which component if applicable, to which node/terminal), if it has already been defined, based on the most recent context in which it was used.
			
			\item [3d)] If a current is starting from the negative terminal of a voltage source, you should clearly note that its direction is OPPOSITE to the current supplied by the voltage source. \\
		\end{itemize}
		
		\item [4.] [Very Important] When calculating the power absorbed by a given source, the following steps must be strictly followed in the reasoning process: 
		
		\begin{itemize}
			\item [4a)] Determine the current supplied by the source. 
			
			\begin{itemize}
				\item [4a1)] The direction of the current supplied by a source is strictly defined as follows: the negative terminal of the source -> the source itself -> the positive terminal. This direction should be clearly and explicitly included in your response. 
				
				\item [4a2)] [Very Important] Double-check this carefully in the reasoning process: the current supplied by a source is NOT: the positive terminal -> the source itself -> the negative terminal. 
			\end{itemize}
			
			\item [4b)] Determine the voltage across the source.
			
			\item [4c)] Calculate the power supplied by the source.
		\end{itemize}
		
		\item [5.] In source transformation, you should follow the following steps strictly:
		
		\begin{itemize}
			\item [5a)] What are the polarities/directions of the original source?
			
			\item [5b)] If the current flows from Node A to Node B through the current source, then the positive terminal of the voltage source will be at Node B (the code is current is FROM) and the negative terminal at Node A, and vice versa. Based on this, what are the polarities/directions of the resulting source? 
			
			\item [5c)] [Very Important] Double-check this carefully in the reasoning process: The positive terminal of the voltage source is located at the node toward which the current from the current source flows. 
		\end{itemize}
		
		\item [6.] [Very Important] When using superposition to calculate the individual contribution of each source, you should first clearly describe the circuit's structure and connections before proceeding with the analysis.
	\end{itemize}
\end{tcolorbox}

\begin{tcolorbox}[title={Step 2: Solution Summarization to Final Answer}, breakable, enhanced, width=\textwidth, opacityfill=1., colback=verylightgray]
	Based on your calculations, summarize the final answer by providing only the numerical values or expressions that directly address the question posed in the problem statement.
\end{tcolorbox}

\subsection{Prompt Templates for \texttt{.cir} File Generation in \texttt{ngspice} Simulations}
\label{C3}
Based on the textual circuit recognition, we generate \texttt{.cir} files for \texttt{ngspice} simulation. The correctness of these \texttt{.cir} files directly affects both their executability and the accuracy of the simulation results. To ensure high-quality \texttt{.cir} files, the prompting process consists of the following three steps:

\begin{itemize}
	\item [1.] \textit{Initial Generation of the \texttt{.cir} File:}
	Using the circuit information and problem statement as inputs, an initial version of the \texttt{.cir} file is generated. The corresponding prompt template includes the task specification, \texttt{.cir} templates for commonly used advanced modules (e.g., switches and operational amplifiers), a simulation control template, and detailed instructions for file generation.
	
	\item [2.] \textit{Format Correction for the Generated \texttt{.cir} File:}
	The initial \texttt{.cir} file may contain syntax errors that prevent execution. This step specifically prompts Gemini 2.5 Pro to check the file format and correct any issues as needed.
	
	\item [3.] \textit{Accuracy Verification for the \texttt{.cir} File:}
	This final step ensures that the generated \texttt{.cir} file accurately reflects the original circuit diagram and problem statement. The prompt template includes specific aspects that Gemini 2.5 Pro is instructed to verify and revise if any inconsistencies or errors are detected.
\end{itemize}

The detailed prompt templates are presented below. Note that the template shown pertains to the ``Circuit Analysis'' category in Table \ref{TableA1}, but it can be readily adapted to the other three categories: ``Circuit Synthesis'', ``Network Function Analysis'', and ``Network Function Synthesis''.

\begin{tcolorbox}[title={Step 1: Initial Generation of the \texttt{.cir} File}, breakable, enhanced, width=\textwidth, opacityfill=1., colback=verylightgray]
	I am using ngspice to simulate a circuit. Your task is to generate the content of a \texttt{.cir} file using the circuit information and problem statement provided below.\\
	
	\# Begin of circuit information
	
	[Circuit Information]
	
	\# End of circuit information\\
	
	\# Begin of problem statement
	
	[Problem Statement]\\
	
	[Important Note for Problem Statement]
	
	\begin{itemize}
		\item [1.] You SHOULD DIRECTLY OMIT the problem-solving methods, such as source transformations, superposition, and Thevenin/Norton equivalent circuits, in the problem descriptions. Instead, always perform the simulation, as instructed in the standard simulation control template below, based on the original circuit.
		
		\item [2.] You SHOULD DIRECTLY OMIT any required formats specified in the problem statement, such as phasors. All variables in the output should always be expressed in the TIME DOMAIN, as instructed in the standard simulation control template below. 
		
		\item [3.] If there are multiple subproblems or cases, you should consider all of them SEPARATELY by printing and plotting all the required variables. You need to ensure that all variables are well-defined before referring to them. DO NOT use the alter command. The file you generate should be executable directly, without requiring any further modifications. 
		
		\item [4.] If there is any discrepancy between the circuit descriptions in the problem statement and the circuit information above, you should treat the information in the problem statement as correct. 
		
		\item [5.] When the open-circuit voltage or the short-circuit current is required, it is important to carefully determine their directions. In the reasoning process, you should first describe how the current flows in the loop formed by the equivalent current source and the short-circuit path and then use the current flow information to determine the direction of the short-circuit current. 
	\end{itemize}
	
	\# End of problem statement\\

	Notes:
	
	\begin{itemize}
		\item [1.] If the circuit contains an operational amplifier or a switch, you may refer to the following examples.
		
		You may adjust the switch timing according to the specific requirements of the provided circuit. However, DO NOT change other parts. 
		\begin{scriptsize}
		\begin{verbatim}
			************************************
			* Switch (Closed to Open at t = 0) *
			************************************
			* Control voltage for switch (opens just after t=0)
			Vctrl ctrl 0 PULSE(5 0 1n 1n 1n 1e6 1e6)
			
			* Switch model
			.model SW1 SW(RON=1m ROFF=1Meg Vt=2.5 Vh=0)
			************************************
			
			************************************
			* Switch (Open to Closed at t = 0) *
			************************************
			* Control voltage for switch (closes just after t=0)
			Vctrl ctrl 0 PULSE(0 5 1n 1n 1n 1e6 1e6)
			
			* Switch model
			.model SW1 SW(RON=1m ROFF=1Meg Vt=2.5 Vh=0)
			************************************
			
			************************************
			*           Ideal Op-Amp           *
			************************************
			.subckt OPAMP_IDEAL + - out
			* +: non-inverting input
			* -: inverting input
			* out: output
			EOP out 0 VALUE = {1e6 * (V(+) - V(-))}
			.ends
			************************************
		\end{verbatim}
		\end{scriptsize}
		
		\item [2.] You should ALWAYS and EXCLUSIVELY use the following TEMPLATE COMMAND to control the simulation and NEVER use fourier or ac analysis to calculate amplitudes and phases. 
		
		\begin{scriptsize}
			\begin{verbatim}
				***********************************************************************
				*   Standard Simulation Control Template (DO NOT ADD OTHER COMMANDS)  *
				***********************************************************************
				.control
				tran [TIMESTEP] [SIMULATION TIME] ; You should always use tran. For DC 
				circuits without a switch, you may set TIMESTEP to 0.1 and 
				SIMULATION_TIME to 1.
				print [REQUIRED VALUE(S) WITHIN THE TIME DOMAIN IN THE QUESTION HERE] ; 
				DO NOT use print tran [REQUIRED VALUE IN THE QUESTION HERE]. The 
				required value(s) should not be expression(s) and need to be 
				pre-defined as variable(s) before using print or plot but after the 
				tran command. 
				plot [REQUIRED VALUE(S) WITHIN THE TIME DOMAIN IN THE QUESTION HERE] ; 
				DO NOT use plot tran [REQUIRED VALUE IN THE QUESTION HERE]. Do not 
				add title, xlabel, ylabel, etc. 
				.endc
				***********************************************************************
			\end{verbatim}
		\end{scriptsize}
		
		\item [3.] The circuit input to ngspice must be properly grounded. Follow the steps below to check and add grounding to the circuit:
		
		\begin{itemize}
			\item [3a)] If the original circuit already includes a ground node, you should not add another one. Simply incorporate the existing ground into the netlist. 
			
			\item [3b)] If the original circuit is not properly grounded:
			
			\begin{itemize}
				\item [3b-i)] If there is at least one independent voltage source in the circuit, select the negative terminal of any one voltage source and connect it to ground.
				
				\item [3b-ii)] If there are no independent voltage sources but at least one independent current source, select the negative terminal of any one current source and connect it to ground.
			\end{itemize}
		\end{itemize}
		
		\item [4.] To incorporate time-varying sources in ngspice, use behavioral sources (B elements) with expressions involving the time variable for custom waveforms. 
		
		\item [5.] You do not need to apply circuit laws, such as KCL or KVL, to determine the required variables; simply measure them instead. By default, ngspice may not save the currents through components OTHER THAN INDEPENDENT VOLTAGE SOURCES. If measuring a current is necessary, insert a 0-volt DC voltage source in series with the measured element; the current through this voltage source can then be measured. 
		
		\item [6.] Always include ``\verb|.PARAM pi = 3.141592653589793|'' in the .cir file. When defining new parameters with .PARAM using existing parameters, use braces in the whole expressions. i.e., ``\verb|.PARAM new_param = {EXPRESSION_IN_TERMS_OF_OLD_| \verb|PARAMETERS}|''.
		
		\item [7.] To define an independent current source in ngspice, use the syntax \verb|I<name> <node the current is from> <node the current| \verb|is to><value or waveform>|. Be careful with the order of the nodes, as it determines the direction of current flow.
		
		\item [8.] Always assume that the units for current, voltage, and resistance are amperes (A), volts (V), and ohms ($\Omega$), respectively. Similar assumptions apply to other variables. However, you should never display the units when defining variables or electrical elements. For example, use 1 instead of 1 A for a current. 
		
		\item [9.] When a circuit contains energy storage elements such as capacitors and inductors, you should not calculate the initial conditions or use the \verb|uic| option in the \verb|.tran| command during simulation if the initial conditions are not explicitly provided. However, if the initial conditions of the energy storage elements are explicitly given in the problem statement, you must specify them both in the element definitions and in the \verb|.tran| command in the simulation control. 
		
		\item [10.] To define a source with a step function $u(t)$ in ngspice, use a behavioral source with a conditional expression that activates the desired waveform only when time >= 0. For example, 
		
		\begin{itemize}
			\item [10a)] The voltage source $15e^{-2t} \cdot u(t) \rm{~V}$ can be defined as: \verb|Bsource out 0 V = {time > 0 ? 15 * exp(-2 * time)| \verb|: 0}|. Note that you should use ``\verb|time > 0|'' in the expression instead of ``\verb|time >= 0|'' in the case of $u(t)$. 
			
			\item [10b)] The current source $30u(-t) \rm{~mA}$ can be defined as \verb|I = {time <= 0 ? 30m : 0}|. Note that you should use ``\verb|time <= 0|'' in the expression instead of ``\verb|time < 0|'' in the case of $u(-t)$. 
		\end{itemize}
	\end{itemize}
\end{tcolorbox}

\begin{tcolorbox}[title={Step 2: Format Correction for the Generated \texttt{.cir} File}, breakable, enhanced, width=\textwidth, opacityfill=1., colback=verylightgray]
	Please double-check and make the necessary adjustments to your generated \verb|.cir| content based on the following steps. Your output should consist of the updated \verb|.cir| content with all necessary corrections applied. \\
	
	You need to follow the steps below one-by-one during the reasoning process.
	
	\begin{itemize}
		\item [1.] When referencing the voltage between any two nodes---for example, nodes a and b---never use \verb|v(a, b)|; instead, use \verb|v(a) - v(b)|. Always assume \verb|v(0)=0| by default, and never explicitly use \verb|v(0)|; use \verb|0| instead. 
		
		\item [2.] \verb|tran| in ngspice does not support evaluating \verb|.PARAM| expressions inside the command itself. You need to evaluate the parameters manually and plug the numerical values into the \verb|tran| command.
		
		\item [3.] Define each node using only a single digit or letter; that is, the length of a node's name should always be exactly one character. When defining new variables using the ``\verb|let|'' command, do not name the new variables as ``\verb|v|'', ``\verb|V|'', ``\verb|i|'', or ``\verb|I|''. Use alternative names instead. Note that you should use ``\verb|let|'' without a dot prefix instead of ``\verb|.let|'' to define new variables. 
		
		\item [4.] If there is any behavioral source (i.e., B-source), ensure that the element name begins with ``\verb|B|'' to indicate a behavioral source.
		
		\item [5.] To define a time-varying voltage or current source with piecewise linear values, you should use a \verb|PWL| (Piecewise Linear) source in ngspice. However, DO NOT modify the \verb|PULSE| function used to define the switch, if it exists.
		
		\item [6.] When using \verb|.PARAM| constants in \verb|.control| blocks, define them again with let as scalar variables (e.g., \verb|let is = 2|) because \verb|.PARAM| values are not recognized as numeric vectors during control block expressions. Note that you should use the numeric values of the \verb|.PARAM| variables on the right-hand side of the let expression, instead of their symbolic names. 
		
		\item [7.] You should also include a dummy 0-volt DC voltage source to facilitate the measurement of current through dependent sources. Be very careful to ensure that the current direction matches the one provided in the circuit information and/or the problem descriptions. 
		
		\item [8.] Double-check the scientific notation of values in the \verb|.cir| file content. 
		
		\item [9.] In scenarios involving a current-controlled voltage or current source, if the current through a voltage source is used to control other dependent sources, you should check again that the correct current direction is used for the controlling current. During the reasoning process, you should follow the guideline outlined below. 
		\begin{itemize}
			\item [9a)] For a voltage source, such as \verb|V1|, the conventional direction of the current \verb|I(V1)| is defined as flowing from the positive terminal of the voltage source, THROUGH THE SOURCE ITSELF, to the negative terminal. 
			
			\item [9b)] The direction of \verb|I(V1)| is OPPOSITE to the direction of the current supplied by the voltage source \verb|V1|.
			
			\item [9c)] The current \verb|I(V1)| is NOT the current supplied by the voltage source \verb|V1|; there should be a MINUS SIGN. 
			
			\item [9d)] You should recheck the direction of the controlled current --- it may not be correctly defined.
		\end{itemize}
	\end{itemize}
\end{tcolorbox}

\begin{tcolorbox}[title={Step 3: Accuracy Verification for the \texttt{.cir} File}, breakable, enhanced, width=\textwidth, opacityfill=1., colback=verylightgray]
	Please double-check and make the necessary adjustments to your generated \verb|.cir| content based on the following steps. Your output should consist of the updated \verb|.cir| content with all necessary corrections applied. \\
	
	You need to follow the steps below one-by-one during the reasoning process.
	
	\begin{itemize}
		\item [1.] Double-check the connections of the components in the generated \verb|.cir| content by referring to the original circuit information. If you find any discrepancies, revise the \verb|.cir| content to ensure alignment with the original. 
		
		\item [2.] When the required values in the problem include power, carefully define the sign of the power---i.e., whether $p = +vi$ or $p = -vi$---by correctly applying the passive sign convention. Use the following rules to determine the appropriate sign:
		\begin{itemize}
			\item [2a)] If the power of a VOLTAGE SOURCE is to be calculated, define the direction of current $i$ through the voltage source as: from the positive terminal of the voltage source -> through the voltage source itself -> to the negative terminal. Then, use $p = -vi$ (NOTE THE NEGATIVE SIGN) to calculate the power SUPPLIED by the voltage source.
			
			\item [2b)] If the power of a CURRENT SOURCE is to be calculated, define the voltage $v$ across the current source as the voltage at the terminal from which the current ENTERS the source (and then flows through it) MINUS the voltage at the terminal into which the current FLOWS OUT OF. Then, use $p = -vi$ (NOTE THE NEGATIVE SIGN) to calculate the power SUPPLIED by the current source.
			
			\item [2c)] If the power of a passive component (such as a resistor, capacitor, or inductor) is to be calculated, define the direction of current $i$ through the passive component as: from the positive terminal of the component -> through the component itself -> to the negative terminal. Then, use $p = +vi$ (NOTE THE POSITIVE SIGN) to calculate the power ABSORBED by the passive component. 
		\end{itemize}
		During the process above, you need to be very careful when reasoning and determining the direction of the current. 
		
		\item [3.] Some components, such as sources and operational amplifiers, have defined polarities. You should double-check that the terminals of these components are correctly connected to the corresponding nodes.
		
		\item [4.] When measuring the current through a voltage source, it is important to double-check the current's direction. During the reasoning process, you should follow the guidelines outlined below. 
		\begin{itemize}
			\item [4a)] [Very Important] For a voltage source, such as \verb|V1|, the conventional direction of the current \verb|I(V1)| is from the positive terminal of the voltage source, through the source itself, and to the negative terminal. This direction is defined as OPPOSITE to the current actually supplied by the voltage source \verb|V1|. 
			
			\item [4b)] By default, ngspice may not save the currents through components OTHER THAN INDEPENDENT VOLTAGE SOURCES. If measuring a current is necessary, insert a 0-volt DC voltage source in series with the measured element; the current through this voltage source can then be measured.
		\end{itemize}
		\item [5.] In scenarios involving a current-controlled voltage or current source, if the current through a voltage source is used to control other dependent sources, you should double-check that the correct current direction is used for the controlling current. Additionally, ensure that the controlled variable is indeed a current. You may refer to 4b) when double-checking the current direction. 
		
		\item [6.] If any current is measured, be very careful to ensure that the current direction matches the one specified in the circuit information and/or the problem description. 
		
		\item [7.] When the open-circuit voltage or the short-circuit current is required, it is important to carefully double-check their directions and make corrections if necessary. Before determining the direction of the short-circuit current, you should first describe how the current flows in the loop formed by the equivalent current source and the short-circuit path. 
	\end{itemize}
\end{tcolorbox}

\subsection{Prompt Template for Generating a Python Script to Convert a Gemini-Generated Solution into a Numerical One}
\label{C4}

This section provides a prompt template for generating a Python script that converts a Gemini-generated solution from natural language into a numerical one, as illustrated in Step 2 of Section \ref{S441}. The template below applies to problems in which the \texttt{ngspice} simulation results are functions of time---specifically, the ``Circuit Analysis'' and ``Circuit Synthesis'' categories in Table \ref{TableA1}. The prompt can be easily adapted for problems in the ``Network Function Analysis'' and ``Network Function Synthesis'' categories, where the \texttt{ngspice} outputs represent the magnitude and phase of a network function.

Note that synthesis problems require determining the values of one or more circuit elements, given certain voltages or currents. As illustrated in Section \ref{S43}, \texttt{ngspice} may treat some of these given voltages or currents as unknowns, using the element values computed by Gemini 2.5 Pro to derive these ``unknown'' quantities. If the simulated values match those specified in the problem description, the Gemini-generated solution is verified as correct. Therefore, we include the full Gemini-generated solution---rather than just the concise final answer---in the prompt below, to provide sufficient context for selecting output variables in the Python script.

\begin{tcolorbox}[breakable, enhanced, width=\textwidth, opacityfill=1., colback=verylightgray]
	You are an expert Python programmer specializing in scientific and numerical computing. Your task is to write a complete, self-contained, and immediately runnable Python script based on the provided circuit analysis solution and simulation parameters.
	
	Do NOT write any explanation, comments, or text outside of the Python code block. Directly begin your response with the ``\texttt{import xxx}''.\\
	
	Objective:\\
	
	The script you generate must perform the following actions:
	
	\begin{itemize}
		\item [1.] Parse the provided time parameters to create a time vector using \texttt{numpy}.
		
		\item [2.] Analyze the ``Circuit Solution Text'' to find the mathematical formulas for each of the ``Target Variables'', you can do it outside the Python code (in your chain-of-thought), only put the analyzed formulas into the generated Python code to reduce the complexity. In most case, the ``Circuit Solution Text'' contains quite a few process towards final answer, which you need to check in detail. You can also get the additional information from the ``Concise Answer'' to learn about the relationship between the variables in ``Target Variables'' and in ``Circuit Solution Text''.
		
		\item [3.] Intelligently map the target variables from the Circuit Solution Text (e.g., $I_{\rm{s}}$, $v_{\rm o}(t)$) to the corresponding keys in the ``Target Variables'' list (e.g., \verb|Current_s|, \verb|Voltage_o|), sometimes they can be very similar, but do let them totally match.
		
		\item [4.] Calculate the numerical values for each target variable over the entire time vector.
		
		\item [5.] Store the time vector (with key name ``time'') and the calculated values (The keys must exactly match the ``Target Variables'' list as mentioned in point 3) in a dictionary.
		
		\item [6.] Save the final dictionary as a \texttt{json} file to the specified ``Output File Path''.
	\end{itemize}
	
	Input Information:
	
	\begin{itemize}
		\item [1.] Circuit Solution Text:
		
		[Gemini-Generated Full Solution]
		
		\item [2.] Simulation Parameters and Target Variables:
		
		[Time Interval of the \texttt{ngspice}-Simulated Solution and the Target Variables]
		
		\item [3.] Concise Answer:
		
		[The Summarized Concise Final Answer to the Gemini-Generated Full Solution]
		
		\item [4.] Output File Path:
		
		[The Output Path for the Generated Python Script]
		
	\end{itemize}
	
	Requirements for the Generated Script:
	
	\begin{itemize}
		\item [--] Must import ``\texttt{numpy}'' and ``\texttt{json}''.
		
		\item [--] Must correctly handle various mathematical expressions: phasor notation (e.g., ``$A\angle B^{\circ}$''), time-domain cosine functions (e.g., ``$A\cos(\omega t + \phi)$''), and exponential functions (e.g., ``$Ae^{Bt}$'').
		
		\item [--] Must correctly identify the angular frequency ($\omega$) if it's stated explicitly or implicitly in a cosine function, and use it for converting phasors to the time domain.
		
		\item [--] Must convert all angles from degrees to radians before using them in trigonometric calculations (``\texttt{numpy.deg2rad}'').
		
		\item [--] The final output must be a single Python code block starting with ``\texttt{import}'' and ending with the file writing operation.
	\end{itemize}
\end{tcolorbox}

\bibliographystyle{plainnat}
\bibliography{references}

\end{document}